%% file: ems_arXiv.tex
\documentclass[apj]{emulateapj}

\usepackage{natbib, epsfig}

\shorttitle{FIRST-2MASS Red Quasars}

\newcommand{\ha}{H$\alpha$}
\newcommand{\hb}{H$\beta$}
\newcommand{\ebv}{$E(B-V)$}
\newcommand{\um}{$\mu$m}

\newcommand{\noprint}[1]{}

\begin{document}
\title {\bf FIRST-2MASS Red Quasars: Transitional Objects Emerging from the Dust}

\author{Eilat Glikman\altaffilmark{1}, Tanya Urrutia\altaffilmark{2}, Mark Lacy\altaffilmark{3}, S. George Djorgovski\altaffilmark{4}, Ashish Mahabal\altaffilmark{4}, Adam D. Myers\altaffilmark{5}, Nicholas P. Ross\altaffilmark{6}, Patrick Petitjean\altaffilmark{7}, Jian Ge\altaffilmark{8}, Donald P. Schneider\altaffilmark{9,10}, Donald G. York \altaffilmark{11}}
\email{eilat.glikman@yale.edu}

\altaffiltext{1}{Department of Physics and Yale Center for Astronomy and Astrophysics, Yale University, P.O. Box 208121, New Haven, CT 06520-8121; email: eilat.glikman@yale.edu}
\altaffiltext{2}{Leibniz Institut fŸr Astrophysik, An der Sternwarte 16, 14482 Potsdam, Germany}
\altaffiltext{3}{National Radio Astronomy Observatory, Charlottesville, VA}
\altaffiltext{4}{Astronomy Department, California Institute of Technology, Pasadena, CA, 91125}
\altaffiltext{5}{Department of Physics and Astronomy, University of Wyoming, Laramie, WY 82071, USA}
\altaffiltext{6}{Lawrence Berkeley National Laboratory, 1 Cyclotron Road, Berkeley, CA 92420, USA}
\altaffiltext{7}{Institut dÕAstrophysique de Paris, UMR7095 CNRS, UniversitŽ Pierre et Marie Curie, 98 bis bd Arago, 75014 Paris, France }
\altaffiltext{8}{Astronomy Department, University of Florida, 211 Bryant Space Science Center, PO Box 112055, Gainesville, FL 32611, USA}
\altaffiltext{9}{Department of Astronomy and Astrophysics, The Pennsylvania State University, University Park, PA 16802}
\altaffiltext{10}{Institute for Gravitation and the Cosmos, The Pennsylvania State University, University Park, PA 16802}
\altaffiltext{11}{Department of Astronomy and Astrophysics, University of Chicago, Chicago, IL 60637, USA}

\begin{abstract}
We present a sample of 120 dust-reddened quasars identified by matching radio sources detected at 1.4 GHz in the FIRST survey with the near-infrared 2MASS catalog and color-selecting red sources.  Optical and/or near-infrared spectroscopy provide broad wavelength sampling of their spectral energy distributions that we use to determine their reddening, characterized by \ebv.  We demonstrate that the reddening in these quasars is best-described by SMC-like dust. This sample spans a wide range in redshift and reddening ($0.1 \lesssim z \lesssim 3$, $0.1 \lesssim E(B-V) \lesssim 1.5$), which we use to investigate the possible correlation of luminosity with reddening.  At every redshift, dust-reddened quasars are intrinsically the most luminous quasars.  We interpret this result in the context of merger-driven quasar/galaxy co-evolution where these reddened quasars are revealing an emergent phase during which the heavily obscured quasar is shedding its cocoon of dust prior to becoming a ``normal''  blue quasar.  
When correcting for extinction, we find that, depending on how the parent population is defined, these red quasars make up $\lesssim 15-20\%$ of the luminous quasar population.  We estimate, based on the fraction of objects in this phase, that its duration is $15-20\%$ as long as the unobscured, blue quasar phase. 
\end{abstract}

\section {Introduction}

\defcitealias{Glikman04}{Paper~I}
\defcitealias{Glikman07}{Paper~II}

Quasars are among the most energetically powerful objects in the universe.  Their luminosities are indicative of how supermassive black holes (SMBHs; $10^6-10^9$ M$_\sun$) grow; namely, through accretion at or near the Eddington limit.   Many observations suggest that SMBH growth and the build-up of their host galaxies are closely linked.  SMBHs appear to be a ubiquitous feature at the centers of all large galaxies.  SMBH masses are proportional to the mass/velocity dispersion of their host spheroid \citep[the M$-\sigma$ relation;][]{Ferrarese00,Gebhardt00,Magorrian98}.  Furthermore, the cosmological evolution of the QSO luminosity density and of the star formation rate are remarkably similar \citep{Shaver96,Wall05}.  Theoretical models of cosmological galaxy evolution require feedback from the SMBH growth as an ingredient in order to reproduce the observed $K$-band number counts and galaxy luminosity function \citep{Kauffmann00,DiMatteo05,Croton06}.  

All of this information forms the basis for quasar-galaxy co-formation models. In these models, SMBHs and galaxies co-evolve through major mergers which fuel both a starburst and accretion onto the nuclear black hole \citep{Sanders88a,Hopkins05c,Hopkins06a}.  The models predict an obscured phase for the young quasar resulting from the large amounts of gas and dust funneled inwards to fuel the quasar.  A luminous blue quasar emerges at the end of this process, shining until it exhausts its fuel supply.  In this picture reddening is correlated with the evolutionary stage of the quasar.

Such scenarios have been invoked to explain the presence of buried AGN seen in ultraluminous infrared galaxies \citep[ULIRGSs;][]{Sanders88a}, a high fraction of which also show evidence of merging and interaction \citep{Sanders96}.  However, luminous, blue quasars show few signs of interaction; their hosts are mostly undisturbed galaxies \citep[e.g.,][but see \citealt{Bennert08}]{Dunlop03,Floyd04,Zakamska06}.  These seemingly conflicting observations suggest that dust that obscures buried quasars might be cleared during a transitional phase. Such transitional objects would represent a key link in the evolutionary path, where the dust that completely obscured the AGN hosted in the ULIRG is cleared, as signs of interaction dissipate, eventually to reveal an unobscured, luminous quasar. This missing link should be a population of highly reddened, but not completely obscured, quasars which have been largely missed by optical quasar surveys.

Obscuration in AGN is often discussed in the context of an orientation-based AGN-unification model \citep[c.f.,][]{Antonucci93,Urry95,Treister04}. This model provides an inherent description of accreting black holes as having an axisymmetric geometry with an accretion disk surrounded by photoionized clouds that give rise to broad emission lines.  Dusty, molecular clouds, often modeled as a torus co-planar with the accretion disk \citep[though the structure may be more complex, e.g.,][]{Elitzur08}, obscure the broad emission lines along certain lines-of-sight.  Farther out lie cooler photoionized clouds emitting narrow forbidden lines.  AGN seen along certain lines of sight intersect the obscuring torus and reveal only narrow-emission lines in their spectra.   

The unification scheme is well-established for low-luminosity AGN, and has been extended to so-called Type 2 (obscured) quasars as well \citep{Zakamska03, Zakamska04, Zakamska05}.  While Type 2 quasars are interesting in their own right, recent research confirms that these objects are not reddened as a result of recent merger and are not tied into the evolutionary picture of galaxy and quasar co-formation \citep{Sturm06, Zakamska06}.  
Observations of low-luminosity AGN selected in X-rays or the mid-infrared show a strong increase in the ratio of obscured/unobscured AGN with decreasing luminosity \citep[e.g.,][]{Treister08,Treister09}.\footnote{However, see \citet{Lawrence10} for a detailed discussion of possible departure from this trend.}
One possible interpretation of this is that the covering fraction from the obscuring material shrinks with increasing AGN luminosity \citep[the so-called ``receding torus''][]{Lawrence91}. 

In this paper, we make a distinction between reddened Type 1 objects (whose spectra show broad-line emission) and Type 2 AGN which have only narrow emission lines and are typically more heavily obscured.  We demonstrate that this distinction selects a unique population of objects whose physical mechanism for the reddening appears to differ from an orientation-based extinction.

Reddened quasars can be missed for several reasons: (1) dust extinction can dim sources below the detection limit of a given imaging survey, and (2) dust-reddening changes the color of a quasar, reddening ultraviolet excess (so-called UVX) selected quasars into the optical color locus of stars as a complex function of reddening and redshift -- particularly at redshifts of around $1.5 < z < 2.5$, where quasars are most abundant.  (3) Morphological selection of point sources excludes quasars whose rest-frame UV and optical emission is dimmed relative to their host galaxies, causing them to appear extended, especially in blue-band images.  To mitigate these obstacles, red quasars must be selected at wavelengths that are less sensitive to dust extinction (e.g., radio, infrared, or X-rays), using color cuts that include dust-reddened objects.

Small samples of red quasars have been found using various methods -- selection in the  radio \citep{Webster95,Gregg02,White03b,Urrutia09}, near-infrared  \citep{Cutri01,Maddox08,Georgakakis09}, mid-infrared \citep{Lacy04,Lacy07,Polletta06,Polletta08}, and  X-rays \citep{Kim99,Polletta06} -- with large variations in the estimated fraction of red quasars in the overall quasar population.  If red quasars represent the transitional phase between hidden accretion and unobscured radiation, then the relative abundance of red quasars constrains the duration of this transitional phase.  Furthermore, catching quasars in the act of this ``clearing out" may inform the underlying processes linking galaxy evolution to black hole growth.

In this paper we present a complete sample of 120 dust-reddened quasars, with well-understood selection criteria -- the largest to date.  We use this sample to constrain the fraction of reddened quasars.  We also study their reddening properties and the evolution of these properties with redshift and luminosity.  

The paper is organized as follows: we review the results from our previous work and motivate our color selection for finding red quasars in Section \ref{sec:finding}.  We describe the candidate selection process as well as followup spectroscopic observations and classification in Section \ref{sec:selecting}.  In Section \ref{sec:sample_properties}, we describe the properties and demographics of this red quasar population compared to optically-selected quasars and in particular UVX-selected quasars.  In Section \ref{sec:reddening}, we explore the nature of the reddening and its effect on estimating the fraction of reddened quasars.  Section \ref{sec:bal} concerns a unique subset of broad absorption line (BAL) quasars whose fraction is significantly higher among red quasars than in the overall quasar population.  Finally, in Section \ref{sec:evolution} we advocate for a scenario in which dust reddened quasars are a short-lived phase in a merger-driven co-evolution of galaxies and AGN.  In Section \ref{sec:summary} we summarize our findings and present ideas for resolving remaining open questions in the future.  Any time we present the results of cosmological calculations we use the following parameters: $H_0=70$ km s$^{-1}$ Mpc$^{-1}$, $\Omega_M=0.30$, and $\Omega_\Lambda=0.70$.

\section{Finding Red Quasars} \label{sec:finding}

\begin{figure}
\plotone{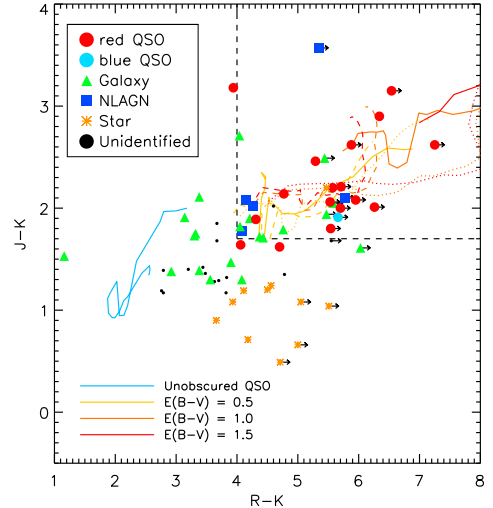}
\caption{An updated version of Figure 4 from \citetalias{Glikman04} showing the colors of FIRST radio sources with 2MASS matches and no optical counterpart in the POSS-I survey.  We find that the color cuts $R-K\ge4$ and $J-K\ge1.7$ select quasars with a 50\% efficiency, avoiding galaxies, which are bluer in $R-K$ and red dwarf stars (e.g., M stars), which are bluer in $J-K$.  We also plot color tracks for quasars at $z=0.1-2.5$ with different amounts of reddening (solid lines) and with the host galaxy contributing 50\% of the flux at 0.9-1\um.  The reddenings are defined in the lower left legend (and see text for details).  We also plot the effect of a host galaxy on the colors of reddened quasars for different amounts of reddening with an elliptical host galaxy (dotted line) and an Sc type galaxy (dashed line).  Although the presence of a star-forming host shifts the colors of reddened quasars blueward in $R-K$ color, especially for more heavily reddened nuclei, their colors remain within our selection box. }\label{fig:nncolor}
\end{figure}

In \citet[][hereafter Paper~I]{Glikman04} we sought to identify the most elusive members of the reddened quasar population by searching for objects that were detected in the radio and near-infrared, but were undetected in a flux-limited optical survey.  We matched the 2000 July Faint Images of the Radio Sky at Twenty Centimeters (FIRST) radio catalog \citep{White97} with the Second Incremental Point-Source catalog from the Two Micron All Sky Survey \citep[2MASS;][]{Skrutskie06} and selected objects that {\em lacked an optical counterpart} in the Automatic Plate Measuring (APM)\footnote{\url{\tt{http://www.ast.cam.ac.uk/$\sim$mike/apmcat/}}} machine scans of the first generation Palomar Observatory Sky Survey (POSS-I) plates.  Figure \ref{fig:nncolor} shows the 69 candidates selected in this effort; 54 objects are spectroscopically identified\footnote{\citetalias{Glikman04} reported 50 spectroscopic observations, subsequently, four additional spectra -- a galaxy and three M stars -- have become available through the Sloan Digital Sky Survey \citep[SDSS;][]{York00}} of which seventeen are red quasars.  Approximately $50\%$ of the objects defined by $R-K\geq4$ and $J-K\geq1.7$ are heavily obscured quasars. The other $\sim 50\%$ are galaxies exhibiting some form of activity such as a starburst or narrow-line AGN activity. Objects with bluer $J-K$ colors tend to be late-type stars; sources with $R-K<4$ tend to be elliptical galaxies due to the large 4000\AA\ break, but relatively flat spectral energy distribution (SED) redward.

To further explore this empirically-determined color selection, we compute the colors tracks of reddened quasars in $J-K$ vs. $R-K$ space in the redshift range $z=0.1-2.5$.  To do this, we combine the {\em Hubble Space Telescope} (HST) UV composite quasar template from \citet{Telfer02} with the optical-to-near-infrared composite spectrum from \citet{Glikman06} to produce a template spectrum that spans 300\AA\ to 3.5\um.  We convolve the template with $R$, $J$ and $K_s$ transmission curves to produce the solid blue line in Figure \ref{fig:nncolor}.  This curve is nearly identical to the quasar colors presented by \citep{Hewett06} in a study of the optical-through-near infrared colors of astrophysical objects in the United Kingdom Infrared Telescope (UKIRT) Infrared Deep Sky Survey \citep[UKIDSS;][]{Lawrence07} photometric system.  We then redden the template using a Small Magellanic Cloud (SMC) reddening law \citep[and see section \ref{ssec:reddening}]{Fitzpatrick99,Gordon98} and convolve the resultant spectra to produce the yellow, orange and red solid curves corresponding to $E(B-V) = 0.5, 1.0$ and 1.5, respectively.  We also explore the effect of host galaxy contribution on the colors of reddened quasars by adding an {\em un}reddened host galaxy template that contributes 50\% of the overall continuum between $0.9 - 1.0$\um\footnote{This assumption is based on the findings from \citet{Glikman06} that a $2L^*$ host galaxy contributes 6\% of the total continuum of an average quasar at these wavelength.  Here we increase the contribution to 50\% to provide a generous limit on the effect of the host.} on top of the reddened quasar.  The dotted lines show the effect of elliptical host galaxy, which are the typical hosts of low redshift quasars \citep[c.f.,]{Dunlop03,Floyd04}.  We also explore the effect of a host with younger stellar populations by adding an Sb-type galaxy (dashed line).  Both composite spectra are taken from the \citet{Mannucci01} optical-through-near-infrared galaxy template library.  We also explored the effect of 70\% host galaxy contribution, which we do not plot so as not to busy the figure.  In all cases, the models stay within our color cuts, including the models of quasar plus Sc host galaxy, which shift the color tracks toward bluer $R-K$ colors, but not by enough to remove the sources from our sample selection box.  

Having found an efficient color selection criterion for reddened quasars, we expanded our search in the same area of sky, but included optical detections.  We expanded the sample from \citetalias{Glikman04} by imposing the aforementioned color cuts on FIRST-2MASS sources but included objects {\em with} optical detections in the Guide Star Catalog II \citep[GSC-II;][]{Lasker08};  this selection addresses the fraction of quasars missed due to optical reddening \citep[][hereafter Paper~II]{Glikman07}.  The 2000 July FIRST catalog has 2716 deg$^2$ areal overlap with the 2nd 2MASS incremental release.  In this area we found 156 quasar candidates.  Spectroscopic identification of 77\% of these objects resulted in a sample of 52 red quasars.  We used this sample to study their reddening properties and estimate the overall fraction of quasars that are reddened by dust.  Because of the small size of that sample, as well as the shallow limit of the 2MASS survey compounded by the spectroscopic incompleteness at the faintest $K$ magnitudes, we could only estimate the fraction of red quasars for objects with {\it unreddened} $K\leq 14.0$.  We found that at this bright end, red quasars make up $56\pm16$\% of broad-line-emitting quasars.  

Taking advantage of the superior depth and image quality of the Sloan Digital Sky Survey \citep[SDSS;][]{York00}, \citet{Urrutia09} developed a red quasar sample using FIRST, 2MASS and the fifth data release of SDSS \cite[DR5;][]{Adelman-McCarthy07} with slightly modified color cuts from \citetalias{Glikman07}: $r' - K>5$ and $J-K>1.3$.  Because of the overlap of the areas and color-selection with \citetalias{Glikman07}, 28.7\% (35/122) of the candidates including 42\% (24/57) of the red quasars in \citet{Urrutia09} were previously identified spectroscopically in \citetalias{Glikman07}.  

In this paper we more than double our sample of red quasars using the full 9033 deg$^2$ overlap between the FIRST and 2MASS surveys.  This sample includes most of the objects in \citetalias{Glikman04}, \citetalias{Glikman07} and \citet{Urrutia09} and has better spectroscopic sampling at the faint end (based on 2MASS $K$ magnitudes) yielding a catalog which includes some of the most  luminous -- yet previously unidentified -- quasars in the Universe.   

\section{Selecting the FIRST-2MASS Red Quasar Sample} \label{sec:selecting}

Here we describe in detail the construction of our red quasar sample, which depends on three wavelength regimes: (1) candidate quasars are radio sources, to avoid confusion with stars; (2) we require that our quasars be bright at 2 \micron, since we are searching for red objects; (3) we use optical data to select objects with red optical-to-near-infrared colors.  In addition, on spectroscopic followup, we define a red quasar as an object whose spectrum shows at least one broad emission line, with $v \ge 1000$ km s$^{-1}$, and whose reddening is measured at $E(B-V)\ge0.1$.  

The FIRST Radio Survey \citep{Becker95} has mapped 9033 deg$^2$ of the sky at 20 cm with the Very Large Array (VLA) in the B-configuration.  Coverage includes $\sim 8400$ deg$^2$ in the north Galactic cap and $\sim 600$ deg$^2$ in the south Galactic cap. With the antennae in this configuration, the survey's resolution is roughly 5\arcsec\  and has 0\farcs5 positional accuracy.  The 3-minute snapshot integration time yields a typical rms of 0.15 mJy.  The 2003 April 11 catalog has $\sim 810,000$ sources brighter than the survey's detection threshold of 1 mJy \citep{White97}.

The 2MASS near-infrared survey \citep{Skrutskie06} used two telescopes, one at Mt. Hopkins, Arizona and one at Cerro Tololo, Chile, to image the sky in three bands, $J$ (1.24 \micron), $H$ (1.66 \micron), and $K_s$ (2.16 \micron), between 1997 and 2001.  This four-year effort produced a Point Source Catalog (PSC) with $\sim 4.70\times 10^8$ sources.  The survey is $99\%$ complete to a $10\sigma$ magnitude limit of $J=15.8$, $H=15.1$ $K=14.3$.  Fainter sources are included in the catalog (down to $K\lesssim 16$) but with lower signal-to-noise ratio and completeness.  The 2MASS survey has a resolution of 4\arcsec\ and an astrometric accuracy better than 0\farcs5.

The Guide Star Catalog II \citep[GSC-II][]{Lasker08} is an all-sky optical catalog produced by scanning the second-generation photographic Palomar Observatory \citep[POSS-II][]{Reid91} and UK Schmidt Sky Surveys at 1\arcsec\ resolution and generating a catalog with positions, magnitudes and morphological classifications.  We employed the GSC 2.2.1 catalog which reaches the POSS-II photographic plate limits of $F < 20.80$\footnote{Photographic $F$ band is equivalent to a $R$ magnitude, peaking at 6750 \AA.} and $J < 22.50$\footnote{Photographic $J$ band is equivalent to a $B$ magnitude, peaking at 5750 \AA.}.  The astrometric accuracy of GSC-II is better than 1\arcsec.  
 
We matched the 2003 April version of the FIRST catalog and the 2MASS All-Sky Point Source catalog using a 2\arcsec\ radius. In \citetalias{Glikman04} we showed that a 2\arcsec\ match between FIRST and 2MASS has a $0.3\%$ false detection rate and is the radius at which background contamination is minimized, while preserving completeness.  This criterion yielded 66,953 matches, including 66,728 unique FIRST sources -- a catalog we refer to henceforth as F2M.  We applied the infrared color cut ($J-K \geq 1.7$) to this list, which reduced the number of  candidates to 8443 sources. We matched these sources, using the FIRST positions, to the GSC-II catalog with a search radius of 2\arcsec.  There were 8242 matches within 2\arcsec\ of the FIRST positions in F2M.  Of these, 8075 are unique FIRST sources, which may have two or more GSC and/or 2MASS matches.  We kept only the closest match to each FIRST source in our candidate list. 

In order to apply the color cuts to the 8075 matches we need to take into account sources that have been detected in one optical band but not in another.  We replaced the undetected magnitudes for such objects with their respective plate limits, $B=22.50$ and $R=20.80$.  In cases where bright objects were detected on plates observed through the Photographic $J$ bandpass but were undetected in Photographic $F$, artificially red colors may have been assigned.  We examined finding charts of all 44 such sources and found that 8 were indeed too faint to be discerned on the Digitized Sky Survey images; we keep them in the candidate list.  We obtained magnitudes for the remaining 36 sources using the SDSS-DR8 catalog \citep{Aihara11} which contributed $r$ magnitudes -- 32 were too blue to remain in our candidate list.  This procedure found 347 FIRST-2MASS sources with GSC II matches within 2\arcsec\ obeying our color criteria. 

In Section \ref{ssec:matches} we fully quantify the matching statistics between F2M and GSC-II, but to finish this section we outline some broad trends.  Another 358 F2M sources had no match in GSC-II within 2\arcsec. We matched these sources to a larger search radius of 6\arcsec\ to find the closest counterpart. We examined the Digitized Sky Survey (DSS) images of these objects and removed 67 extended objects (e.g., saturated stars and large galaxies) whose FIRST and GSC-II coordinates were separated by more than 6\arcsec, but which were clearly associated with the FIRST source.  Thirty-six sources had no match within 6\arcsec, showing only blank sky on the DSS image; eleven of these objects were previously reported in \citetalias{Glikman04}.  We include these sources in our candidate list, adopting the plate limits as lower limits to their $B$ and $R$ magnitudes.  

We examined the DSS images of the 247 objects that had GSC II matches within 6\arcsec\ but not within 2\arcsec, to search for objects whose real optical counterparts may be fainter than the POSS-II plate limit, but which may have had chance coincidence matches to another source within 6\arcsec\ of their FIRST positions.  Many of these objects were large galaxies or blended groups of sources that overlapped the radio and infrared positions but whose centroids implied optical positions separated by 2\arcsec\ and 6\arcsec. We removed these sources and retained the remaining 94 optical objects which appeared isolated.  We describe below the matching statistics between F2M and GSC-II which we use to determine the reliability that a F2M - GSC-II association is real.

\subsection{F2M Matches to GSC-II sources} \label{ssec:matches}

\begin{figure}
\epsscale{0.75}
\plotone{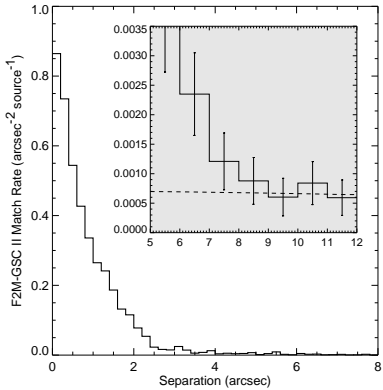}
\caption{Histogram of separations for F2M-GSC II matches, binned by 0\farcs2 and normalized by the annular area of each bin and number of F2M sources (8443).  Only the closest matches are included.  The inset shows the chance coincidence rate ({\em dashed line}) at large radii.}\label{fig:sephist}
\end{figure}

To determine the completeness and contamination statistics of our sample,  we follow the procedure of \citet{McMahon02} in determining the chance coincidence rates as a function of separation between F2M and GSC II.  We matched our 8443 F2M sources to GSC II out to 12\arcsec, keeping only the nearest match (there were 8354 unique matches).  To model the background coincidence rate, we shifted the 8443 F2M positions by 5\arcmin\ to the north and rematched this list to GSC-II, again out to 12\arcsec.  We determine the mean source density, $\bar{\rho}_{\mathrm{opt}}= 7.09\times 10^{-4}$ arcsec$^{-2}$, and the variance in the density, $\sigma^2_{\mathrm{opt}}=1.45\times 10^{-7}$ arcsec$^{-4}$, using the formalism outlined in \S5 of \citet{McMahon02}.  Figure \ref{fig:sephist} shows the distribution of match separations plotted in 0\farcs2 bins.  Integrating under the histogram of real matches and the chance coincidence curve to 2\arcsec\ determines that $99.8\%$ of these matches are physically associated; there are 8075 real matches and 14 random associations. This result implies that of the 343 objects in our candidate list, 0.59 (or $\sim 1$) objects are chance coincidences. Integrating between 2\arcsec\ and 12\arcsec\ implies that 279 real and 35 coincidence matches would be added, suggesting our completeness is $95.6\%$ (8075/8443).  

\begin{figure*}
\epsscale{0.75}
\plottwo{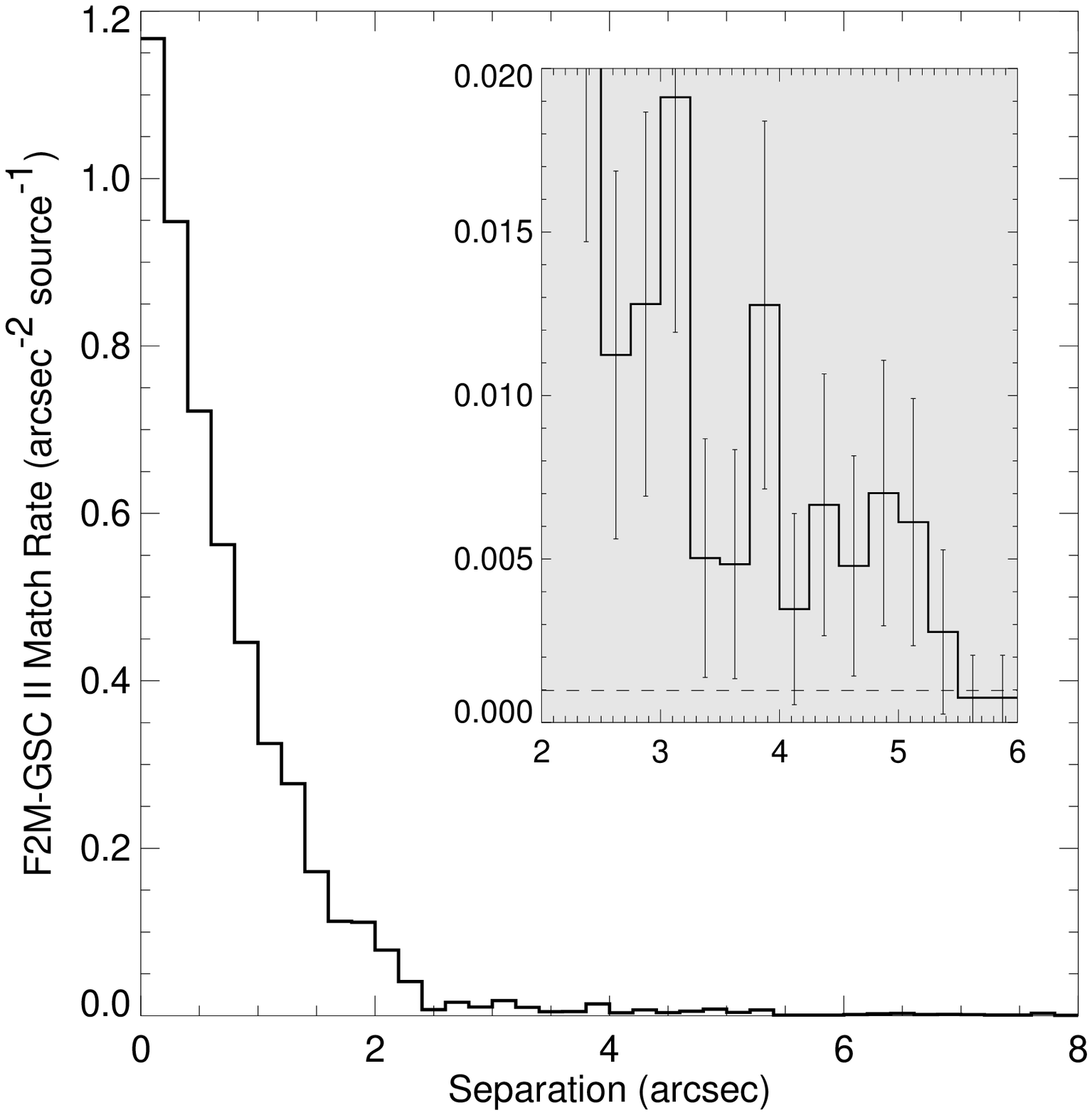}{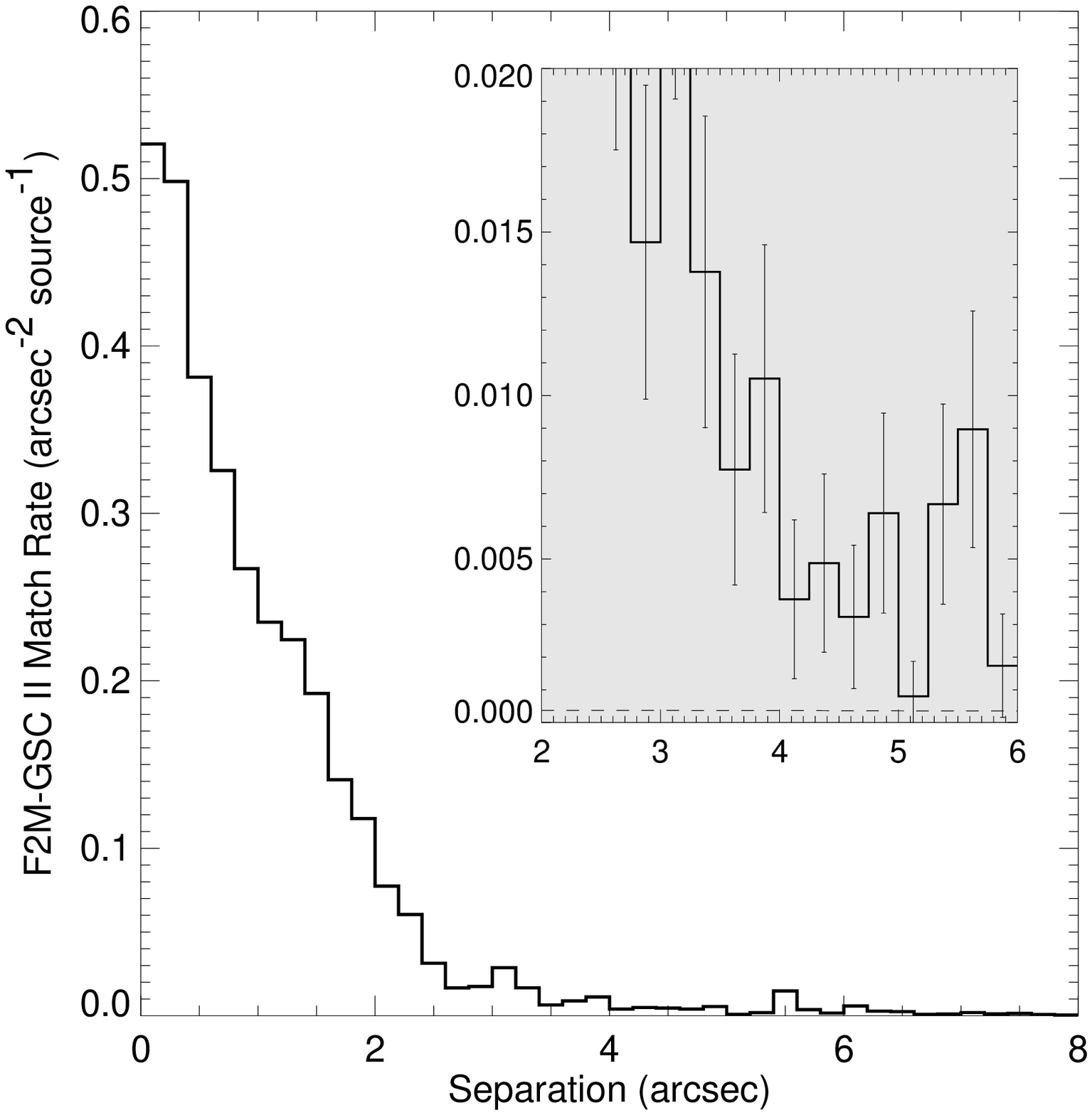}
\caption{Left -- radio point source matches as a function of separation, binned by 0\farcs2.  Right -- extended radio source matches as a function of separation.  Only the closest matches are included.  The inset shows the chance coincidence rate ({\em dashed line}) between 2\arcsec\ and 6\arcsec.
}\label{fig:sephist_pt_ext}
\end{figure*}

Figure \ref{fig:sephist_pt_ext} shows the separation distribution for pointlike radio morphologies (left), and extended radio morphologies (right).  We define ``pointlike'' to be FIRST sources with $S_{Peak}/S_{Integrated}>0.9$.  The inset shows the chance coincidence rate (dashed line) compared to the matched sources.  These figures demonstrate that extended radio sources are more likely to have real matches at larger radii than point sources.  We use these results to determine the probability of chance coincidence for each of the 94 sources whose nearest GSC II match lies between 2\arcsec\ - 6\arcsec\ away.  We ascribe a probability to each source for being a chance coincidence based on its radio morphology and separation from the FIRST position.  Adding the probabilities of the first ten objects (sorted from lowest to highest probability) in this list sum to $891\%$, or $\sim 9$ chance coincidences.  We assume that these objects are {\it unmatched} sources, and so we adopt the POSS-II plate limit for their $B$ and $R$ magnitudes.

\begin{figure}
\begin{center}
\includegraphics[scale=.6,angle=0]{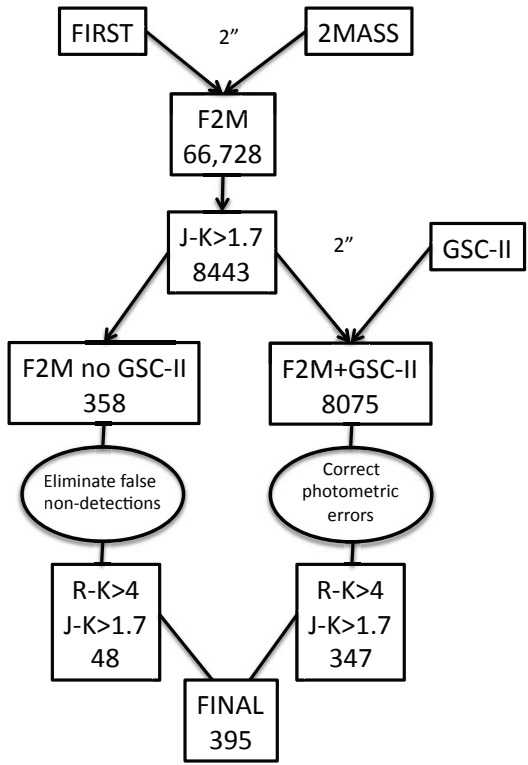}
\caption{Schematic diagram of our selection process.}\label{fig:flowchart}
\end{center}
\end{figure}

The final candidate list, therefore, contains 395 sources, 347 of which have detections in the GSC-II catalog, and 48 of which are fainter than the sensitivity limit of GSC-II.  This list includes 31 objects from \citetalias{Glikman04} that had no detection in the APM catalog and have $J-K\geq1.7$ and $R-K\geq 4$.  A flowchart of our selection process is shown in Figure \ref{fig:flowchart}, and the list of quasar candidates and their attributes is provided in Table \ref{tab:candidates}.

\subsection{Spectroscopic Observations}

Spectroscopic observations of our red quasar candidates were performed in the optical and/or near-infrared over the timespan of ten years with additional spectra from archival and published sources in the literature.  In total we  obtained spectroscopy for 316 red quasar candidates.  The nature of these spectra is highly heterogeneous, and we use the spectra for two main purposes (1) object identification (i.e., galaxy, quasar) and redshift determination; (2) for the quasars, reddening determination (see \S 5.1 below and \S 5 of \citetalias{Glikman07}).
  
A total of 209 optical spectra were obtained at the W. M. Keck Observatory with the Echellette Spectrograph and Imager \citep[ESI;][]{Sheinis02} and the Low Resolution Imaging Spectrograph \citep[LRIS;][]{Oke95}, the Palomar observatory's 5-meter Hale telescope with the Double Spectrograph, and the Lick Observatory's 3m Shane telescope with the Kast spectrograph.  Additional spectra were taken from the SDSS and SDSS-III archives \citep{Gunn98,Gunn06,Eisenstein11}, including the Baryon Oscillation Spectroscopic Survey quasar survey \citep[BOSS;][]{Ross11}.   Column (14) of Table \ref{tab:candidates} lists the origin of the optical spectroscopy.

The 170 near-infrared spectra (covering the wavelength range $\sim 0.9 - 2.5\ \mu$m) were obtained at the NASA IRTF with SpeX \citep{Rayner03}, the Palomar 5-meter Hale with TripleSpec \citep{Herter08} or with TIFKAM on the 2.4 m MDM telescope.  We reduced the SpeX and TripleSpec data using the Spextool software  which is designed for cross-dispersed near-infrared spectroscopy \citep{Cushing04}.  The procedure includes flat fielding, sky subtraction, extraction and co-addition of individual exposures, and wavelength calibration.  The spectra are also corrected for telluric absorption using the spectra of nearby A0V stars that were obtained directly before or after each target observation \citep{Vacca03}.  Column (15) lists the origin of the near-infrared spectroscopy.  

A total of 80 objects have both an optical and a near-infrared spectrum.  Having both spectral regions is especially useful in cases where the nature of an object is not readily identifiable.  For quasars, a near-infrared spectrum enables identification of heavily dust-obscured objects whose rest-frame ultraviolet emission lines are beyond detection due to extinction.  On the other hand, the optical spectrum better-constrains our estimates of $E(B-V)$, since the rest-frame UV is more sensitive to dust than longer wavelength light.  For galaxies, a near-infrared spectrum has fewer identifiable features than in the optical and is often difficult to classify.  Thirty-four candidates with only a near-infrared spectrum could not be classified because they lacked strong emission lines or other features.  

\subsection{Spectroscopic Identifications}

\begin{figure}
\plotone{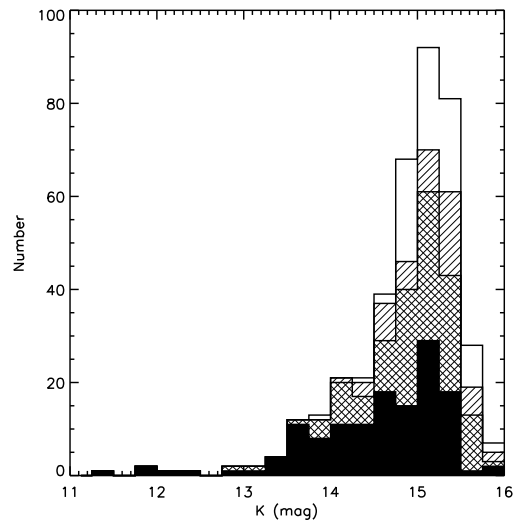}
\caption{Histogram of the $K$ magnitudes for the F2M candidates ($unshaded$).  All spectroscopically observed objects are overplotted in the forward-hashed areas.  Objects whose spectrum yielded an identification are overplotted in the cross-hashed histogram, and the quasars are overplotted in black. Our overall spectroscopic completeness is 80\%, but rises to 96.6\% for $K\le 14.75$. }\label{fig:khist}
\end{figure}

Of the 316 objects with spectra (out of 395 candidates) we have been able to identify 263, thus our survey is  80\% spectroscopically complete (67\% when considering only identified spectra).  The distribution of  spectroscopic completeness as a function of $K$ magnitude is shown in Figure \ref{fig:khist}.  The distribution of all F2M candidates is shown in the unshaded histogram.  Overplotted are histograms of spectroscopically observed objects (forward hash), objects with spectra whose type is identified (cross hash), and quasars (black) found in our survey.  We are 96.6\% spectroscopically complete for candidates brighter than $K=14.75$; our completeness drops to 72.8\% for $K>14.75$.
 
\begin{figure}
\plotone{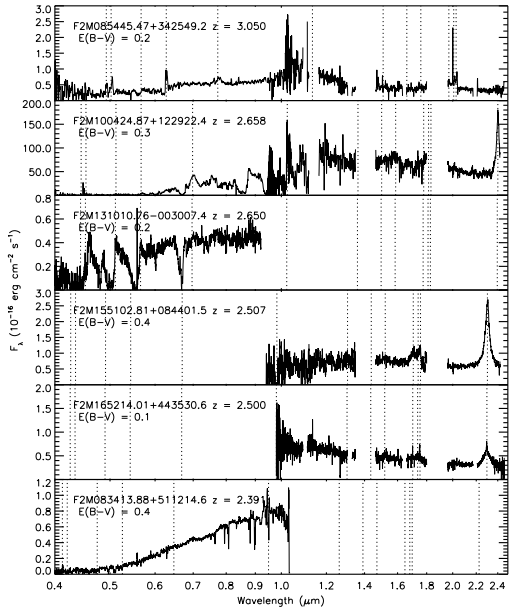}
\caption{F2M candidates identified as red quasars, having at least one broad emission line and $E(B-V)\ge0.1$, ordered by redshift.  The dotted lines show expected positions (in $\mu$m) of prominent emission lines in the optical and near-infrared: Ly$\alpha$~1216,
N~V~1240,
Si~IV~1400,
C~IV~1550,
C~III]~1909,
Mg~II~2800,
[O~II]~3727,
H$\delta$~4102,
H$\gamma$~4341,
H$\beta$~4862,
[O~III]~4959,
[O~III]~5007,
H$\alpha$~6563,
Pa$\gamma$~10941,
Pa$\beta$~12822 and
Pa$\alpha$~18756\AA. The full atlas is presented at the end of the paper. }\label{fig:spectra}
\end{figure}

There are 135 spectroscopically confirmed quasars in our survey, defined as having at least one broad emission line, with a width of $v\ge 1000$ km s$^{-1}$ based on a Gaussian fit.  We use the emission lines to assign a redshift to each quasar.  Three additional candidates are identified as quasars in the literature; we list their references in Table \ref{tab:candidates}.  We present the spectra of  120 red quasars, defined as having $E(B-V)\ge 0.1$\footnote{The determination of $E(B-V)$ is outlined in Section \ref{ssec:reddening}  of this paper as well as \S 5 of \citetalias[][]{Glikman07}.} in Figure \ref{fig:spectra} in order of decreasing redshift.  

 As noted above, as well as in \citetalias{Glikman04}, \citetalias{Glikman07} and, e.g., \citet{Rawlings95}, reddened objects may only reveal narrow lines in their optical spectra (which show rest-frame UV emission, depending on the redshift), while a near-infrared spectrum may be needed to reveal broad emission lines confirming the presence of a quasar.   For example, nearly all of the red quasars in \citet{Urrutia09} were identified from optical spectra.  F2MJ104043.66+593409.55 is listed as a narrow-line AGN (NLAGN) at $z=0.147$ in Table 1 of \citet{Urrutia09}; however, our near-infrared spectroscopy of this object reveals strong, broad Paschen $\alpha$ 18756\AA, Pa$\beta$ 12822\AA, and Pa$\gamma$ 10941\AA\ atop a very red continuum, confirming that it is a Type 1 quasar.  In fact, of the eleven objects labeled NLAGN based only on optical spectroscopy in \citet{Urrutia09}, six are re-classified as quasars in this work based on near-infrared spectroscopy.  

This strongly suggests that many of the objects classified here as NLAGN from only optical spectra will reveal broad emission in the near-infrared and argues for near-infrared spectroscopy of {\em all} NLAGN with $J-K\gtrsim1.7$ to recover any missed red quasars and to better determine the fraction of quasars that show no broad lines in the optical.   On the other hand, there are five red quasars in the \citet{Urrutia09} sample that lie outside our color cuts, with $1.3 < J-K < 1.7$.  This amounts to $\sim 9\%$ of that sample (5/56), which, if extended to our sample, suggests that we are missing $\sim 11$ red quasars because of our $J-K$ color cut.

We also find nine BL Lac objects: highly variable, featureless blue objects with bright radio flux densities ($\gtrsim 100$ mJy).  Two of these AGNs are repeats from \citetalias{Glikman07}.  The remaining seven have been identified as BL Lacs elsewhere; references are listed in column (16) of Table \ref{tab:candidates}.  As demonstrated in \S8 of \citetalias{Glikman07}, these blue objects are selected in a survey targeting red objects because of significant variability between the epochs of optical and infrared observations, where the former detected the source in a faint state and the latter was viewed with the object in a bright state.  We exclude these objects in our analysis for the remainder of this paper.

\begin{figure}
\plotone{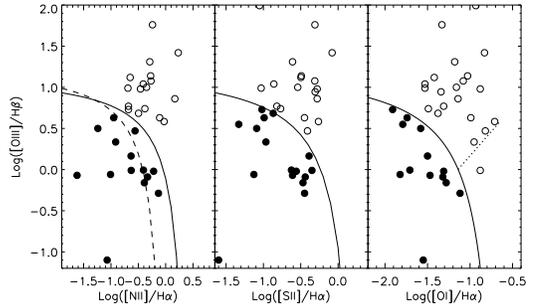}
\caption{Line diagnostics for the narrow-line-emitting spectra, plotted based on the classifications of \citet{Kewley06}. The pure-star formation boundary is shown with a dashed line in the leftmost panel, while the Seyfert-LINER divide is shown with a dashed line in the rightmost panel; LINERs have lower [\ion{O}{3}]/H$\beta$ values than Seyferts.
The left panel shows the [\ion{O}{3}]/H$\beta$ vs. [\ion{N}{2}]/H$\alpha$ diagnostic. In this panel, objects below and to the left of the dotted line are pure starbursts. The middle panel shows the [\ion{O}{3}]/H$\beta$ vs. [\ion{S}{2}]/H$\alpha$ diagnostic. The right panel shows the [\ion{O}{3}]/H$\beta$ vs. [\ion{O}{1}]/H$\alpha$  diagnostic. Objects to the left and below the solid line are starburst dominated (filled circles), while objects above and to the right of the line are AGN dominated (open circles).
}\label{fig:bpt}
\end{figure}

Fifty-two objects reveal narrow line emission in their spectra and are identified as either AGN and/or star-forming galaxies.  For objects with an optical spectrum, we model the H$\beta$, [\ion{O}{3}], [\ion{O}{1}], [\ion{N}{2}], H$\alpha$, and [\ion{S}{2}] lines with Gaussian profiles and plot their line ratios on Baldwin-Phillips-Terlevich \citep[BPT;][]{Baldwin81} line-ratio diagrams to determine the source of the ionizing flux (AGN, star formation, composite object showing contribution from both AGN and star formation) for each spectrum (Figure \ref{fig:bpt}).  This includes seventeen objects from \citetalias{Glikman07}.  Three narrow-line emitters, e.g., F2MJ104902.95+401031.6 at $z=0.715$, only have spectroscopic coverage of H$\beta$ and [\ion{O}{3}] from an optical spectrum.  With $\log$([\ion{O}{3}]/H$\beta$)=0.75, the nature of this object's emission is not clear, but it is likely to have some contribution from star formation and is probably a composite object.  We list the line diagnostics from our measurements in Table \ref{tab:diagnostics}, including, for completeness, the sources already analyzed in \citetalias{Glikman07}.

Thirteen of the narrow-line-emitting objects have only a near-infrared spectrum.  We cannot distinguish between their source of ionization based only on narrow Paschen lines, for which there are no BPT-diagram-like diagnostics.  However, the presence of bright radio emission suggests that they may host AGN.

Fifty-six of our objects are galaxies with no obvious emission lines.  We determine their redshifts by cross correlating their spectra with elliptical and S0 galaxy templates from \citet{Kinney96}, following the method of \citet{Tonry79}.  

There are five stars in our sample, which is a fraction consistent with the previous F2M red quasar samples.  We assign stars a redshift of 0 in Table \ref{tab:candidates}.  

\section{The Observed Properties of Red Quasars} \label{sec:sample_properties}

\subsection{Comparison Samples} \label{sec:fbqs}

Our final F2M red quasar sample consists of 120 objects spanning $z\simeq 0.1 - 3$.  In contrast to optically-selected quasars, these objects have optical-to-near-infrared SEDs that rise toward longer wavelengths.  Are we simply finding the tail end of an intrinsic distribution of quasar slopes \citep[e.g., the red quasars discussed in][]{Richards03} or axisymmetrically distributed nuclear dust \citep[e.g., type 1.5 quasars presented in][]{Smith02} or do these heavily reddened quasars represent a separate population of objects, possibly associated with a phase in quasar evolution \citep[i.e., the ``blowout'' phase][]{Hopkins08}?  To address this question, we must compare the F2M quasars with an appropriate optically-selected radio-detected quasar sample.

The most natural comparison set for the F2M red quasars is the optically-selected FIRST Bright Quasar Survey \citep[FBQS;][]{Gregg96}.   This survey searched for quasars in matches between FIRST radio sources and sources in the APM scans of the first generation Palomar all sky survey (POSS-I).  The main survey contains 636 quasars over 2682 deg$^2$ and imposed an $E=17.8$ magnitude limit on the APM sources as well as an $O-E \leq 2$ color criterion \citep[FBQS II;][]{White00}.  A deeper segment of the survey covered 589 deg$^2$ and found 321 quasars with the same color cut, but with $E\leq18.9$ \citep[FBQS III;][]{Becker01}.  Isolating the FBQS quasars that are detected in 2MASS forms an optically-selected sample of quasars that exist in the $K$-band flux-limited survey; there are 503 and 134 2MASS matches to FBQS II and III, respectively. 

We also construct a subsample of quasars from the SDSS \citep{Schneider10} that are detected in both FIRST and 2MASS to compare the efficiency of the SDSS quasar selection, which is more sensitive to reddened objects than simple UVX color selection because it includes FIRST-targeted sources as well as outliers from the stellar locus \citep{Richards02}.   We estimate the effective area for radio-detected SDSS quasars to be the area of the 2003 April 11 FIRST catalog release (9033 deg$^2$) minus $\sim 300$ deg$^2$ which appear to be lacking SDSS coverage\footnote{We estimate this area by computing the spherical area of this lune from its four corners.}.   We then apply the 77.4\% spectroscopic completeness \citep{Richards06} to this area for an effective area of 6770 deg$^2$.

\subsection{Observed Surface Density}

\begin{figure}
\plotone{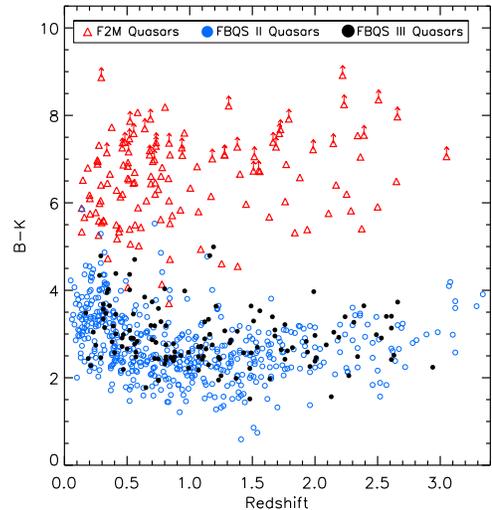}
\caption{Redshift distribution of as a function of $B-K$ color of F2M quasars ({\em red triangles}) compared with FBQS II ({\em blue open circles}) and III ({\em filled circles}).  The F2M and FBQS surveys reach the same redshift ranges, making them good comparison samples.  }\label{fig:z_dist}
\end{figure}

\begin{figure}
\plotone{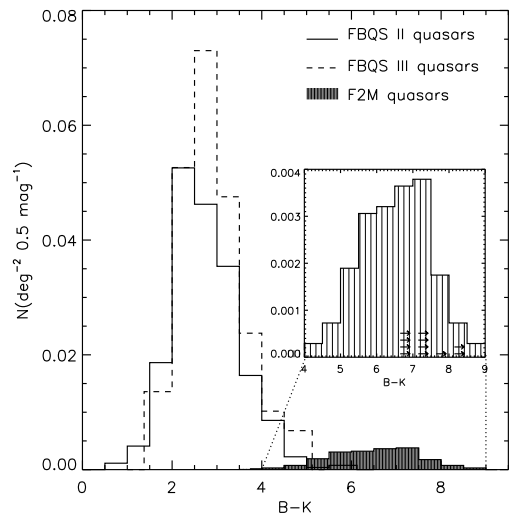}
\caption{$B-K$ distribution of F2M quasars compared with FBQS II and III; arrows denote bins which are lower limits.  The distribution of the optically selected (FBQS) samples peak near $B-K\sim 2.5$, though the distribution of  FBQS-III (dsahed line), which goes 1.1 magnitudes deeper than FBQS-II, is slightly shifted redward demonstrating that a fainter optical flux limit will picks up some redder sources.  The F2M quasars appear to have a separate distribution, and not just an extension of the red tail of FBQS quasars, suggesting that these objects are not merely a tail end of the intrinsic distribution of quasar colors. }\label{fig:bkhist}
\end{figure}

Figure \ref{fig:z_dist} plots $B-K$ color versus redshift for F2M and FBQS quasars and shows that the quasars found in the F2M survey span the same redshift range as FBQS quasars.  In Figure \ref{fig:bkhist} we plot a histogram showing the distribution of the $B-K$ color for FBQS II and III (solid and dashed line histograms, respectively) next to the same for the F2M quasars (shaded histogram).  The histograms have been normalized to each survey's coverage area for ease of comparison. \citet{Webster95} and \citetalias{Glikman04} showed that optically-selected quasars have typical $B-K\simeq 2.5$ color, which is not a strong function of redshift, as can be seen in Figure \ref{fig:z_dist}.   Therefore $B-K$ color can be used as a rough proxy for reddening.  

Figure \ref{fig:bkhist} shows the distribution of the $B-K$ colors of FBQS and F2M quasars.  The overlap between the two samples is small, by construction, but it appears that the red quasar color distribution (shown in more detail in the inset) does not simply fill in the tail end of the optically selected $B-K$ color distribution.  This would be expected if there were no reddening in quasars; the colors would then represent the intrinsic variation of their optical continuum slopes.    A similar observation was made by \citet{Richards03} who examined the distribution of the relative color\footnote{Defined as the residual color of a quasar at a given redshift after subtracting the median quasar color at the same redshift \citep{Richards01}, e.g., $\Delta(g-i)$ and correcting for Galactic extinction.} of SDSS quasars.  In their Figure 3, \citet{Richards03} note that there is an excess of sources with red $\Delta(g-i)$ colors compared to a Gaussian fit to the distribution defined by its peak and blue wing.  They conclude that that there are some quasars that are intrinsically red because their continuum has a flatter spectral index, but exist within the normal distribution of spectral indices for quasars.  The quasars far in the tail of the $\Delta(g^*-i^*)$, however, are {\em reddened} by dust.  We address this in the context of red quasars representing a phase of quasar evolution in \S \ref{sec:evolution}.

\begin{figure}
\plotone{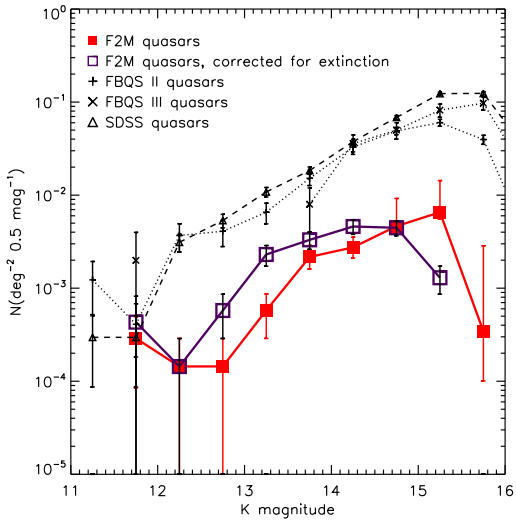}
\caption{Spatial density of quasars on the sky of F2M red quasars (filled red squared) compared with FBQS II and III (dotted lines) and SDSS quasars with FIRST detections \citep[dashed line;][]{Schneider10}. 
Radio plus optical selection for quasars misses $11\pm2\%$ of quasars because of reddening.
The open violet squares show the spatial density F2M red quasars after correcting for $K$-band absorption. Beyond $K=14.5$ incompleteness from the 2MASS flux limit begins to become apparent.  Comparing the space distributions of F2M quasars with FBQS we find that red quasars make up $21\pm2\%$ of radio-selected quasars with $K\leq14.5$.  }\label{fig:space_dist}
\end{figure}

Figure \ref{fig:space_dist} shows the observed number counts of the F2M quasars (red squares and line) compared with the number counts of FBQS quasars (plus sign for FBQS II and ``$\times$'' symbol for FBQS III connected by a dotted line).  Table \ref{tab:surfdens} lists the number counts for the F2M quasars plotted in Figure \ref{fig:space_dist}.  To determine the fraction of quasars that are red, we integrate the F2M curve and an average of the FBQS curves out to $K=15.5$ and compute their ratios:
\begin{equation}
\frac{N(K \le 15.5)_{\rm F2M}}{N(K \le 15.5)_{\rm FBQS}} \times 100.
\end{equation}
Comparing the total areal densities, we find that FBQS quasars miss $11\pm2\%$ of radio-selected quasars because of their red colors.  

We also plot the number counts of the radio-detected SDSS quasars described above with a dashed line.  Since many F2M quasars had SDSS spectroscopy, there is considerable overlap between the samples.  Therefore, the fraction of red quasars missed by SDSS is smaller. Despite this, \citet{Urrutia09} showed that $\sim 60\%$ of the red quasars in their sample are not selected for the main $z<3$ sample \citep[e.g., see][]{Richards02,Richards04}.  Furthermore, the completeness of SDSS drops significantly in redshift ranges that correspond to colors overlapping the stellar locus, which is most pronounced for $2.7 \lesssim z \lesssim 3.2$ -- at the edge of our survey -- and $1.5 \lesssim z \lesssim 3.2$ for Broad Absorption Line (BAL) quasars with large absorption troughs (we elaborate more on this in Section \ref{sec:bal}).  For the F2M quasars in this work, we find that 89 of our candidates have SDSS spectra.\footnote{Note that we list SDSS as the main source of spectroscopy for only 61 candidates in Table \ref{tab:candidates} because we only list SDSS as the source of the optical spectroscopy in the absence of superior quality spectra from our own observations.}  

It is the targeting of FIRST matches in SDSS that increases the probability of a spectroscopic selection in SDSS \citep{Richards02}.  This feature means that there is a radio-dependent bias of red quasar spectra in the SDSS spectroscopic database.  Radio-quiet red quasars (which fall below the FIRST detection threshold) would therefore be less likely to be spectroscopically targeted by SDSS, resulting in a large missed fraction of red quasars in the analyses of quasar properties (e.g., luminosity functions, clustering) determined from this, and other optically-selected samples.  

\subsection{Radio Properties}

Since the quasars in this survey are selected in the radio, we examine their radio properties to determine if radio-selection introduces a bias.  Although quasars were originally discovered as optical counterparts to radio-sources \citep{Schmidt63} it was not long before it was realized that only $\sim 10\%$ of quasars are strong radio emitters \citep{Sandage65}.  The radio properties of quasars are often described using the radio-loudness parameter defined as the ratio of radio to optical power \citep[$R\equiv f({\rm 1.4 GHz})/f(B_{\rm int})$;][]{Stocke92}.   The traditional definition of a ``radio-loud'' quasar is one having $\sim R^* \ge 10$, where $R^*$ is the $K$-corrected radio-loudness parameter.  The majority of FBQS quasars exist in $3 \lesssim R \lesssim 100$ and span a ``radio intermediate'' regime \citep{White00}.  

In \citetalias{Glikman04} we computed the radio-loudness parameter for the seventeen F2M quasars in that sample using the extinction corrected $B$-band magnitude, $B_{\rm int}$.  We found that these sources also lay in the radio intermediate regime, with only two quasars having $R>100$.

Here we examine the radio-loudness for the full F2M quasar sample.  We use a more current definition of the radio-loudness parameter, as defined by \citet{Ivezic02}, which uses the difference between the magnitude equivalent of the FIRST flux density ($t = -2.5 \log(F_{\rm int}/3631{\rm Jy})$) and the SDSS magnitude in a given bandpass, $m$, both on the AB magnitude system:
\begin{equation}
R_m = \log(F_{\rm radio}/F_{\rm optical}) = 0.4(m-t). \label{eqn:rl}
\end{equation}
We use the $g$-band, which is the closest to the previously-used $B$-band, for this calculation.  This definition of $R_m$ does not include a $K$-correction, which avoids adding uncertainties introduced by the unknown radio spectral indices of these FIRST-selected sources, as well as the already-uncertain reddening corrections (see Section \ref{sec:reddening}) which would be compounded by the uncertainty of adding a $K$-correction to the optical magnitude (see \citealt{Glikman11} and \citealt{Croom09} for discussions of the uncertainties introduced by $K$-corrections in quasars).

\begin{figure*}
\plottwo{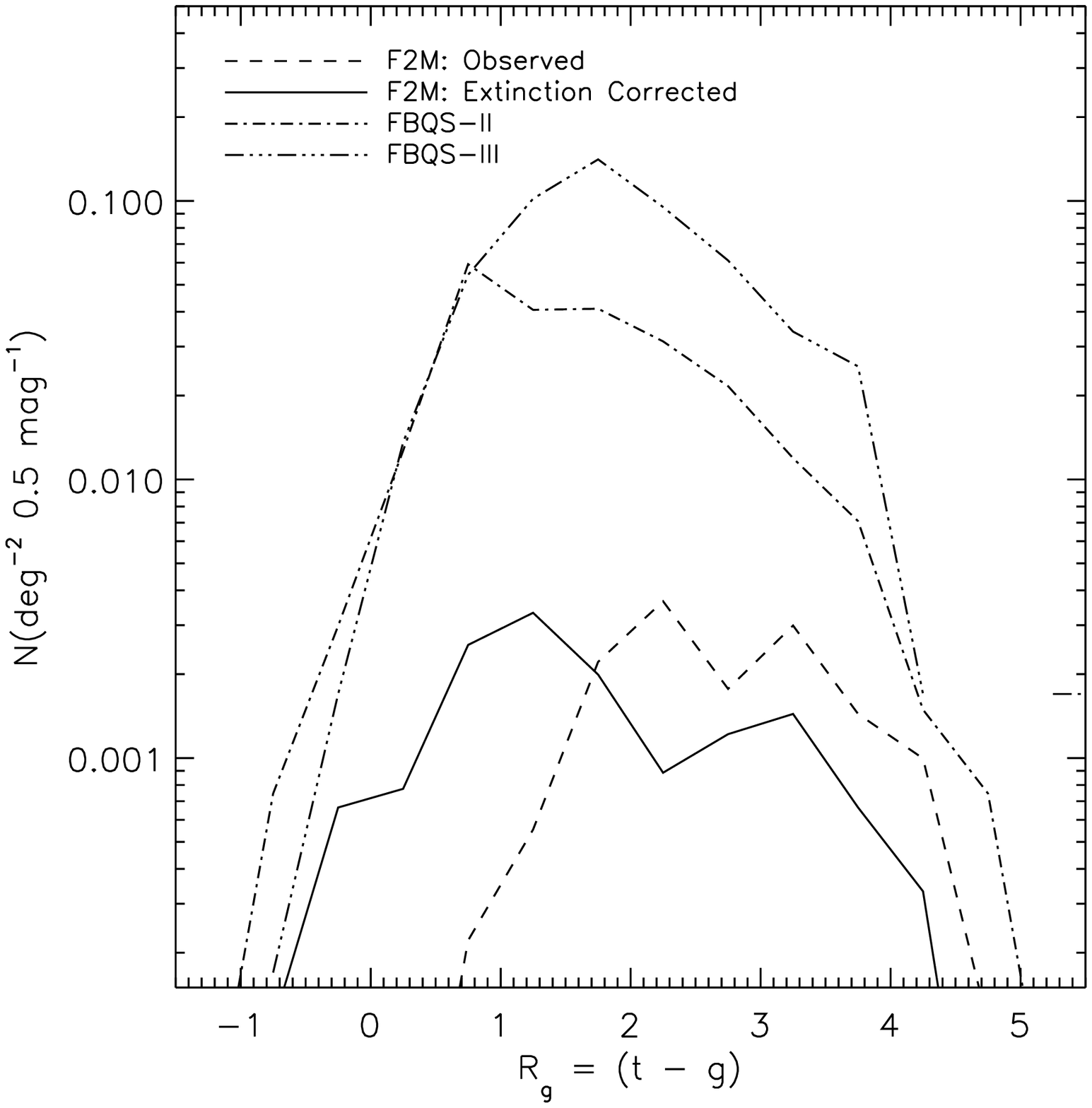}{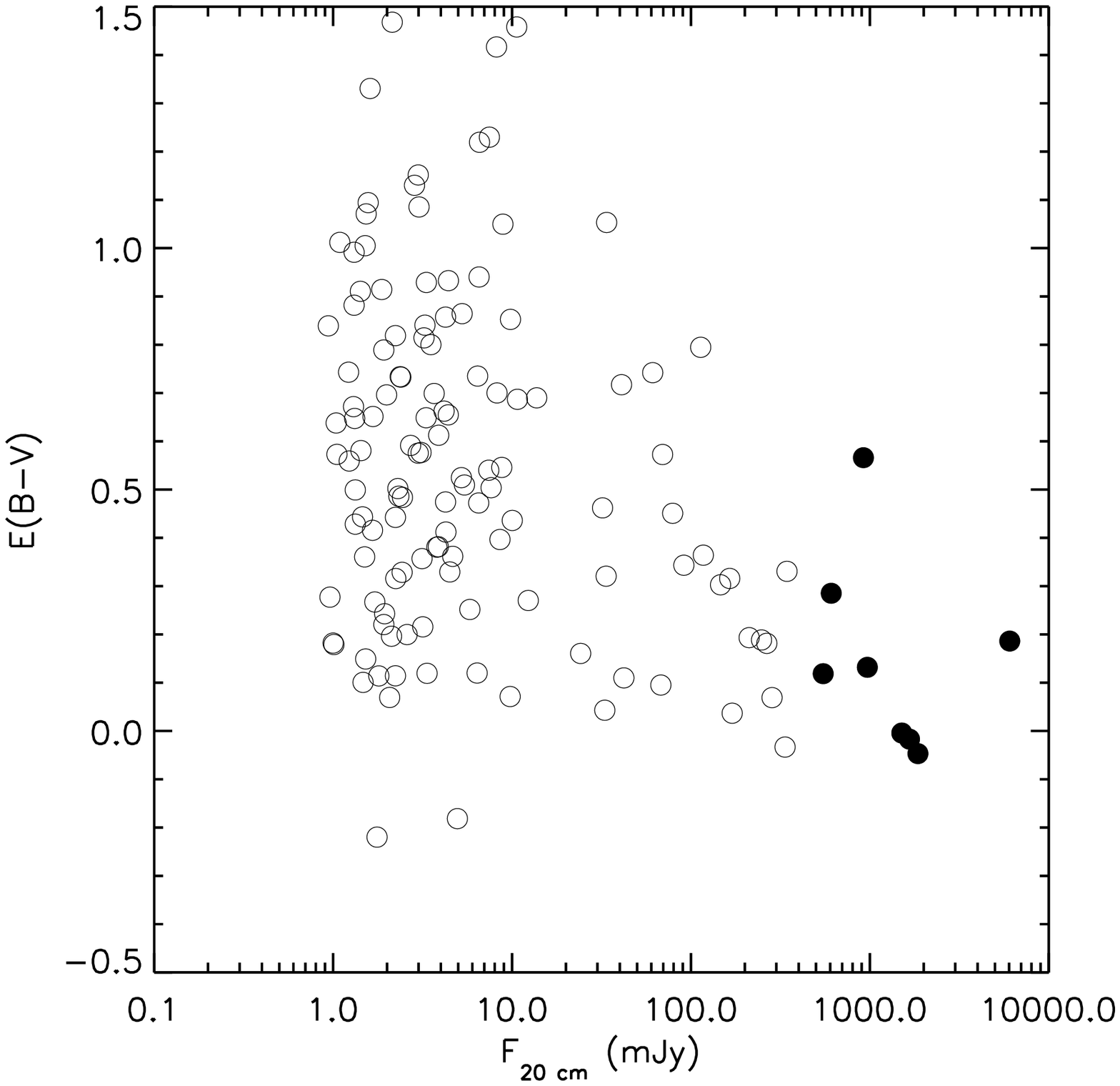}
\caption{{\em Left -- } Radio loudness distribution for F2M quasars based on observed SDSS $g$-band magnitudes (dashed line) compared with FBQS quasars (dot-dahsed line), as defined in Equation \ref{eqn:rl}.  Both distributions have been normalized by their respective survey areas.  Based on the observed optical magnitudes, which are dimmed because of extinction, the F2M quasars appear to be radio ``louder''  than the FBQS. Once we correct the $g$-band for extinction, however, the radio-loudness for F2M quasars decreases and appears to be distributed similarly to the FBQS quasars. {\em Right--} A plot of reddening \ebv vs.\ FIRST 20 cm flux density  shows that the most radio bright quasars (filled circles have $F_{20\rm~cm} > 500$ mJy) are the least reddened. This suggests that a red synchrotron component is unlikely to be the cause of the red colors of F2M quasars.}\label{fig:rl}
\end{figure*}

Figure \ref{fig:rl}, left, shows the distribution of $R_g$ for the F2M red quasars compared with FBQS quasars.  The dashed line shows $R_g$ for the F2M red quasars using their apparent $g$-band magnitudes (uncorrected for reddening) from SDSS while the dot-dashed and dot-dot-dot-dashed lines are from FBQS-II and FBQS-III.  The uncorrected F2M distribution suggests that red quasars are skewed toward radio-loud sources. However this is because dust extinction suppresses the denominator in Equation \ref{eqn:rl} artificially enhancing $R_g$.  Once we correct for extinction (our methodology for dereddening is described in Section \ref{sec:reddening}) the $R_g$ distribution for the F2M quasars (solid line) has the same distribution as the FBQS quasars (dash-dot line).   We see that for both the FBQS and F2M samples many  are bona-fide ``radio-quiet'' quasars with $R_g < 1$.  Still, most quasars lie below the FIRST detection threshold and they are even ``quieter'' in the radio; only 16\% of the quasars in the SDSS quasar catalog \citep{Schneider10} overlap the FIRST survey area and are detected in FIRST (and have 2MASS detections).  

In \citetalias{Glikman07} we reported contemporaneous 3.6 cm and 20 cm flux density measurements with the VLA for 44 F2M quasars.  The distribution of spectral indices for F2M quasars showed that there are few flat spectrum sources ($\alpha_{\rm 1.4 GHz/8.3 GHz}>-0.5$, where $S_\nu \propto \nu^\alpha$) and that the fraction of steep-spectrum sources rises with decreasing 20 cm flux density.  
The dearth of flat radio spectral indices in the F2M quasar sample suggests that their red colors are not caused by synchrotron contamination.  This is in contrast to the red quasars found in the Parkes Half-Jansky Flat-Spectrum quasar sample \citep[PHFS\footnote{The PHFS radio flux limit is 500 times higher than FIRST, so the contribution of red synchrotron emission in the optical is only an issue at high radio luminosities.}; ][]{Drinkwater97} where \citet{Whiting01} do find a red synchrotron component contributing to their red colors.  To see if very bright radio sources contain some red synchrotron emission that may masquerade as dust-reddening, we plot \ebv\ vs.\ 20 cm FIRST flux density for all F2M quasars in Figure \ref{fig:rl}, right.  We see that below $\sim 20$ mJy there is no correlation with reddening and that the most radio-bright sources tend to be the least reddened.  In particular, seven of the 
eight F2M quasars with $F_{20~\rm cm} > 500$ mJy have \ebv$\lesssim 0.4$ suggesting that a red synchrotron enhancement is not important in these sources.   
We argue, therefore, that the red colors of F2M quasars are the result of dust extinction.
The range of spectral indices in the F2M quasars also suggests that we are viewing the F2M quasars at a variety of orientation angles.  

If the dust that reddens the quasars is located in their host galaxies as part of a merger-driven picture for quasar/galaxy co-evolution  (see Section \ref{sec:evolution}) then we would not expect the radio emission from the quasar to be related to dust located parsecs to kiloparsecs away from the black hole.   Therefore, our findings here should extend to the overall quasar population, regardless of their radio properties.  

Recent results, however, suggest a possible correlation between reddening in quasars and radio emission. \citet{White07} found, in radio stacking of quasars (the overwhelming majority of which are radio-quiet) that redder
quasars have higher median radio fluxes. In addition, \citet{Georgakakis09} studied the ten brightest red quasars in 2MASS, selected without any radio constraint, and found that 60\% were detected in the radio (compared with $\sim10\%$ for all quasars).  They also reported evidence for high levels of star formation based on Spitzer photometry suggesting that these QSOs are ÒyoungÓ.  
Follow-up work by \citet{Georgakakis12} find that such bright red quasars have spectral properties in the radio indicative of young radio jets.
A radio-independent selection of red quasars \citep[e.g., with WISE;][]{Wright10} is needed to determine whether we can extend the conclusions drawn from this radio-selected sample to the entire quasar population. 

\subsection{Morphological Properties}

In addition to their blue, UV excess SEDs, unobscured quasars outshine their host galaxies and have a stellar morphology.   Many quasar samples have exploited this feature, requiring that the images of candidate quasars be unresolved point sources \citep[e.g., FBQS, 2QZ, ][]{Croom01,Glikman10}.  Reddened quasars, however, can appear extended in their rest-frame UV and optical images as the quasar emission is attenuated relative to their host galaxies.  

\begin{figure}
\plotone{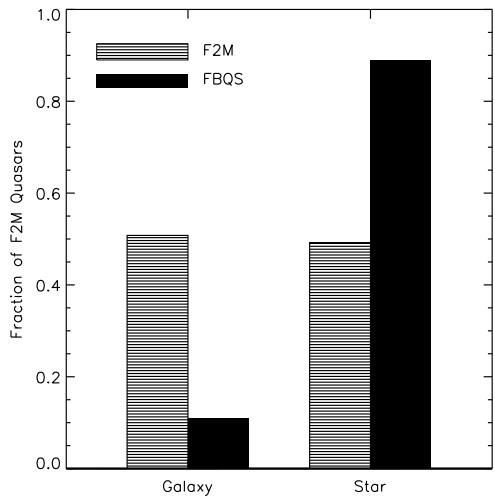}
\caption{Morphological distribution of F2M quasars based on the global SDSS morphological classification (shaded bar).  Half of these reddened quasars appear extended, making them elusive in surveys that select for point sources.
For comparison, we show the distribution of the same for the FBQS quasars (black bar).  
Both distributions are normalized by the total number of quasars in that sample.}\label{fig:morph}
\end{figure}

The SDSS assigns a morphology for all sources detected in its five imaging filters ({\tt type\_u, type\_g, type\_r, type\_i,type\_g}).  A global morphology, ({\tt type}), which is based on the combined flux from the five photometric bands, is also assigned \citep{Stoughton02}.
Figure \ref{fig:morph} shows the distribution of F2M quasar morphologies, shown as a fraction of the total number of sources (shaded bar).  For comparison, we also plot the SDSS morphologies of the FBQS quasars, which were selected to have stellar morphologies in the digitized POSS-I data (black bar).  We see that roughly half of the F2M quasars have an extended morphology, implying that surveys for red quasars that apply a morphological restriction, especially in the optical, can miss $\sim 50\%$ of the red quasars.  

We also checked for a color dependence on morphology by examining the classification of F2M quasars in the individual filters.  We find that the ratio of 'STAR' and 'GALAXY' morphologies are distributed similar to the global distribution in the $u$, $g$, $r$ and $i$ filters.  In the $z$ band, however, $\sim 60\%$ of F2M quasars have a stellar morphology, consistent with  dust-reddening in these quasars. 

\section{Reddening in Quasars} \label{sec:reddening}

\subsection{Estimates of Reddening in Quasars} \label{ssec:reddening}

In \citetalias{Glikman07} we analyzed the reddening properties of 56 quasars, a subset of the sample in this paper obeying identical selection criteria.  We applied a standard extinction law to an optical-to-near-infrared quasar template \citep[$f_0;$][]{Brotherton01,Glikman06} using an SMC reddening law  \citep{Fitzpatrick99,Gordon98}, which has been shown to fit reddening in quasars more effectively than a Large Magellenic Cloud (LMC) or Milky Way reddening law \citep{Hopkins04,Richards03}, ignoring emission lines and noisy regions.  Here we follow this method for the full F2M quasar sample and use the resultant $E(B-V)$ values throughout the paper.  Since we obtained additional spectra for some of the quasars reported in \citetalias{Glikman07}, adding in a missing optical or near-infrared spectrum, we re-compute $E(B-V)$ for those quasars along with the expanded sample.

In \citetalias{Glikman07}, we noted several caveats to consider when determining reddening:  (1) The uncertainty in the scaling between the optical and near-infrared spectra can result in poorly constrained fits, and likely erroneous $E(B-V)$ values (e.g., Figure 10 of \citetalias{Glikman07}).  The scaling is performed in the overlapping ends of each spectrum's wavelength range, which are the noisiest and worst-calibrated.
(2) Reddening determined from fits to the combined spectrum sometimes differed significantly from fits to the optical or near-infrared spectrum of the same object.  In such cases we typically rely on the fit to the optical spectrum, since it is always sampling the shortest wavelength light, which is most sensitive to extinction, and therefore best constrains the absorption parameters.  (3) The presence of host galaxy light, especially in low-redshift, low-luminosity sources, where  the 4000\AA\ break is occasionally visible, can affect our reddening estimates.  We checked for this issue in \citetalias{Glikman07} and found this effect to be negligible.  Furthermore, removing the galaxy in some sources but not others adds an inconsistency to our analysis.  After all, the galaxy is {\em there} in all cases, but our ability to quantify its contribution to the light depends on many factors, e.g., signal-to-noise ratio of our spectrum around the rest-frame 4000\AA\ break, the wavelength range of a given spectrum, etc.  We therefore do not remove the galaxy in our reddening analysis.  

In HST images of a subsample of F2M quasars \citet{Urrutia08} performed detailed morphological analysis and point-spread function (PSF) subtraction for these sources and found that the $g-I$ colors  of the point sources were even redder than their total aperture colors.  The host galaxies also showed evidence for merging and interactions, which suggests that star-formation in the host galaxies of these sources added blue light to the lower-resolution observations (this is corroborated by the color-tracks shown in Figure \ref{fig:nncolor}; we discuss the implications of these HST observations in Section \ref{sec:evolution}).  We may therefore be {\em underestimating} of $E(B-V)$ in our quasars.  Table \ref{tab:ebv} lists the extinction parameters for 131 quasars in our sample that have spectra.   

We also investigated, in \citetalias{Glikman07}, the correlation between reddening derived from Balmer decrements (i.e., the ratio of \ha\ to \hb) measured from our quasar spectra.  We only conduct this analysis in cases where both lines appear in a single spectrum, to avoid uncertainties introduced from scaling.  We found that \ebv\ measured from Balmer decrements was consistent with \ebv\ measured from continuum fitting, on average, but varied significantly from object to object (with a scatter of $\sim 0.5$ mag).  

Although Balmer decrements have been used historically to measure reddening along the line of sight to AGN and quasars \citep[e.g.,][]{Maiolino01} we argue that using the full continuum to model the extinction to a source is a far more reliable approach.  First, in both approaches, $E(B-V)$ is effectively derived by determining the relative extinction from a flux ratio at different wavelengths compared to an intrinsic flux ratio,
\begin{equation}
E(B-V) = -\frac{1.086}{k(\lambda)} \log\Bigg[\frac{f(\lambda)}{f_0(\lambda)}\Bigg]. \label{eqn:ebv}
\end{equation}
The same dust law, $k(\lambda)$, is used for measurements of Balmer decrements, replacing $k(\lambda)$ with $k(\rm H\alpha)-k(\rm H\beta)$,  $f(\lambda)$ with the measured ratio of H$\alpha$/H$\beta$ line fluxes and $f_0(\lambda)$ with an {\em intrinsic} ratio of H$\alpha$/H$\beta$ in Equation \ref{eqn:ebv}.  Therefore, in both methods the same formulation is used to determine $E(B-V)$ except that more wavelengths are sampled when the full continuum is used, which makes it a more robust measurement.

Secondly, both methods require an assumption about the intrinsic flux ratios of parts of a quasar's spectrum as a function of wavelength ($f_0(\lambda)$ versus H$\alpha$).  The intrinsic shape of quasars' continua are known to have intrinsic variation that follows a Gaussian distribution \citep[$f_\nu \propto \nu^\alpha$, with $\sigma_\alpha = 0.30$;][]{VandenBerk01,Richards03}.  This means that an intrinsically redder-than-average quasar may be over-corrected in a continuum fit.  
We partially mitigate this issue by using the FBQS quasar composite spectrum \citep{Brotherton01} as the rest-frame UV to optical portion of the template that we use to fit our quasars.  Since this composite is derived from the same quasar sample to which we are comparing our F2M quasars, our analysis remains internally consistent.  In addition, the \citet{Brotherton01} composite was shown to be slightly redder ($\alpha=-0.46$) than quasar templates derived from optically-selected quasars \citep[$\alpha = -0.32$;][]{Francis91} making it less likely that we are overestimating the reddening in our quasars.
Furthermore, in \citetalias{Glikman07} we explored the effect of varying the intrinsic slope of the quasar composite template to the $\pm 1\sigma_\alpha$ on the values of $E(B-V)$ derived from continuum fits; the variation is $\lesssim 0.1$ magnitudes in $E(B-V)$ (see \S 5.3 and Figure 14 of that paper).

On the other hand, the variation in the intrinsic Balmer decrement of the broad line region can have a wide range of values spanning at least a factor of two.  This large range is a result of collisional excitation of the Balmer lines and radiative transfer effects at the high densities ($N_e \sim 10^{8-10}$ cm$^{-3}$) of the broad line region \citep{Netzer75,Rees89}. 
The traditionally-used intrinsic H$\alpha$/H$\beta$ ratio, derived from Case B recombination, is 2.88 \citep{Osterbrock89}.  However, Case B recombination is likely ruled out in the broad line region, as shown by the Paschen line ratios in the near-infrared \citep[e.g.,][]{Glikman06,Oyabu09}.
We measure a Balmer decrement for the FBQS template spectrum of 4.091 for the total line profile.  
With such high variations in the intrinsic Balmer decrements of quasars, it is not surprising that there is a large scatter when comparing Balmer decrement-derived reddenings to continuum-based reddening measurements.  We therefore  consider the Balmer decrement to be a far less reliable reddening indicator compared with fitting a reddened template to a full spectrum. 

\begin{figure}
\plotone{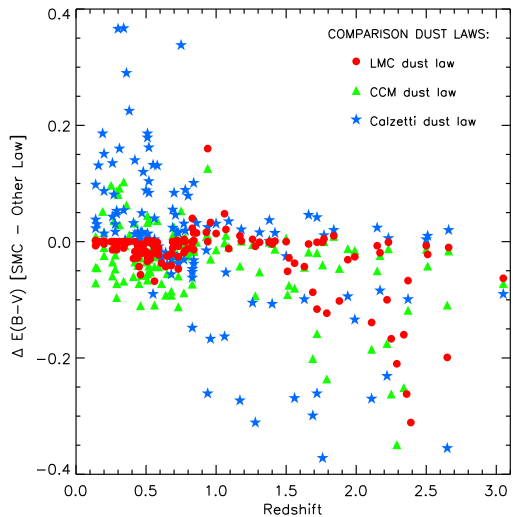}
\caption{Comparison of \ebv\ values derived from different dust laws applied to reddened template fits to the F2M quasars.  We plot the difference between the SMC and the LMC, Milky Way (CCM) and \citet{Calzetti94} starburst dust laws [$\Delta$\ebv] as a function of redshift.  We see that the discrepancy increases toward higher redshifts where more rest-frame UV light -- where the dust laws begin to diverge --  is used in the fitting.  At $z>1$ the discrepancy between the derived extinctions always measure the lowest \ebv\ for the SMC dust law.}\label{fig:ebv_compare}
\end{figure}

We also explored whether the SMC dust law is the appropriate reddening law to use for the F2M quasars.  To do this we fit our spectra with an average Large Magellenic Cloud (LMC) dust law from \citep{Misselt99} as well as the Milky Way dust law of \citet{Cardelli89} (CCM, henceforth) updated in the near-UV by \citet{Odonnell94}.  In addition, we fit the \citet{Calzetti94} dust law for starburst galaxies to explore its effectiveness at fitting the continuum of red quasars.   \citet{Gordon03} performed a comparison between the SMC, LMC and Milky Way dust laws and found that at rest frame near-UV ($\sim 2500$\AA) through near infrared ($\sim 1$\um) the three dust laws are extremely similar.  The extinction curves begin to diverge shortward of $\sim 2000$\AA.  Figure \ref{fig:ebv_compare} shows the difference between \ebv\ derived using an SMC dust law and the other dust laws, as a function of redshift.  
We find excellent agreement between the values of \ebv\ derived using the SMC, LMC and CCM for most objects, with the standard deviation between \ebv\ derived using the SMC dust law and the LMC and CCM laws are $\sigma_{E(B-V)}=0.055$ and 0.089 magnitudes, respectively.  The extinctions measured with the \citet{Calzetti94} dust law have much larger scatter and poorer agreement with a $\sigma_{E(B-V)}=0.189$.  
We see that the discrepancy between the dust laws increases at increasing redshifts since we are sampling more of the rest-frame UV part of the quasars' spectra. This is consistent with the findings of \citet{Gordon03}.

\begin{figure}
\plotone{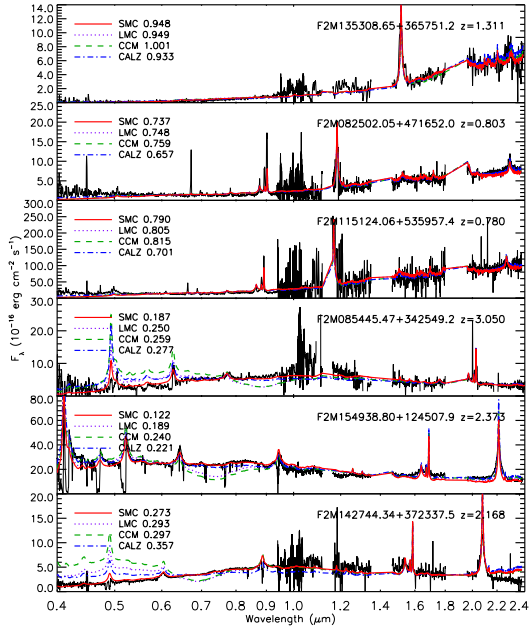}
\caption{Comparison fits of a quasar template reddened by four dust laws, SMC, LMC, Milky Way (CCM) and the \citet{Calzetti94} law for starburst galaxies to F2M quasars.  The top three panels show cases where all four dust laws produce similar satisfactory fits, and \ebv\ values that agree to within $\sim 0.1$ magnitudes.  The bottom three panels show that in cases where the discrepancy between the fits is large,  the SMC dust law produces the most consistent fits to the spectra, especially at high redshifts where the 2175\AA\ ``bump'' characteristic of the SMC and LMC dust laws is not seen. }\label{fig:ebv_spec_compare}
\end{figure}

We note that in cases where there is a large discrepancy between these two dust laws, the SMC dust law yields a {\em lower} value of \ebv.  This means that the SMC law is a conservative choice when estimating reddening in these quasars.  We plot in Figure \ref{fig:ebv_spec_compare} example fits to F2M quasar spectra showing instances where the four dust-laws produce good fits and excellent agreement (top three spectra) and cases where the SMC dust law produces superior fits to the others, especially around rest-frame $\lambda=2175$\AA\ (bottom three spectra). 

In addition, quasars can sometimes be reddened by intervening absorbers with CO, H$_{2}$,  and HD \citep{Noterdaeme10} as well as by dust in intervening \ion{Mg}{2} absorption systems, whose average extinction curves are very similar to an SMC dust law \citep{York06}.  \citet{Srianand08} find an LMC extinction law in \ion{Mg}{2} absorbers in two quasars with $E(B-V)\gtrsim 0.3$.  Among the F2M sources, we find quasars with narrow absorption-line systems whose $z_{abs} \lesssim z_{em}$.  Absorption lines that are within 5000 km~s$^{-1}$ of the emission lines are considered either 'intrinsic' or 'associated' absorbers \citep{Foltz86}.  In some of the objects the redshifts of the emission lines and the absorption lines are identical within their respective errors.  Recently, \citet{Shen11} have demonstrated that quasars with associated absorbers have signatures of enhanced star formation.  They suggest that these absorbers are possible large-scale outflows indicative of a phase in a merger-driven evolutionary scenario for quasars (see Section \ref{sec:evolution}).  We will explore the sub-population of F2M red quasars that show broad, associated absorption in their spectra in Section \ref{sec:bal}.

\subsection{Trends With Reddening} \label{fig:ebv}

\begin{figure}
\plotone{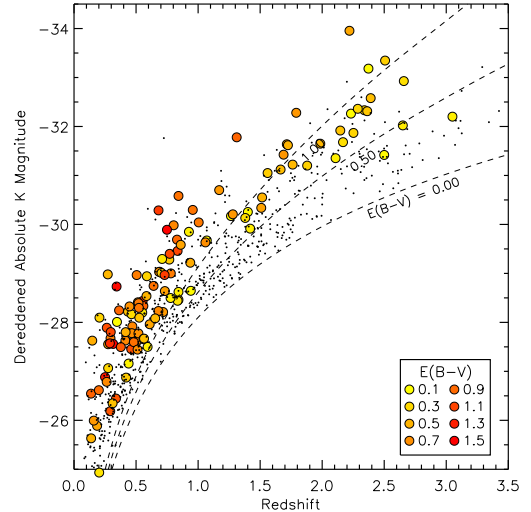}
\caption{Dereddened $K$-band absolute magnitude as a function of redshift.  The colors of the circles correspond to the amount of extinction, ranging from low extinction (yellow) to heavily reddened (red).  The dotted lines indicate the survey limit ($K<16$) for increasing amounts of extinction.  The small dots are FBQS-II and FBQS-III quasars, which we assume are unabsorbed.  At every redshift, red quasars are the most luminous.}\label{fig:eb_min_v_color}
\end{figure}

Following  \citetalias{Glikman07} and \citet{Urrutia09}, we deem a quasar ``red'' if its reddening is $E(B-V)>0.1$; 120 of the F2M quasars satisfy this criterion.  In Figure \ref{fig:eb_min_v_color} we plot the de-reddened $K$-band luminosity for our quasars as a function of redshift.  We color-code each F2M quasar by the amount of reddening, with yellow circles representing lightly reddened quasars, $E(B-V)=0.1$, and red circles representing heavily-reddened quasars, $E(B-V)\ge1.5$.  Shades of orange correspond to increasing amounts of reddening, as annotated in the legend.  We also plot the locations of FBQS quasars (assuming their $K$ magnitudes do not experience any extinction) with black points.  

Several obvious trends become apparent.  At every redshift, F2M red quasars are the {\em most luminous} sources.  
In fact,  at the most luminous end, the space density of red quasars rivals and even exceeds that of blue, unobscured quasars.  

\citet{White03b} predicted this result with a sample of five red quasars out of 35 FIRST-detected quasars selected in the $I$-band.  Most of them were normal, unreddened quasars.  However, they found five heavily reddened objects at $z<1.3$ which, after correcting for reddening, were the most luminous.  The interpretation of this was  that they were ``detecting only the most luminous tip of the red quasar iceberg.''  We began to uncover this trend in \citetalias{Glikman07} and in \citet{Urrutia09} but these samples were not large enough to trace the full range of reddenings across this wide redshift range.  We discuss the implication of this result in \S \ref{sec:evolution} where we interpret red quasars as a short-lived phase in the context of merger-driven QSO ignition and evolution.

In addition, there appears to be a correlation between reddening and redshift in Figure \ref{fig:eb_min_v_color}.  We see heavily reddened objects (\ebv$ \ge 1.5$) at the lowest redshifts, but they become scarce at higher redshifts.  This is largely a selection effect, which has been noted elsewhere \citepalias[e.g., \S 6 of ][]{Glikman04}.  At higher redshifts, what is observed as near-infrared light is rest-frame optical emission, which is more sensitive to dust extinction.  The dashed lines in Figure \ref{fig:eb_min_v_color} show this explicitly: at a given $K$-band flux limit (e.g., $K\lesssim 15.5$ for 2MASS) in order to detect heavily reddened objects at higher redshifts, they must be that much more luminous. To reach the redder objects at higher redshifts and overcome this bias, a more sensitive near-IR survey is needed.

\subsection{The Fraction of Quasars That Are Reddened}

We overplot the surface density of F2M quasars as a function of their {\em intrinsic} $K$-band magnitude, after correcting their apparent $K$-band magnitudes, in Figure \ref{fig:space_dist} with open violet squares.   We find that after correcting for extinction F2M quasars comprise $21\pm2\%$ of radio-selected quasars with $K<14.5$ magnitudes.  The $14.5< K<15$ bin begins to show incompleteness, but we can estimate a lower limit to the fraction of red quasars with $K<15$ at $15\pm1\%$.  This fraction is remarkably consistent with the result from \citet{Richards03} who find that SDSS misses $\sim 15\%$ of reddened quasars that would otherwise be detected in the imaging survey.  

This comparison, however, produces a lower limit to the fraction of red quasars since the sensitivity of our survey to red quasars is strongly luminosity and redshift dependent, as seen by the dashed lines in Figure \ref{fig:eb_min_v_color}.  We therefore should only compare the space density  of red quasars to FBQS quasars which, when reddened, would still be detected in our survey.  Our color cuts, and the definition we use for ``red quasar'', restrict our sample to $E(B-V) > 0.1$.  Therefore, we ought not count FBQS quasars whose luminosities are too small, at a given redshift, to be detected by 2MASS when reddened by \ebv=0.1.  In other words, we can only compare the densities of quasars in the region of $M_K$ and $z$ space where the red quasars overlap the FBQS points.  There may be (and probably are) red quasars at lower intrinsic luminosities, but we cannot detect them in 2MASS. 

\begin{figure}
\plotone{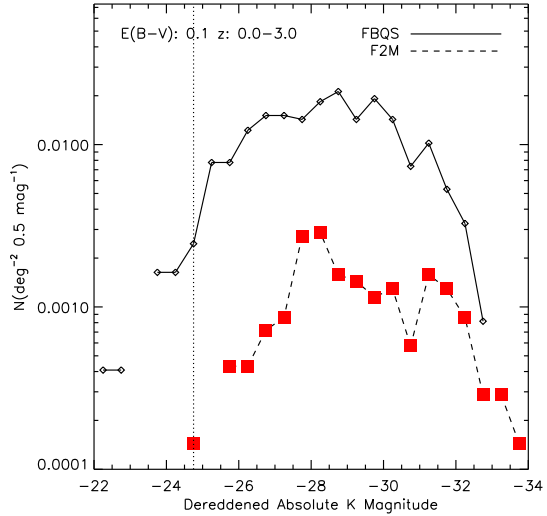}
\caption{Distribution of red quasars as a function of de-reddened absolute $K$-band magnitude compared with FBQS quasars above the 2MASS limit sensitive to \ebv=0.1 mag ($M_K = -25.75$ dotted line).  Including all quasars above this limit, the fraction of quasars that are reddened is 9.9\%.}\label{fig:lumfunc_all}
\end{figure}

In Figure \ref{fig:lumfunc_all} we plot the surface density as a function of $M_K$ of all F2M quasars and compared with FBQS quasars that would be detectable in 2MASS if reddened by \ebv = 0.1.  We caution that these curves are {\em not} luminosity functions since we are plotting a surface density (deg$^{-2}$) not a volume density (Mpc$^{-3}$) and we have not corrected these number counts by the selection functions of either survey\footnote{We do scale the survey areas by their respective spectroscopic completeness.}.  However, since both surveys rely on the same flux-limited radio and near-infrared detection we assume that they are sufficiently comparable for the purposes of measuring the fraction of quasars that are reddened.   We compare only quasars with $M_K \le -25.75$ (indicated by the vertical dotted line), since reddened quasars with lower intrinsic luminosities would drop out of 2MASS.  When comparing these two populations, the fraction of red quasars is 9.9\%.  This is, of course, a lower limit because of the strong redshift-dependent sensitivity of our survey to finding reddened quasars.  

\begin{figure*}
\plottwo{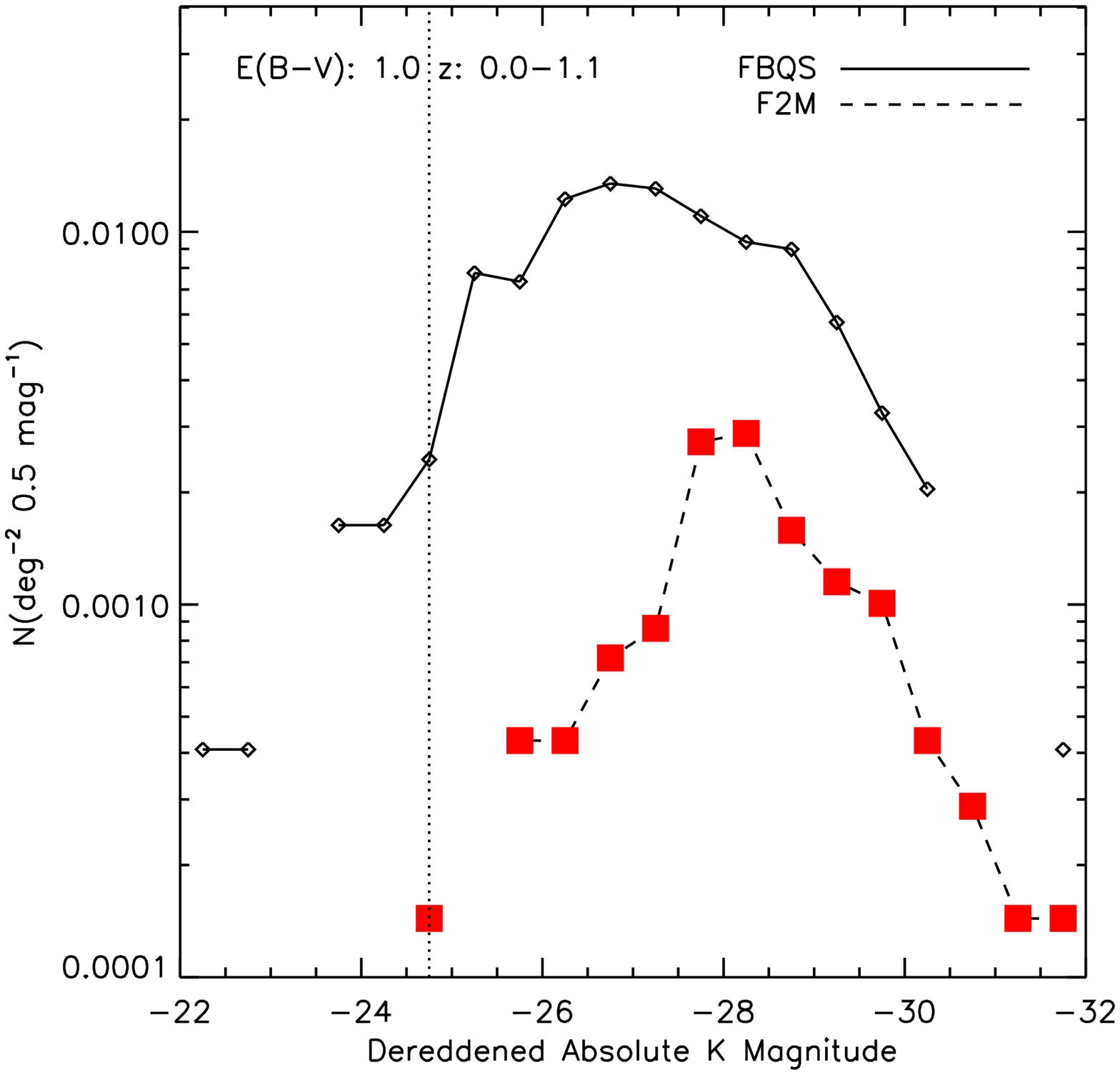}{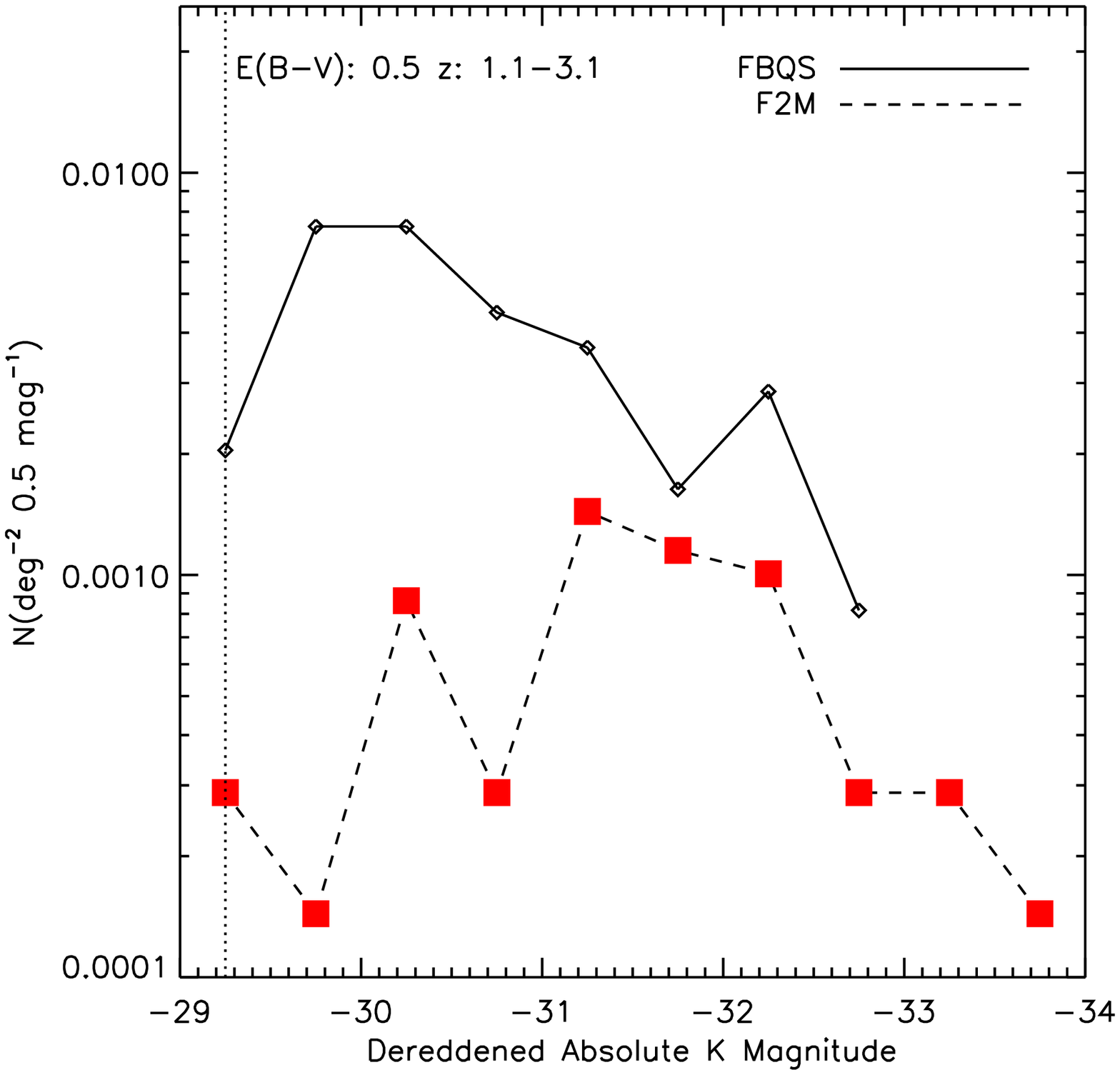}
\caption{Distribution of red quasars as a function of de-reddened absolute $K$-band magnitude in two redshift bins:  (left) $0 < z \le 1.1$ above the \ebv=1.0 sensitivity limit; (right) $1.1<z < 3.1$ above the \ebv=0.5 sensitivity limit.  The dotted vertical line marks the sensitivity limit above the $E(B-V)$ for each plot.  We integrate under each curve to this line to compare the space densities of each sample.  The space density of red quasars in each redshift-reddening bin is 13.3\% (left) and 19.5\% (right).}\label{fig:lumfunc_vz_z}
\end{figure*}

Since our ability to find heavily reddened sources is a strong function of redshift, we divide our sample into two redshift ranges and compare the distributions of quasars with sensitivity limits appropriate to those redshift ranges.  In the left panel of Figure \ref{fig:lumfunc_vz_z} we plot the dereddened $M_K$ distributions of red and blue quasars with $0 < z < 1.1$ and applying the sensitivity limit of \ebv $=1.0$ mag, which Figure \ref{fig:eb_min_v_color} shows is valid for these lower redshifts.  On the right we plot the same distributions but for quasars with $1.1<z<3.1$ and which obey the \ebv$= 0.5$ sensitivity limit.  Comparing the areas under the curves, we find that the surface density of the low redshift red quasars is $13.3\%$ of the FBQS counterpart, while at higher redshifts the surface density of red quasars is $19.5\%$ of the FBQS quasars.  In both cases, and as we noted earlier, we see that red quasars make up a higher fraction of quasars at higher luminosities.  

This trend is in direct contradiction to the ``receding torus'' seen in lower luminosity AGN in which a the fraction of obscured (and reddened) quasars increases with {\em decreasing} luminosity.  It is hard to envision an orientation-based explanation for an increase in the dust covering fraction with increasing luminosity.  We note further that the absence of blue FBQS quasars at the very high luminosities seen for F2M quasars is more difficult to explain as a selection effect.  If the space densities of equally luminous unobscured blue quasars is comparable to or higher than that of red quasars, they would be difficult to miss in a blue radio-selected quasar survey.   We instead interpret the very high de-reddened luminosities of F2M quasars in the context of an evolutionary picture in Section \ref{sec:evolution}.

\section{Broad Absorption Line Quasars} \label{sec:bal}

Broad absorption line (BAL) quasars show broad absorption features (with velocities $>2000$ km s$^{-1}$) blueshifted by $\sim 3000 - 25000$ km s$^{-1}$ from either \ion{C}{4} $\lambda 1549$ and other high ionization UV lines, e.g., \ion{N}{5} 1240\AA, and/or \ion{Mg}{2} $\lambda 2800$ \citep[e.g.][]{Weymann91}.  BAL quasars lacking \ion{Mg}{2} absorption are classified as high-ionization BALs (HiBALs) and those with \ion{Mg}{2} absorption are low-ionization BALs (LoBALs).  An even rarer class of LoBALs show absorption from metastable iron species, \ion{Fe}{2} and \ion{Fe}{3} \citep[FeLoBALs;][]{Hazard87}.  Initially BALs were thought to exist only in radio-quiet quasars, although this was likely due to selection effects. In radio stacking studies of quasars,  \citet{White07} found that the median 20 cm flux density from LoBAL quasars is 2-3 times higher than for non-BAL quasars.   \citet{Becker00} found 29 BALs in the FBQS II survey, amounting to $18\% \pm 4\%$ of the quasars whose redshift range allows their identification.   Low-ionization BALs are far less abundant than the HiBALs, making up between $10\%$ and $20\%$ of BALs depending on the selection method of their parent quasar sample \citep[e.g., ][]{Menou01,Trump06}.  

Thirty-three of the F2M red quasars have spectra with coverage of \ion{Mg}{2} with sufficiently high signal-to-noise ratio in the countinuum to measure the presence of broad absorption features.  This corresponds to objects with an optical spectrum and in the redshift range $0.9 \lesssim z$.  
Following \citet{Urrutia08}, we measure the strength of the \ion{C}{4} and \ion{Mg}{2} absorption lines using the Balnicity Index \citep[BI; Appendix A][]{Weymann91}, a sort of modified equivalent width in units of km s$^{-1}$,  
\begin{equation}
{\rm BI} = -\int^{3000}_{25,000} \bigg[1-\frac{f(V)}{0.9}\bigg] C dV, \label{eqn:bi}
\end{equation}
where $f(V)$ is the normalized flux as a function of velocity displacement from the line center, $V$.   This measurement ignores absorption with velocity widths less than 2000 km s$^{-1}$  and may miss so-called mini-BALs.

To overcome these drawbacks \citet{Hall02b} defined a a more liberal measure of the absorption trough the Absorption Index (AI).  We use a modified version of this measurement, defined by \citet{Trump06}:
\begin{equation}
{\rm AI} = -\int^{0}_{29,000} [1-f(V)] C' dV, \label{eqn:ai}
\end{equation}
where $f(V)$ is the normalized template-subtracted flux as a function of velocity displacement from the line center, $V$ \citep[see Appendix A of][for details on the AI]{Hall02b}.  

In addition, \citet{Trump06} use a reduced chi-squared measurement to determine if an object in their sample is a BAL:
\begin{equation}
\chi^2_{\rm trough} = -\sum \frac{1}{N} \bigg[\frac{1-f(V)}{\sigma}\bigg]^2. \label{eqn:chi2}
\end{equation}
They require that BALs have $\chi^2_{\rm trough}>10$.

We compute all three BAL-determining indices for both \ion{C}{4} and \ion{Mg}{2} for the thirty-three quasars that may reveal a BAL.  We present these measurements in Table \ref{tab:bals}.

Of these thirty-three quasars, twenty show broad absorption lines in their spectra, amounting to $61\%$ of the eligible quasars in our sample.  They are labeled 'BAL' in Column (9) of Table \ref{tab:bals}.  Only {\em one} of the BALs found in our survey is of the typically more common HiBAL type; the rest show absorption in \ion{Mg}{2} and are LoBALs.

Why do we find so many LoBALs in our survey and few HiBALs?  It has long been known that LoBAL quasars have redder colors than HiBAL quasars \citep{Weymann91}.  \citet{Sprayberry92} compared composite spectra of six LoBALs and 34 HiBALs and found that the LoBALs appear to be reddened by an SMC-like dust-extinction law with $E(B-V) = 0.1$.  Dust extinction preferentially excludes these objects from flux-limited quasar surveys and they may be underrepresented by up to a factor of ten.  \citet{Becker00} find that $\sim 3\%$ of quasars are LoBALs; adjusting this figure by a factor of ten as suggested by \citet{Sprayberry92}, implies that out of thirty-four eligible quasars there may be up to 19 LoBALs.  This estimate is consistent with our result of twenty-one.  In fact, the typical extinction in LoBALs may be much higher than that discussed in \citet{Sprayberry92} and \citet{Weymann91} since their objects are optically selected.  The extinctions found for many of these objects from their continuum fits suggest much higher reddening than $E(B-V) = 0.1$ and may represent the large population of red LoBALs predicted by \citet{Becker00}. This result, however, does not explain the deficit of HiBAL quasars.

Since \ion{C}{4} $\lambda 1549$ has a shorter wavelength than \ion{Mg}{2}, we can only detect HiBALs in $z \gtrsim 1.7$ quasars.  Of the 34 quasars able to reveal a BAL, 19 have \ion{C}{4} coverage in their spectra. Sixteen of these are LoBALs, one is a HiBAL.  Many of the remaining objects  are so heavily reddened that there is no significant flux at or blueward of \ion{C}{4}.  It is possible that there may be HiBALs among these 11 objects, but even so they would still be in the minority, far outnumbered by the LoBALs.  

LoBALs and FeLoBALs appear to be fundamentally  different from normal BALs \citep{Becker00,White07,Urrutia08,Farrah10}.  The SEDs of HiBALs do not appear to be different from non-BAL quasars suggesting that they are simply viewed at an orientation along a disk-wind, which is ubiquitous to all quasars \citep{White00,Gallagher07}, although they may be slightly reddened by dust and gas in the wind.  The F2M survey targets red, and selects against blue, objects.  We may be excluding HiBALs along with the blue quasars.  Since enough continuum flux must be detected at \ion{C}{4} to detect absorption, we both select against HiBALs and have difficulty identifying the ones that are selected.  

\citet{Urrutia08} found an anomalously high fraction (37\%) of LoBALs (and FeLoBALs in particular) in a sample of FIRST-2MASS-selected red quasars -- a subset of this sample -- and suggest these are young objects in a ``blowout'' phase.   In our sample, above $z\sim 1.7$, all but one (F2M0134$-$0931) of the eligible red quasars are BALs, and, with one exception (F2M0854+3425), all are LoBALs and most are FeLoBALs.  And while some of these absorption systems may not be the classical broad absorption type (defined by the BI), but they are all displaying some sort of outflows.  In Section \ref{sec:evolution} we argue that this population of objects represents a transitional evolutionary stage in quasar evolution, which has been suggested by \citet{Shen11} for quasars exhibiting associated \ion{Mg}{2} absorbers.

\section{Red Quasars as an Evolutionary Phase} \label{sec:evolution}

Since the discovery of this population of red quasars, the evidence suggesting that they are ``young'' has been mounting.  Merger-driven models of quasar and galaxy co-evolution predict an obscured phase for quasars which precedes the familiar, well-measured, luminous blue quasar phase \citep[e.g.,][]{Sanders88a,Hopkins05}.  Thirteen F2M red quasars (selected from \citetalias{Glikman04} and \citetalias{Glikman07}) were imaged with HST and revealed a large fraction (85\%) of mergers \citep[many of which appear to be major ``train wrecks'';][]{Urrutia08}, consistent with the merger-driven scenario.   These thirteen objects have also been observed with {\em Spitzer} and have low-resolution IRS spectroscopy and MIPS photometry, which, combined with their spectroscopy, have allowed a measurement of their black hole masses and bolometric luminosities and thus estimates of their Eddington ratios.  These quasars have unusually high Eddington ratios, and appear to fall below the back hole mass - bulge luminosity relation \citep{Marconi03}, indicating that they are rapidly growing to ``catch up'' with their hosts (Urrutia et al., in preparation).  \citet{Georgakakis09} studied the mid-infrared properties of a small sample of red quasars from {\em Spitzer} photometry and found that they had higher levels of star formation than blue, unobscured quasars, also suggesting that they are ``young''.   Finally, a large sample of red quasars reported by \citet{Urrutia08} contain an anomalously large fraction of LoBALs which they interpret as a ``blowout'' phase representing feedback from the quasar, expelling dust and quenching star formation.  We find an even larger fraction of LoBALs in this, larger, sample of red quasars.

If the red quasars in our survey are a distinct phase in quasar evolution then we can place a rough estimate on the duration of the phase in comparison to the length of the blue quasar phase, which is estimated at $\sim 10^7$ yr \citep{Yu05,Martini04,Porciani04,Grazian04,Jakobsen03,Haiman02,Bajtlik88}.  From Figure \ref{fig:space_dist} we integrate the space density of FBQS-II Quasars to $K=14.5$ magnitudes to be is $6.4\pm0.5\times 10^{-2}$ deg$^{-2}$.  The extinction-corrected surface density of red quasars to $K=14.5$ magnitudes is $1.2\pm0.1\times10^{-3}$ deg$^{-2}$,  which is a lower limit since we are not sensitive to the most heavily reddened quasars at higher redshifts.  Nevertheless, the ratio of these space densities suggest a rough estimate of the ratio of each phase's duration.  This is the most direct comparison of quasars in these two phases since the FBQS and F2M samples obey the same radio and near-infrared flux limits and differ only in the color selection.  We find that the ratio of red to blue quasars using these samples is $21\pm2\%$, implying that this emergence phase where the fully obscured quasar is shedding its cocoon lasts a few $\times 10^6$ yr.

\citet{Hopkins05} present a model for the evolution of a quasar's light curve during a major merger.  This model computes the bolometric luminosity, $B$-band luminosity and column density $N_H$ averaged over many lines-of-sight as a function of simulation time.  In these simulations, during the merger, the quasar bolometric luminosity peaks twice.  The first peak occurs during a heavily obscured phase, where $N_H$ peaks and the $B$-band luminosity is completely attenuated.  This completely obscured phase lasts for $\sim 3-5\times10^7$ years.  The second peak in bolometric luminosity corresponds with the maximum observed $B$-band luminosity and lower $N_H$, which is interpreted as the ``normal'', unobscured, blue quasar phase.  

The F2M red quasars are {\em not} completely obscured, i.e., they are not Type 2 quasars; their spectra show broad lines (by definition), and some rest-frame $B$-band flux from the quasar.  Their extreme luminosities, disturbed morphologies and moderate $E(B-V)$ values suggest that they are emerging from the first peak in the quasar's light curve.  An estimate of the duration of this ``emergence'' phase, based on the \citet{Hopkins05} model (e.g., Figure 2), is roughly $5\times 10^6$ years, consistent with our estimate.

We caution that this merger-driven quasar/galaxy co-evolution picture should not necessarily be extended to moderate and low luminosity AGN.  Recent observations of the host galaxies AGN selected by their X-ray emission in deep fields whose areas are too small to find the rare high luminosity sources presented here (e.g., CDFS, COSMOS) find no evidence for major mergers in these (mostly obscured) sources.  In fact, the host galaxies of these AGN tend to be mostly disk-dominated with some spheroids and the fraction of major mergers ($\sim 16\%$) does not differ from the inactive galaxy population \citep{Schawinski11,Kocevski11}.  Furthermore,  low luminosity X-ray selected AGN at $z \sim 1$, which are mostly obscured, have extremely low Eddington ratios ($L/L_{\rm Edd} \leq 0.1$), suggesting that they are in a different, ``slow'' phase of growth than the quasars presented here \citep{Simmons11}.

\section{Summary and Conclusions} \label{sec:summary}

We have identified a complete sample of 120 radio-detected, infrared-bright, dust-reddened quasars over the 9033 deg$^2$ area of the FIRST radio survey.  The quasars are selected from a candidate list of 395 FIRST-2MASS matches with optical counterparts from the digitized POSS-II catalog whose colors obey $R-K>4$ and $J-K>1.7$. Objects that lack an optical counterpart are also included and the POSS-II $R$-magnitude limit of 20.80 is used.  Reddened quasars are defined as as objects with a spectrum that reveals at least one broad emission line whose width implies $v\ge 1000$ km s$^{-1}$ as well as reddening of $E(B-V)\ge0.1$ based on fits to a reddened quasar template spectrum.  

The sample spans a broad range of redshifts, $0.13 < z < 3.1$, and reaches reddenings as high as $E(B-V) =1.55$.  Compared to radio-plus-optical-selected quasar samples such as FBQS and radio-detected quasars in SDSS, we find that red quasars make up $11\pm2\%$ of the apparent $K$-band-selected quasars.  However, once we correct for extinction, we find that, depending on how the parent population is defined, red quasars make up $\sim 15-20\%$ of the radio-emitting, luminous quasar population.  We also find that, at every redshift, red quasars are the most intrinsically luminous objects suggesting that they are in a state of high accretion.  This will be tested in a future paper where we will estimate black hole masses and accretion rates for these quasars.  We also reproduce the result from \citet{Urrutia08} that the red quasar population contains a large fraction of LoBALs, which, evidence suggests, may be young objects with strong outflows.

We therefore interpret dust-reddened quasars as a brief evolutionary phase that traces the transition from a heavily enshrouded ULIRG-like phase of black hole growth to the blue, unobscured quasars found in optically-selected samples.  If this red quasar population is interpreted as an evolutionary phase in the lifetime of a quasar, then based on their fraction we estimate the red quasar phase to be $15-20\%$ of the luminous blue quasar lifetime, or a few million years.

\acknowledgements

We thank Meg Urry for a careful reading of the manuscript and helpful comments. We are grateful to the staff of W. M. Keck observatory for their assistance during our observing runs.  EG, TU and ML acknowledge support from {\em Spitzer} grant GO4-PID40143. SGD and AAM acknowledge a partial support from the NSF grants AST-0407448 and AST-0909182.  ADM acknowledges support from the NASA-ADAP program through grants NNX12AI49G and NNX12AE38G.

The National Radio Astronomy Observatory is a facility of the National Science Foundation operated under cooperative agreement by Associated Universities, Inc.

The Digitized Sky Surveys were produced at the Space Telescope Science Institute under U.S. Government grant NAG W-2166. The images of these surveys are based on photographic data obtained using the Oschin Schmidt Telescope on Palomar Mountain and the UK Schmidt Telescope. The plates were processed into the present compressed digital form with the permission of these institutions.

The Second Palomar Observatory Sky Survey (POSS-II) was made by the California Institute of Technology with funds from the National Science Foundation, the National Geographic Society, the Sloan Foundation, the Samuel Oschin Foundation, and the Eastman Kodak Corporation.

The Guide Star Catalogue -- II is a joint project of the Space Telescope Science Institute and the Osservatorio Astronomico di Torino. Space Telescope Science Institute is operated by the Association of Universities for Research in Astronomy, for the National Aeronautics and Space Administration under contract NAS5-26555. The participation of the Osservatorio Astronomico di Torino is supported by the Italian Council for Research in Astronomy. Additional support is provided by  European Southern Observatory, Space Telescope European Coordinating Facility, the International GEMINI project and the European Space Agency Astrophysics Division.

This publication makes use of data products from the Two Micron All Sky Survey, which is a joint project of the University of Massachusetts and the Infrared Processing and Analysis Center/California Institute of Technology, funded by the National Aeronautics and Space Administration and the National Science Foundation.

Funding for the creation and distribution of the SDSS Archive has been provided by the Alfred P. Sloan Foundation, the Participating Institutions, the National Aeronautics and Space Administration, the National Science Foundation, the U.S. Department of Energy, the Japanese Monbukagakusho, and the Max Planck Society. The SDSS Web site is http://www.sdss.org/.

The SDSS is managed by the Astrophysical Research Consortium (ARC) for the Participating Institutions. The Participating Institutions are The University of Chicago, Fermilab, the Institute for Advanced Study, the Japan Participation Group, The Johns Hopkins University, the Korean Scientist Group, Los Alamos National Laboratory, the Max-Planck-Institute for Astronomy (MPIA), the Max-Planck-Institute for Astrophysics (MPA), New Mexico State University, University of Pittsburgh, University of Portsmouth, Princeton University, the United States Naval Observatory, and the University of Washington.

Funding for SDSS-III has been provided by the Alfred P. Sloan Foundation, the Participating Institutions, the National Science Foundation, and the U.S. Department of Energy Office of Science. The SDSS-III web site is http://www.sdss3.org/.

SDSS-III is managed by the Astrophysical Research Consortium for the Participating Institutions of the SDSS-III Collaboration including the University of Arizona, the Brazilian Participation Group, Brookhaven National Laboratory, University of Cambridge, University of Florida, the French Participation Group, the German Participation Group, the Instituto de Astrofisica de Canarias, the Michigan State/Notre Dame/JINA Participation Group, Johns Hopkins University, Lawrence Berkeley National Laboratory, Max Planck Institute for Astrophysics, New Mexico State University, New York University, Ohio State University, Pennsylvania State University, University of Portsmouth, Princeton University, the Spanish Participation Group, University of Tokyo, University of Utah, Vanderbilt University, University of Virginia, University of Washington, and Yale University.

{\it Facilities:} \facility{Keck:I (LRIS)}, \facility{Keck:II (ESI), \facility{Sloan}, \facility{VLA}, \facility{IRTF (SpeX)}, \facility{Hale (TripleSpec)}, \facility{Hiltner (TIFKAM)}, \facility{Shane (Kast Double spectrograph)}, }
\pagebreak

\input{tab1}
\input{tab2}

\input{tab3}
\input{tab4} 
\input{tab5}

\begin{figure}
\epsscale{1}
\figurenum{6.1}
\plotone{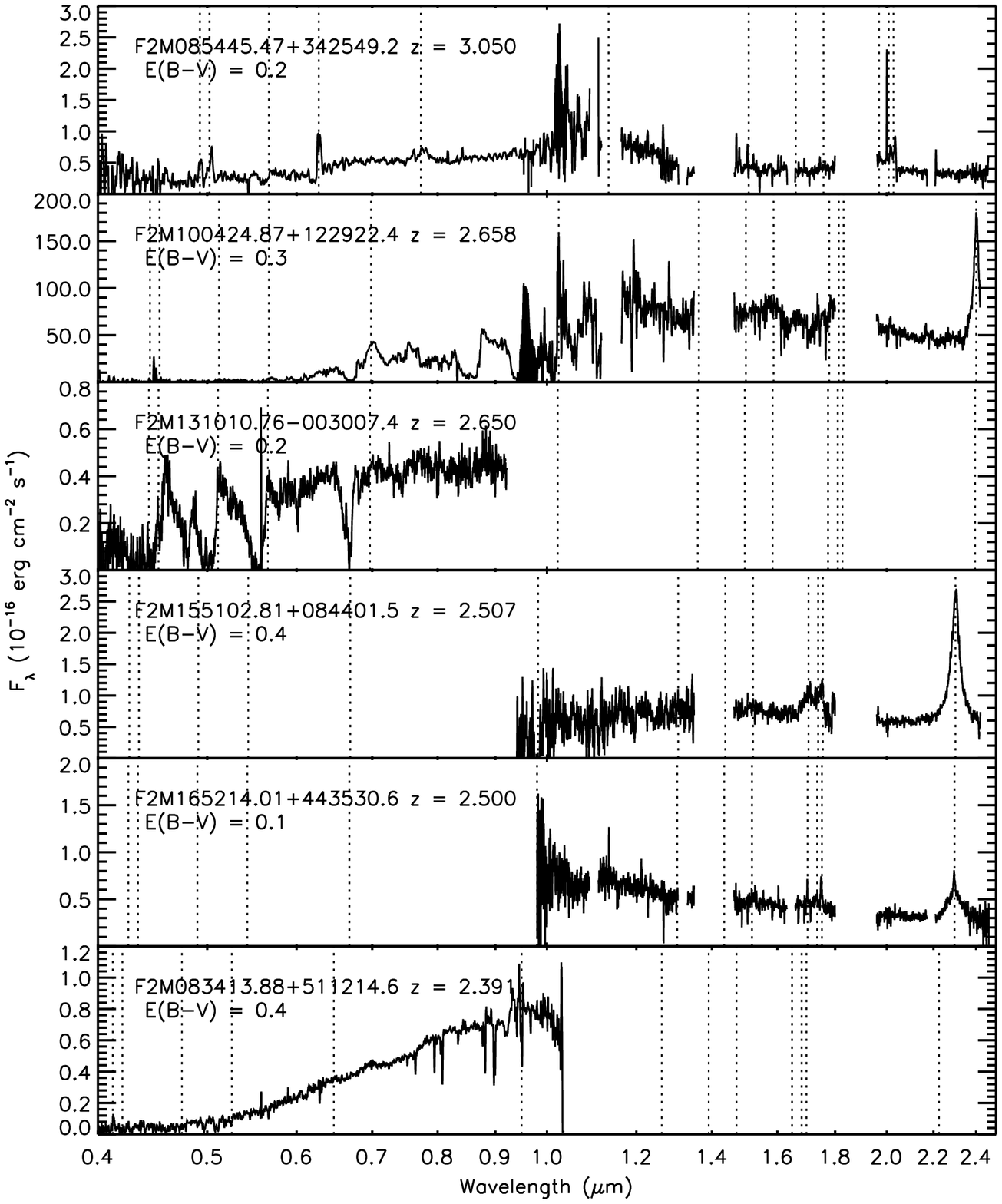}
\caption{F2M candidates identified as red quasars, having at least one broad emission line and $E(B-V)\ge0.1$, ordered by redshift.  The dotted lines show expected positions (in $\mu$m) of
prominent emission lines in the optical and near-infrared:
Ly$\alpha$~1216,
N~V~1240,
Si~IV~1400,
C~IV~1550,
C~III]~1909,
Mg~II~2800,
[O~II]~3727,
H$\delta$~4102,
H$\gamma$~4341,
H$\beta$~4862,
[O~III]~4959,
[O~III]~5007,
H$\alpha$~6563,
Pa$\gamma$~10941,
Pa$\beta$~12822 and
Pa$\alpha$~18756\AA.}\label{fig:spectra_all}
\end{figure}

\begin{figure}
\figurenum{6.2}
\plotone{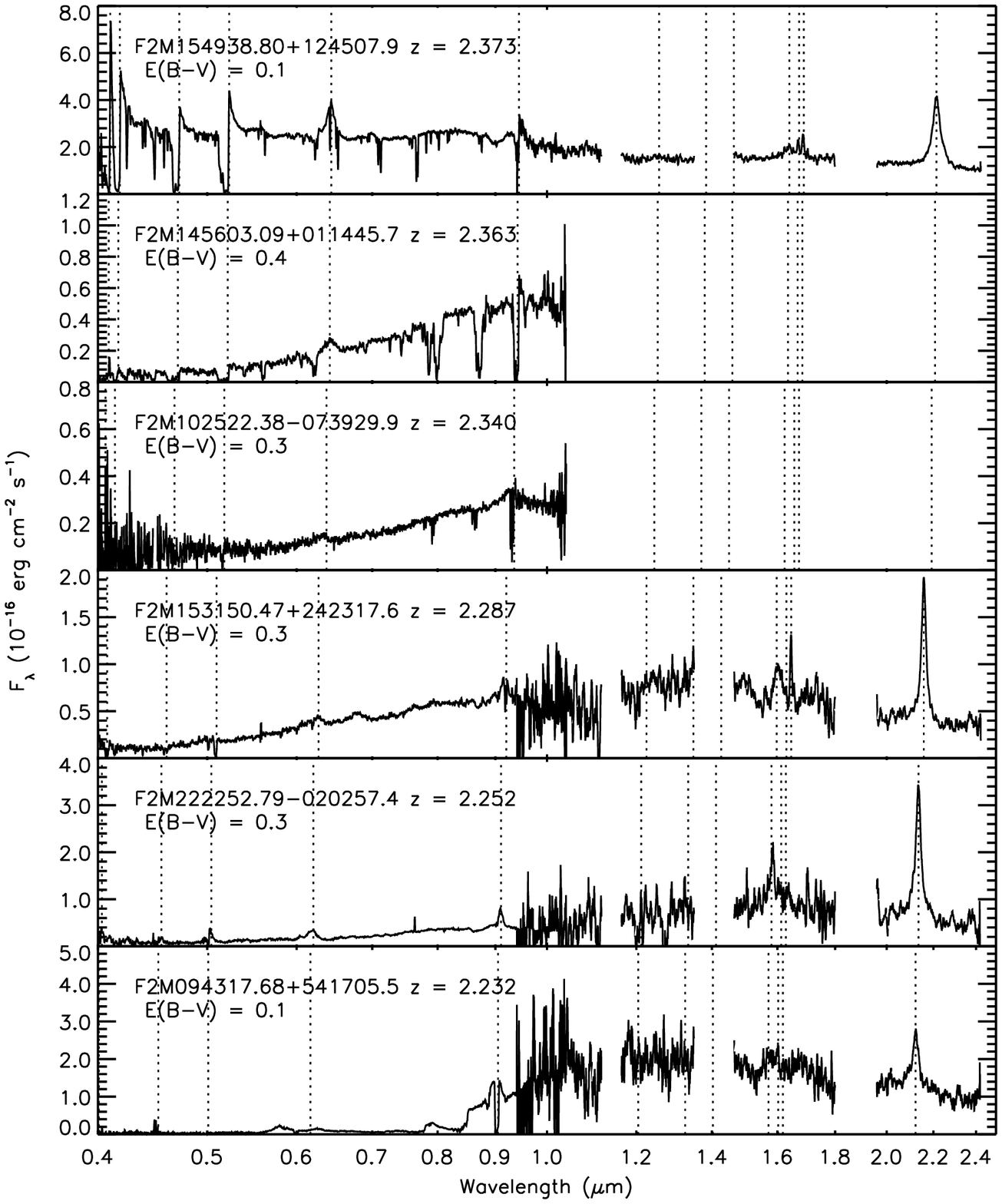}
\caption{{\it Continued.} Spectra of F2M quasars.}
\end{figure}

\begin{figure}
\figurenum{6.3}
\plotone{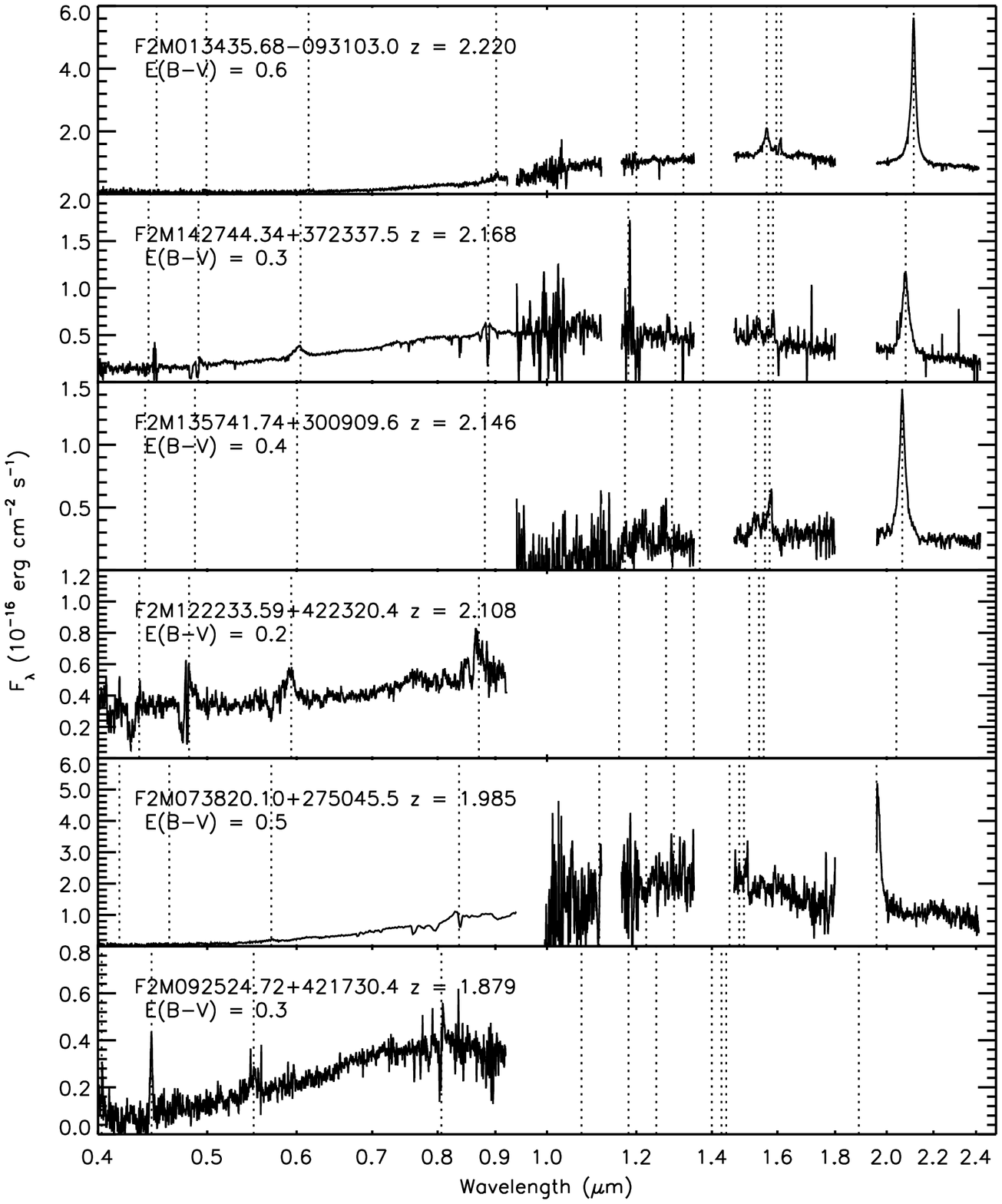}
\caption{{\it Continued.} Spectra of F2M quasars.}
\end{figure}

\begin{figure}
\figurenum{6.4}
\plotone{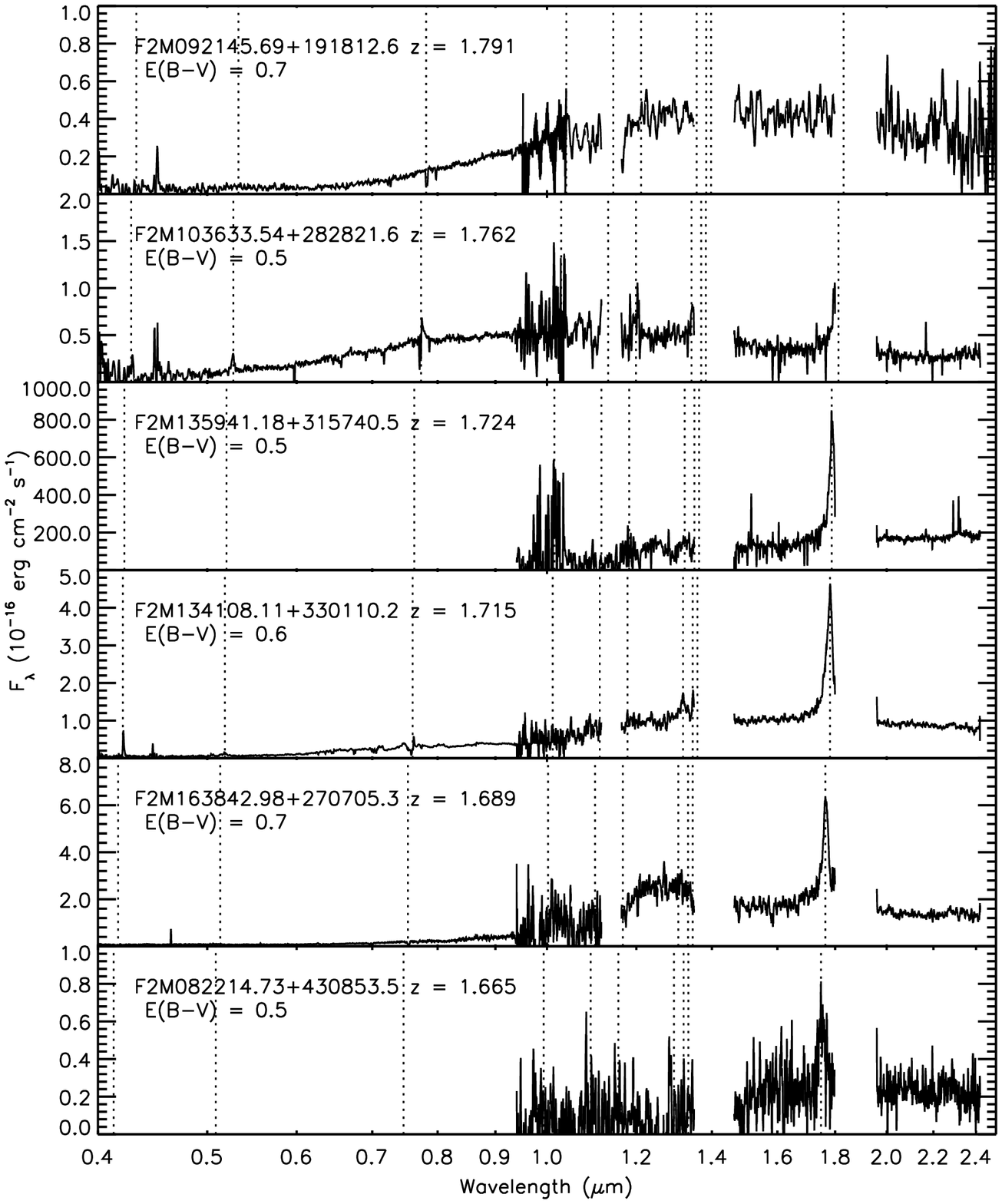}
\caption{{\it Continued.} Spectra of F2M quasars.}
\end{figure}

\begin{figure}
\figurenum{6.5}
\plotone{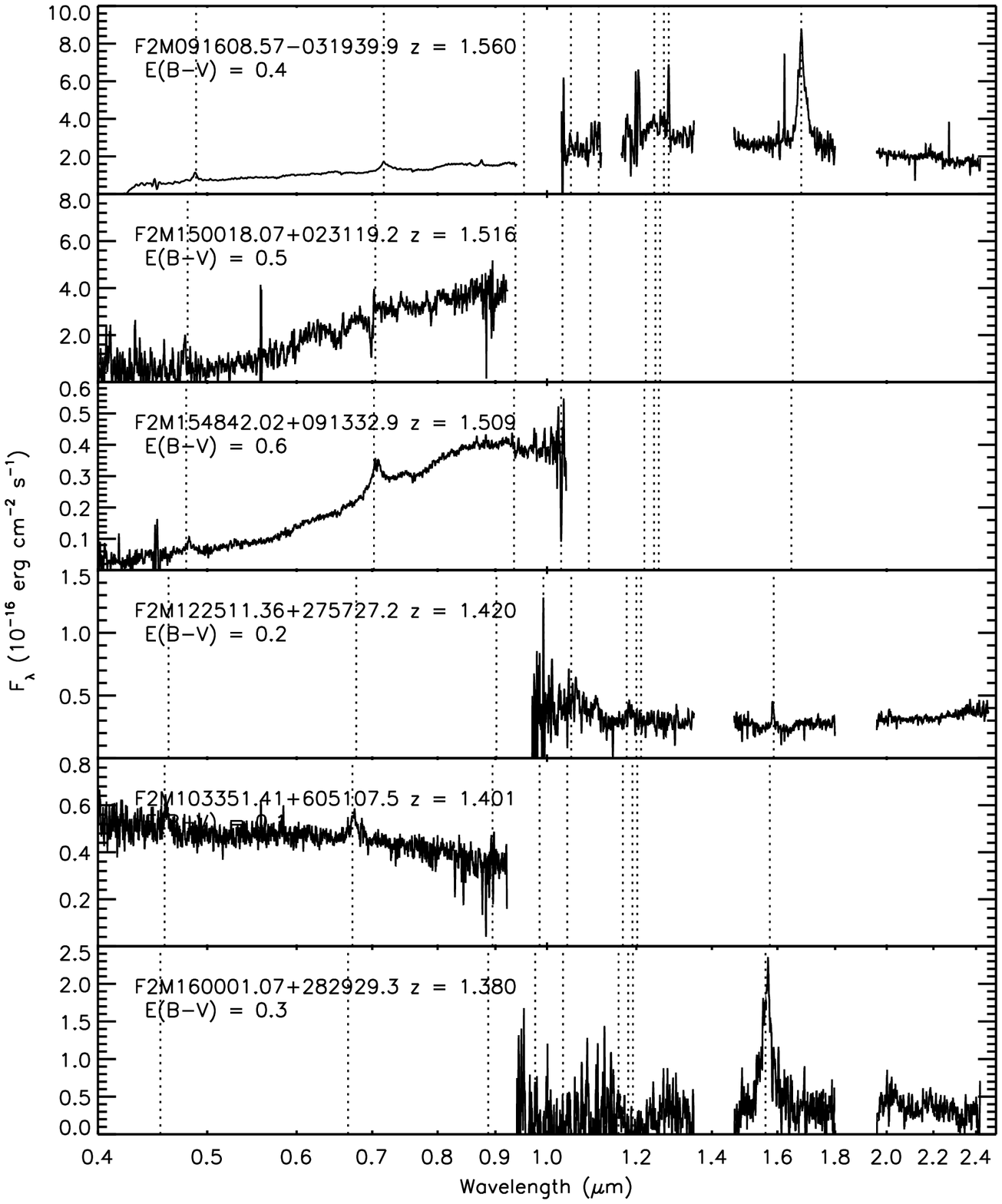}
\caption{{\it Continued.} Spectra of F2M quasars.}
\end{figure}

\begin{figure}
\figurenum{6.6}
\plotone{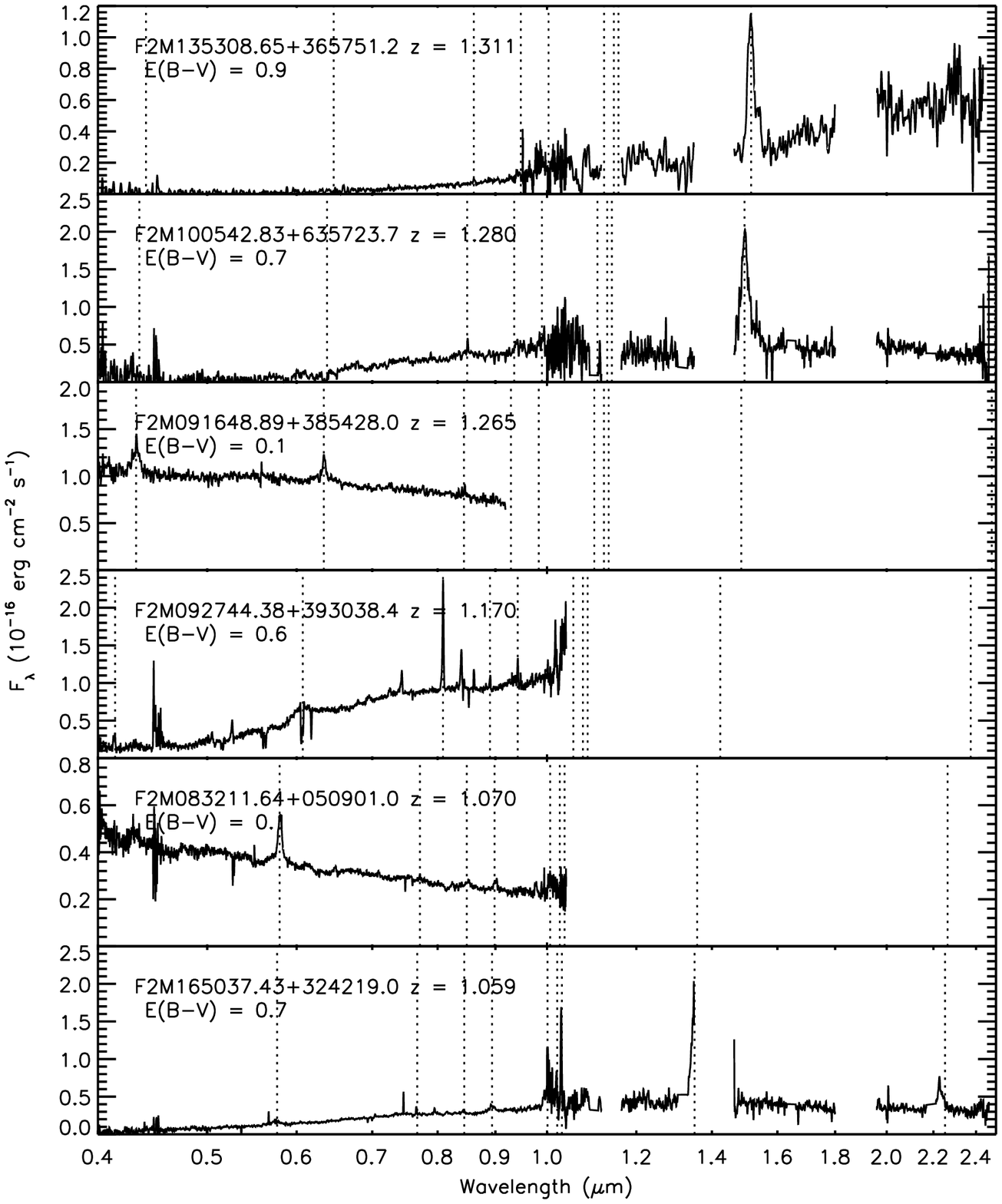}
\caption{{\it Continued.} Spectra of F2M quasars.}
\end{figure}

\begin{figure}
\figurenum{6.7}
\plotone{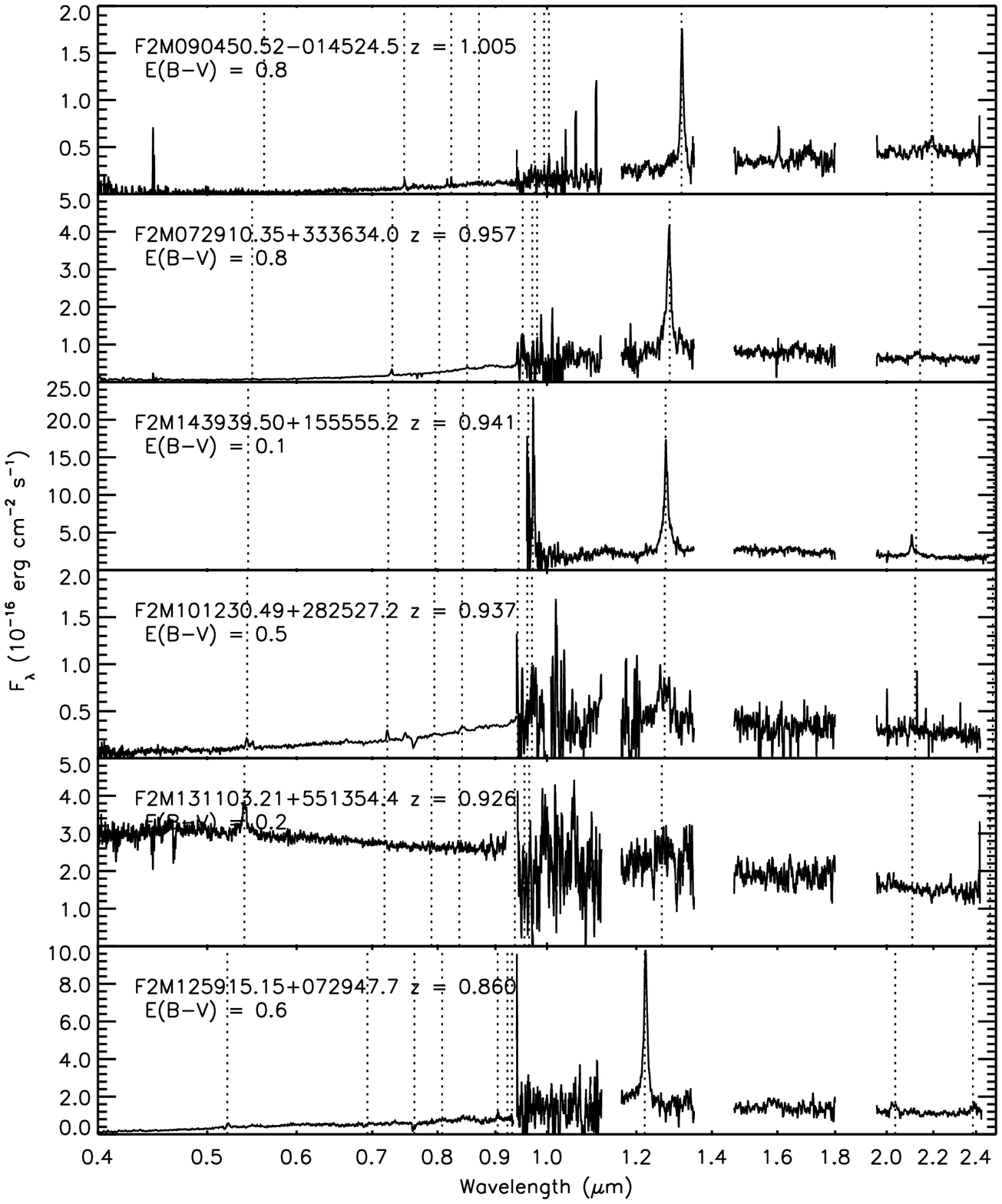}
\caption{{\it Continued.} Spectra of F2M quasars.}
\end{figure}

\begin{figure}
\figurenum{6.8}
\plotone{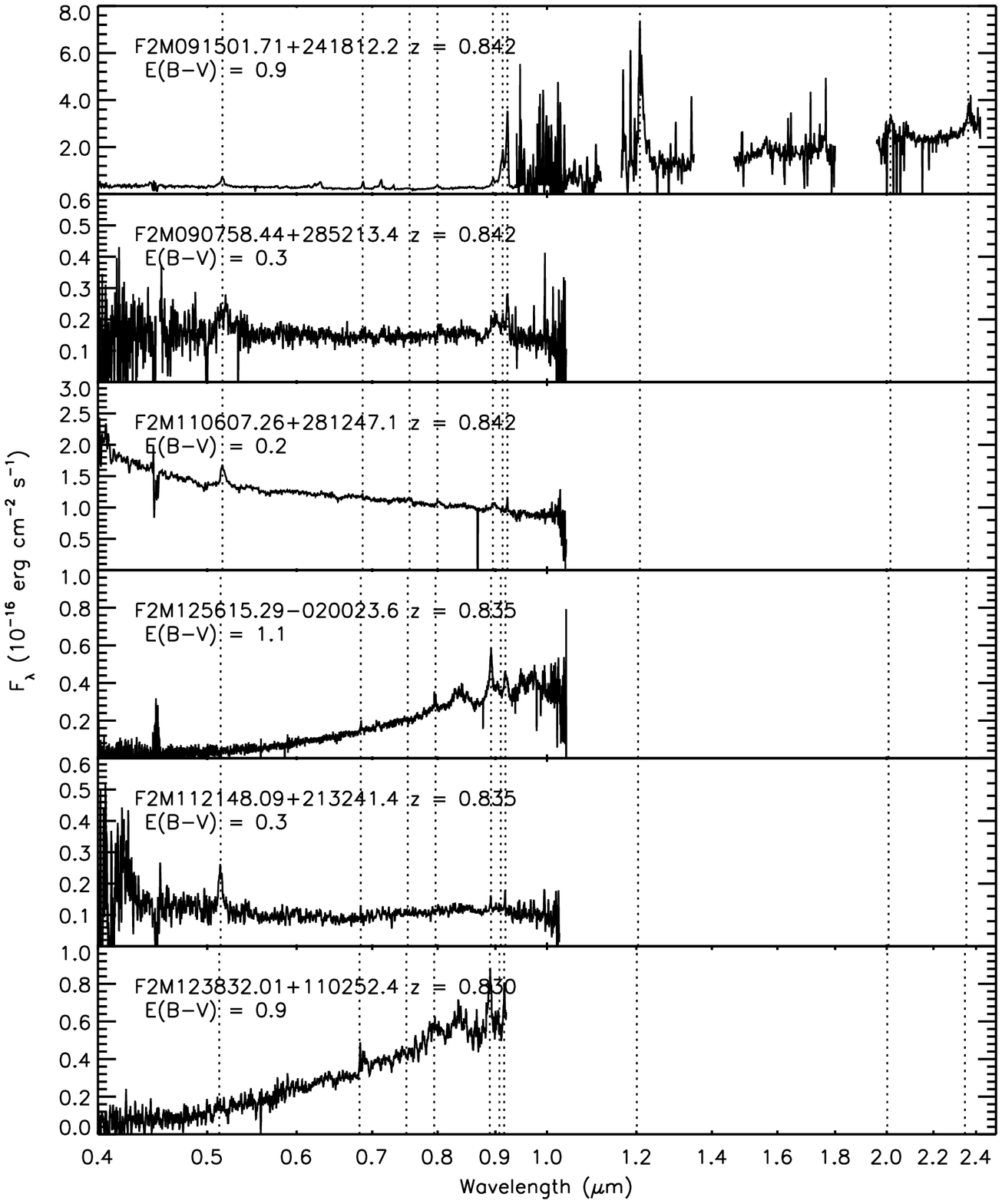}
\caption{{\it Continued.} Spectra of F2M quasars.}
\end{figure}

\begin{figure}
\figurenum{6.9}
\plotone{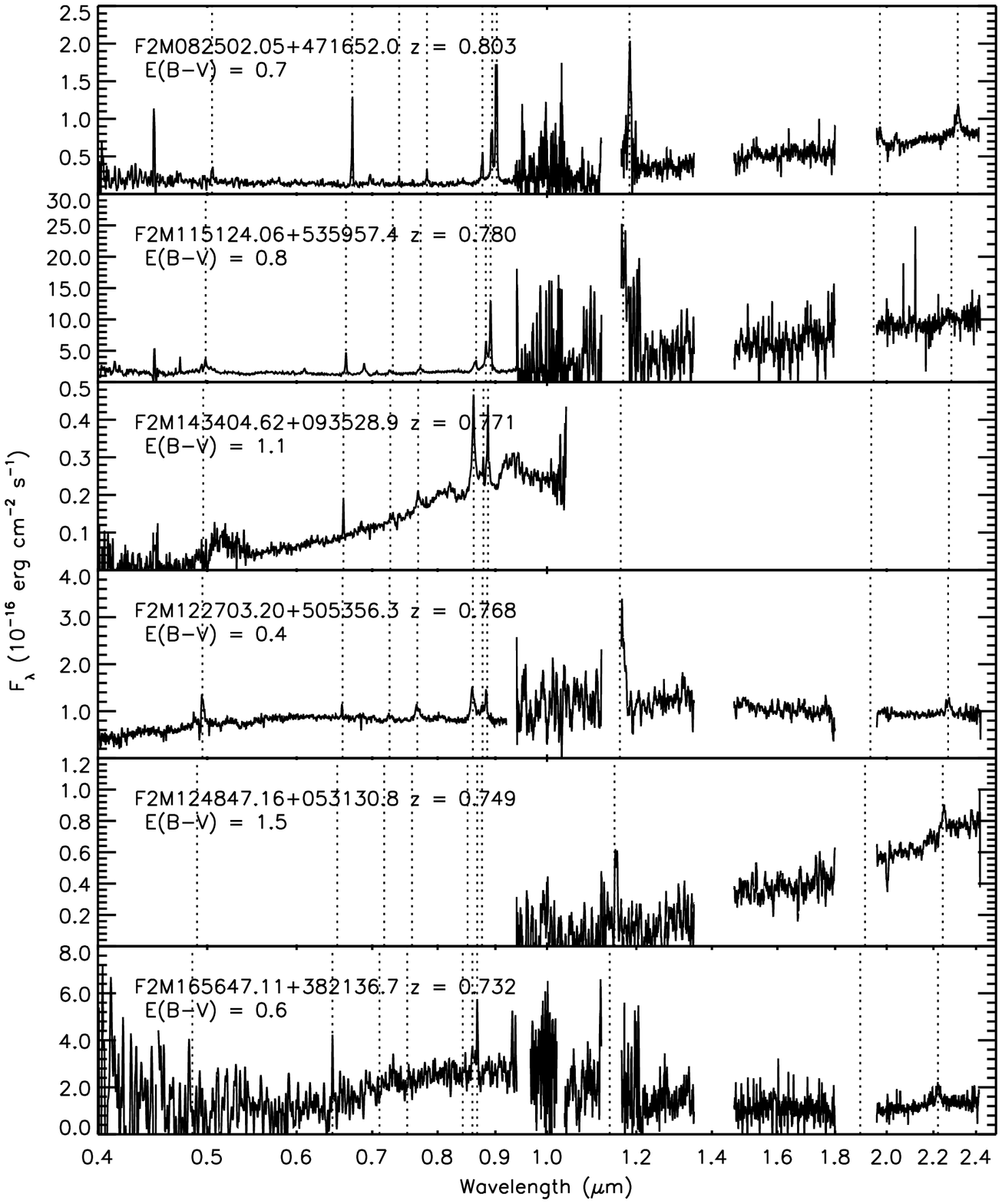}
\caption{{\it Continued.} Spectra of F2M quasars.}
\end{figure}

\begin{figure}
\figurenum{6.10}
\plotone{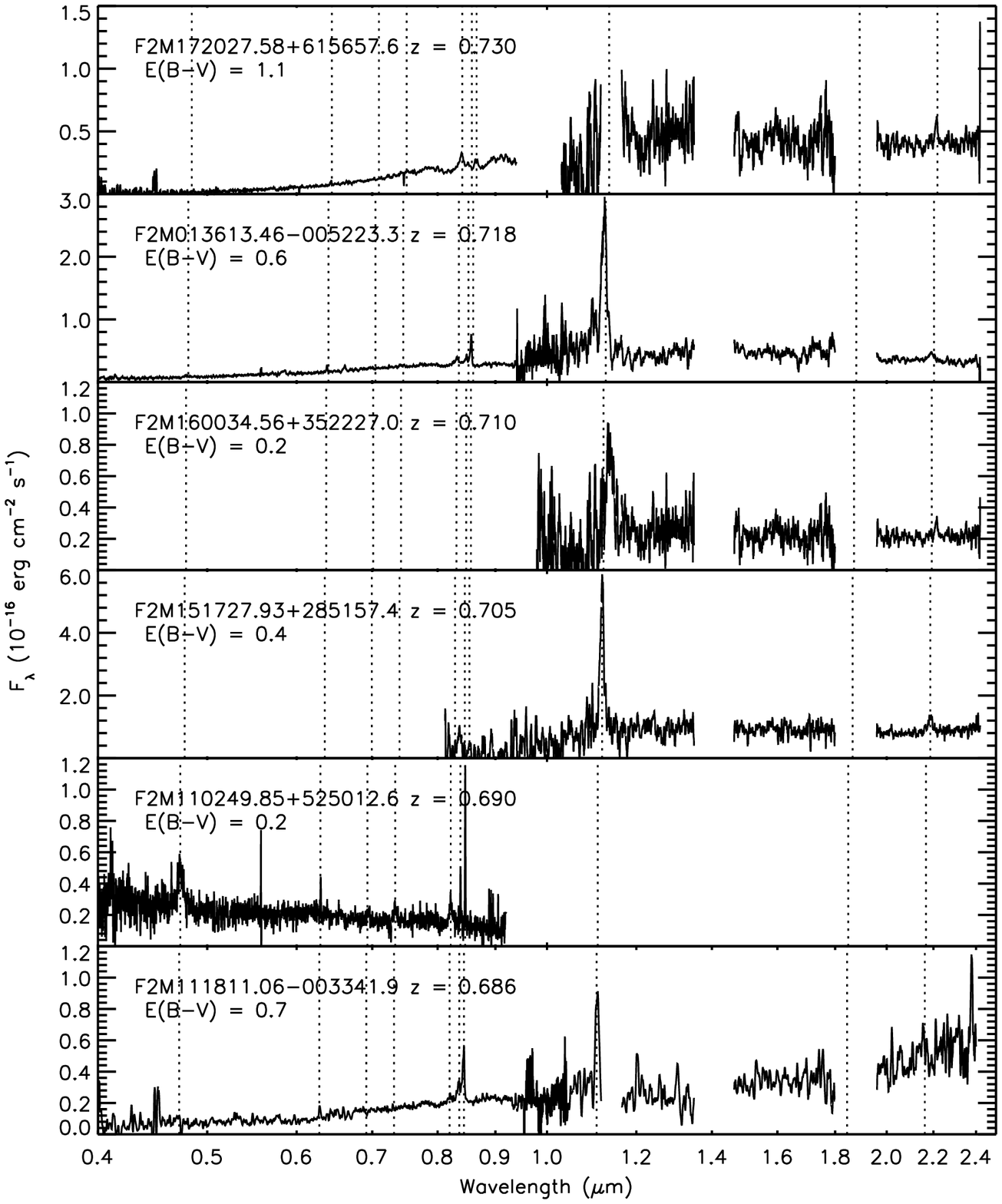}
\caption{{\it Continued.} Spectra of F2M quasars.}
\end{figure}

\clearpage

\begin{figure}
\figurenum{6.11}
\plotone{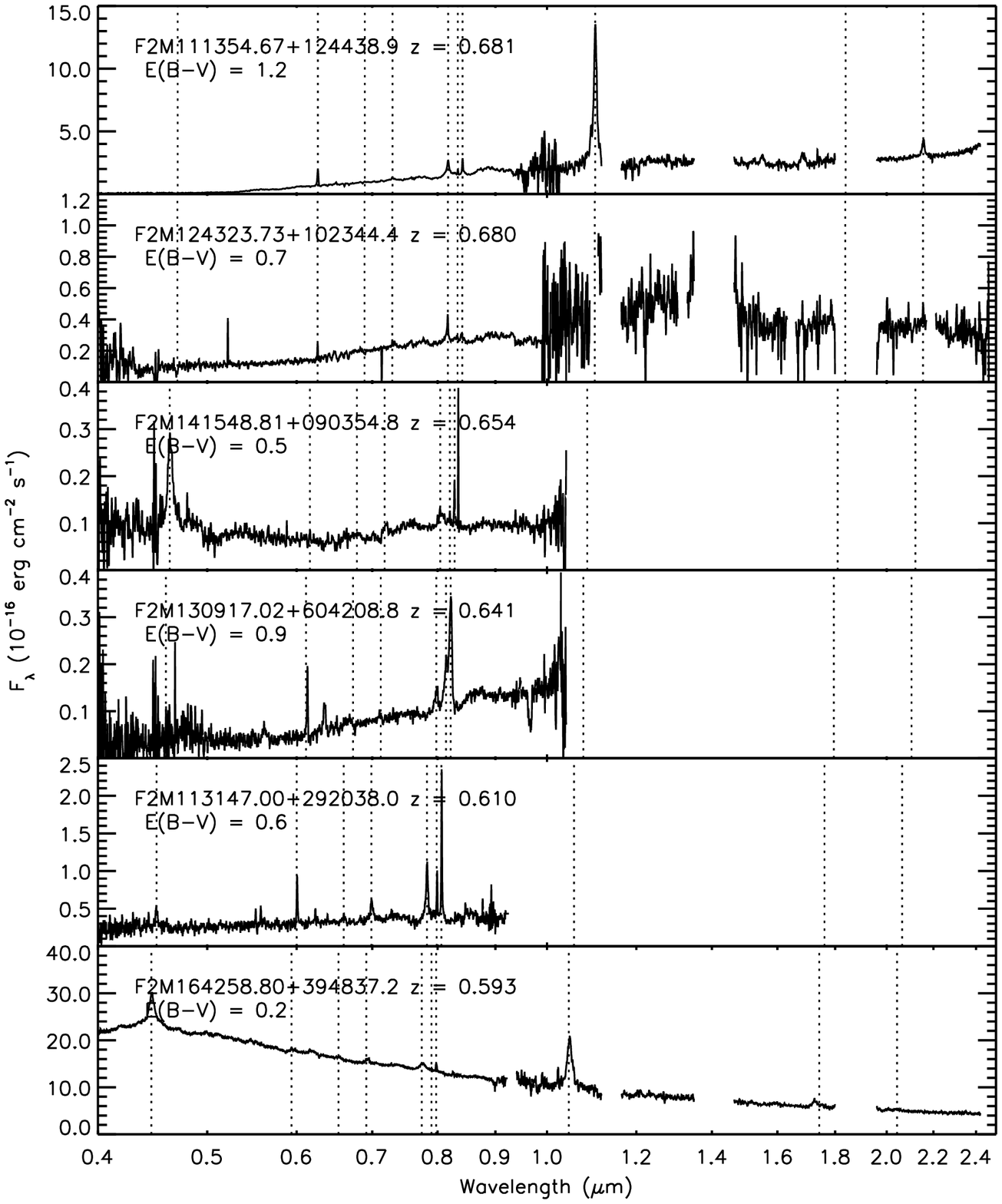}
\caption{{\it Continued.} Spectra of F2M quasars.}
\end{figure}

\begin{figure}
\figurenum{6.12}
\plotone{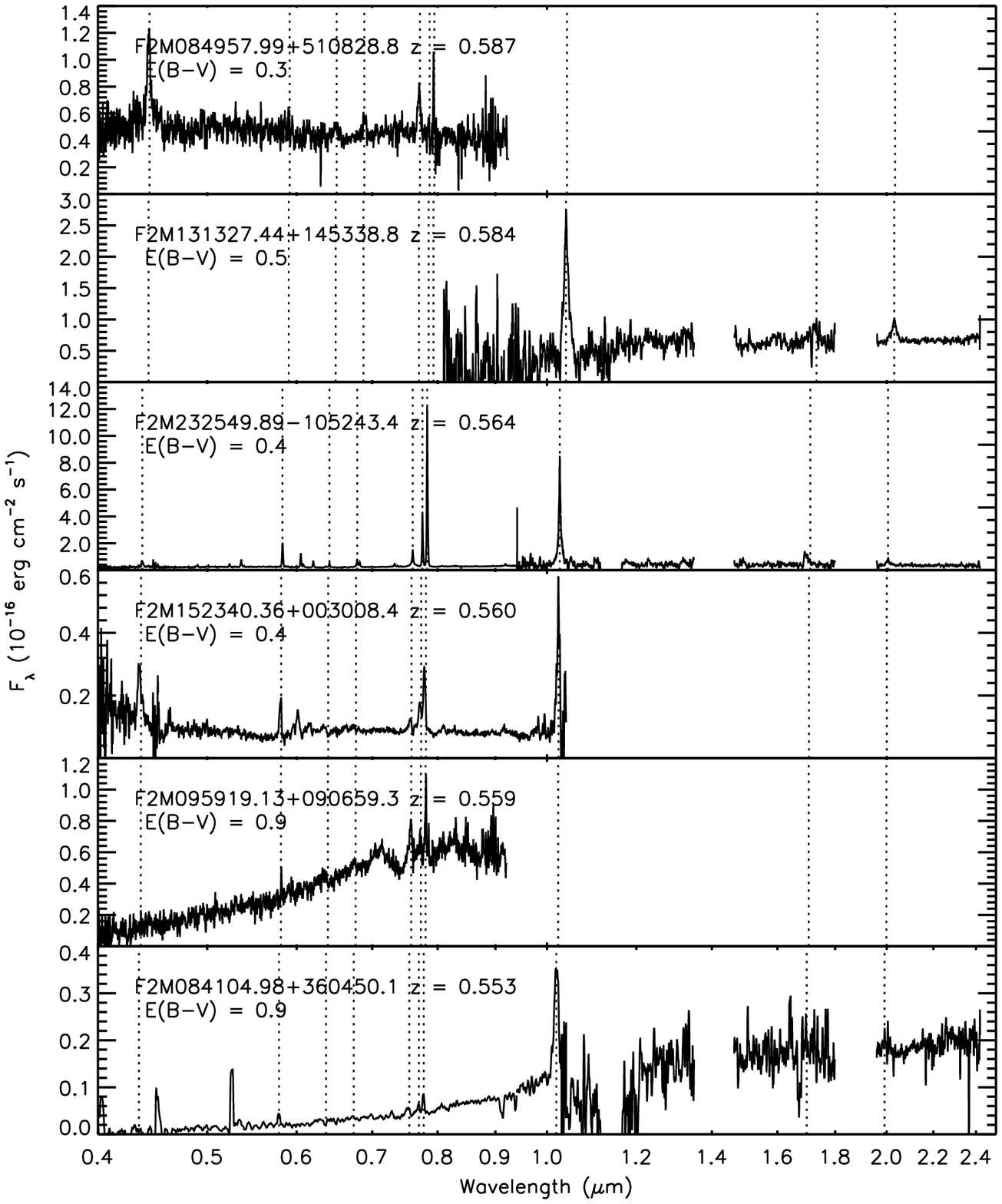}
\caption{{\it Continued.} Spectra of F2M quasars.}
\end{figure}

\begin{figure}
\figurenum{6.13}
\plotone{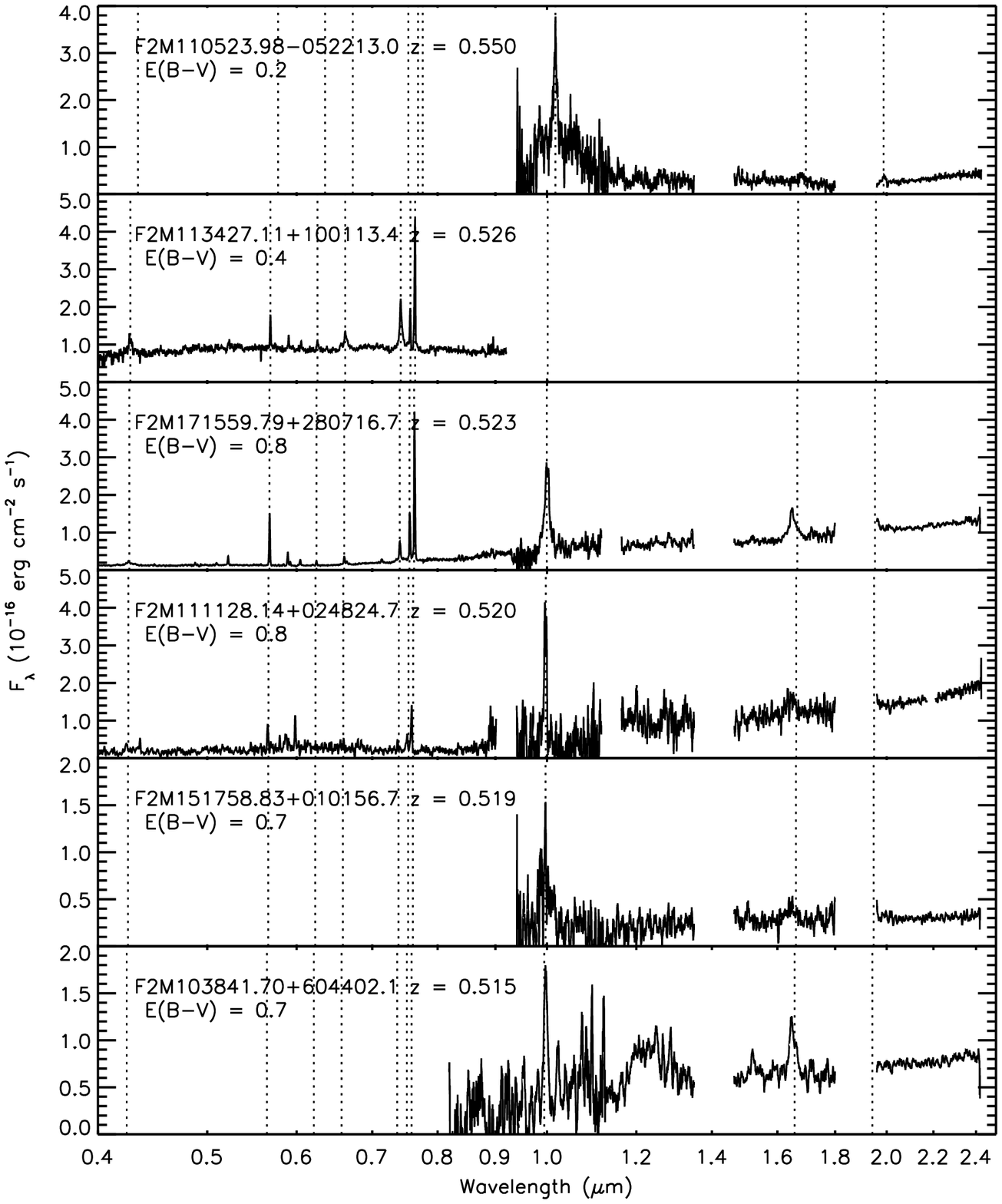}
\caption{{\it Continued.} Spectra of F2M quasars.}
\end{figure}

\begin{figure}
\figurenum{6.14}
\plotone{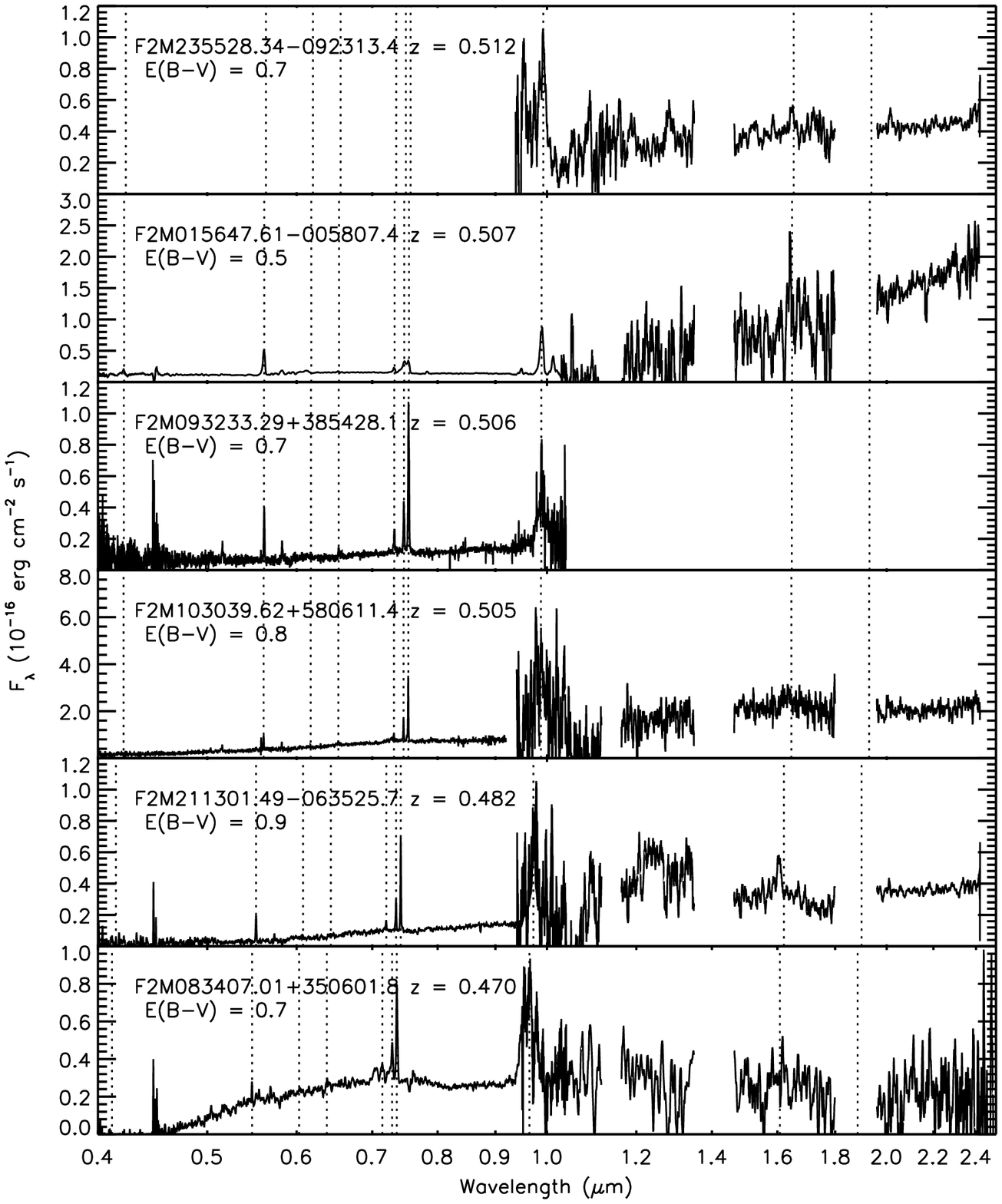}
\caption{{\it Continued.} Spectra of F2M quasars.}
\end{figure}
\clearpage

\begin{figure}
\figurenum{6.15}
\plotone{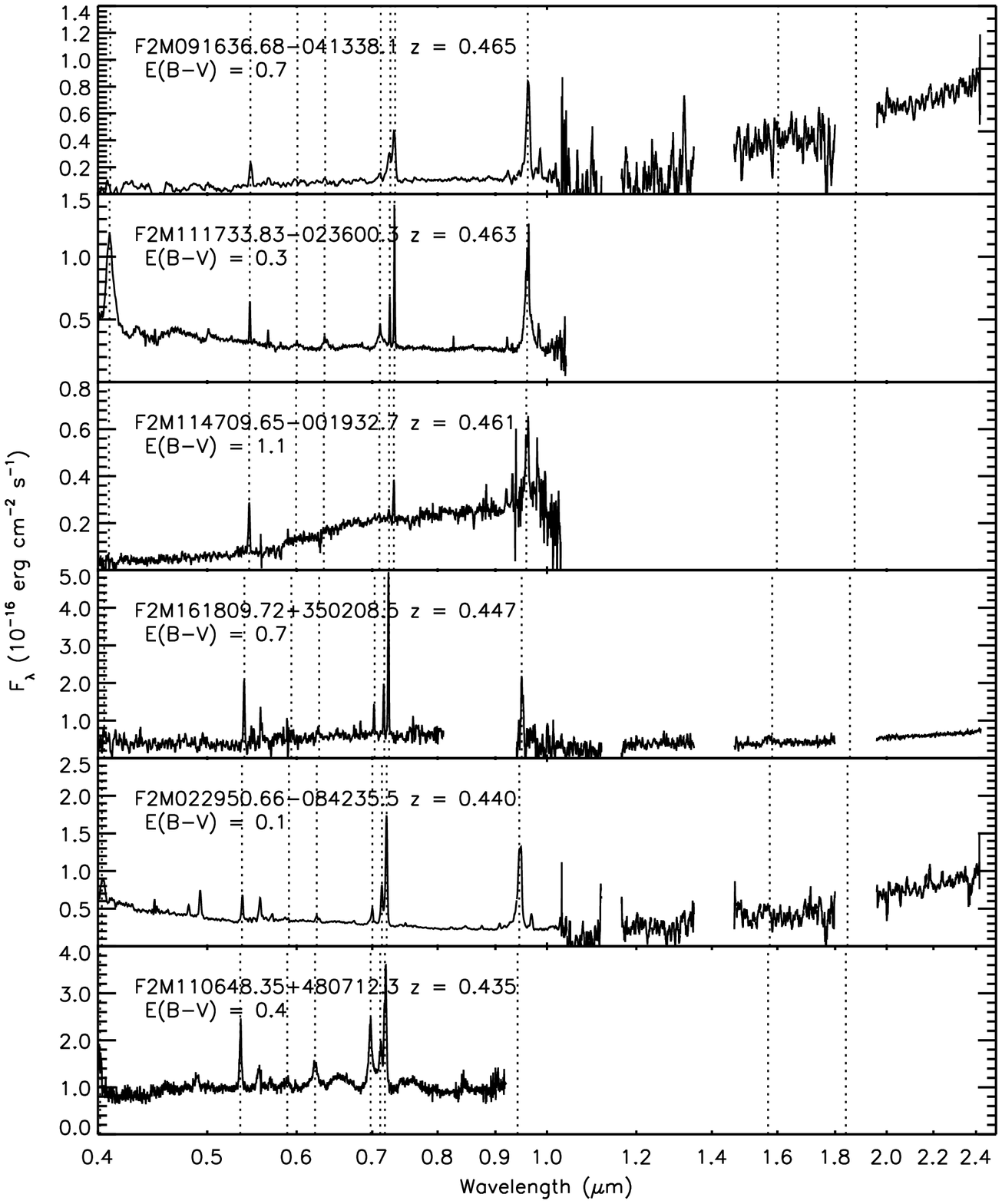}
\caption{{\it Continued.} Spectra of F2M quasars.}
\end{figure}

\begin{figure}
\figurenum{6.16}
\plotone{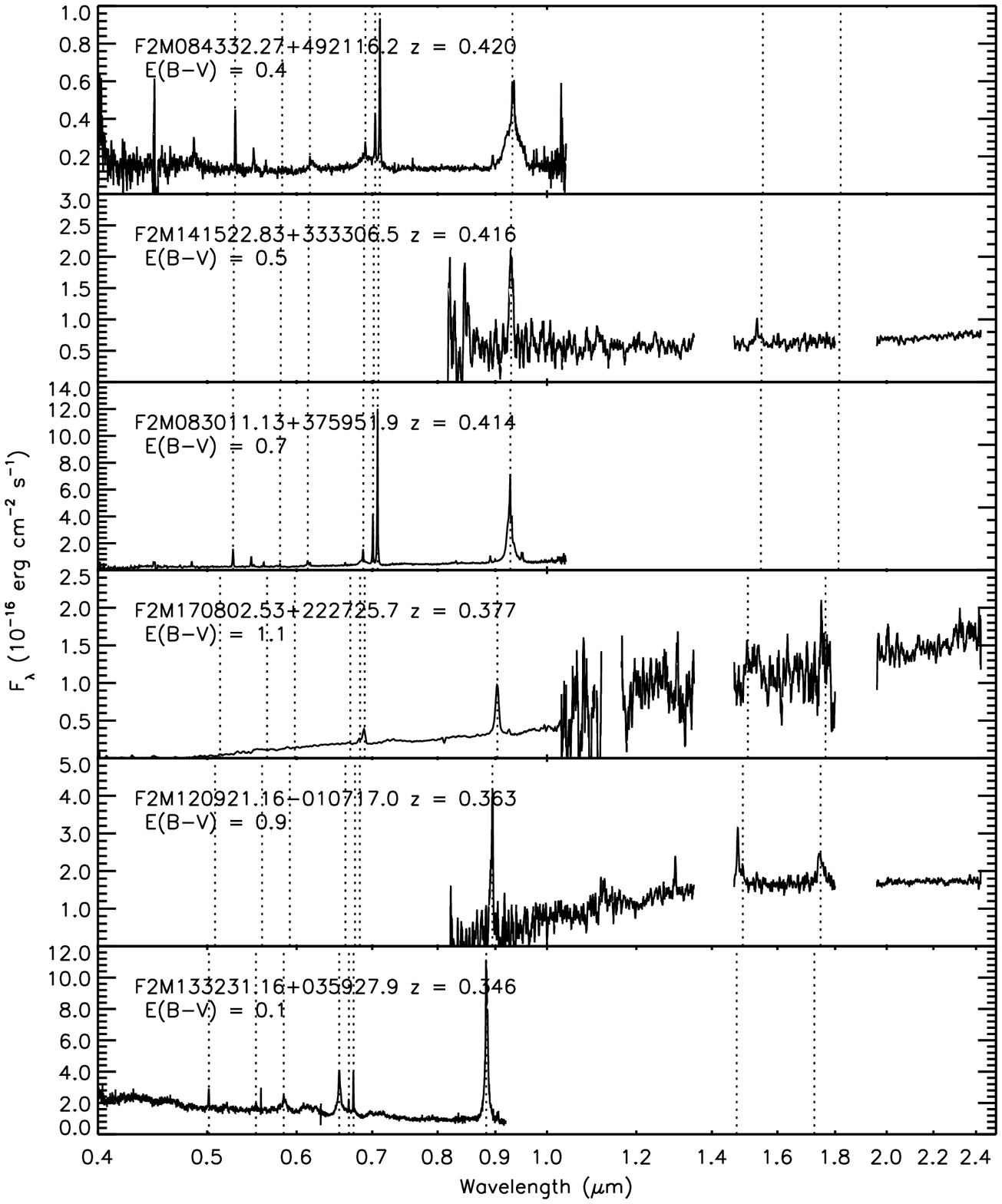}
\caption{{\it Continued.} Spectra of F2M quasars.}
\end{figure}

\begin{figure}
\figurenum{6.17}
\plotone{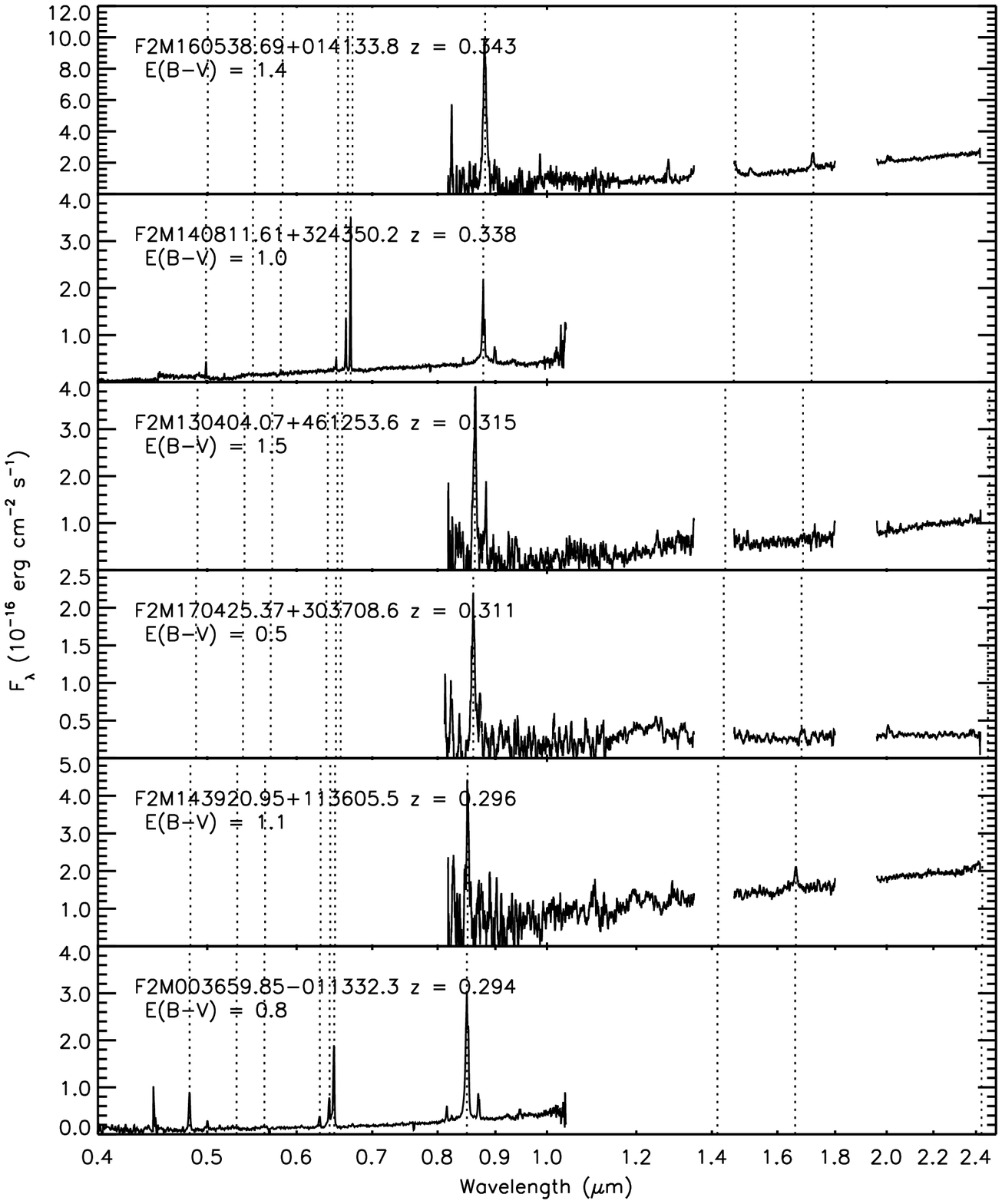}
\caption{{\it Continued.} Spectra of F2M quasars.}
\end{figure}

\begin{figure}
\figurenum{6.18}
\plotone{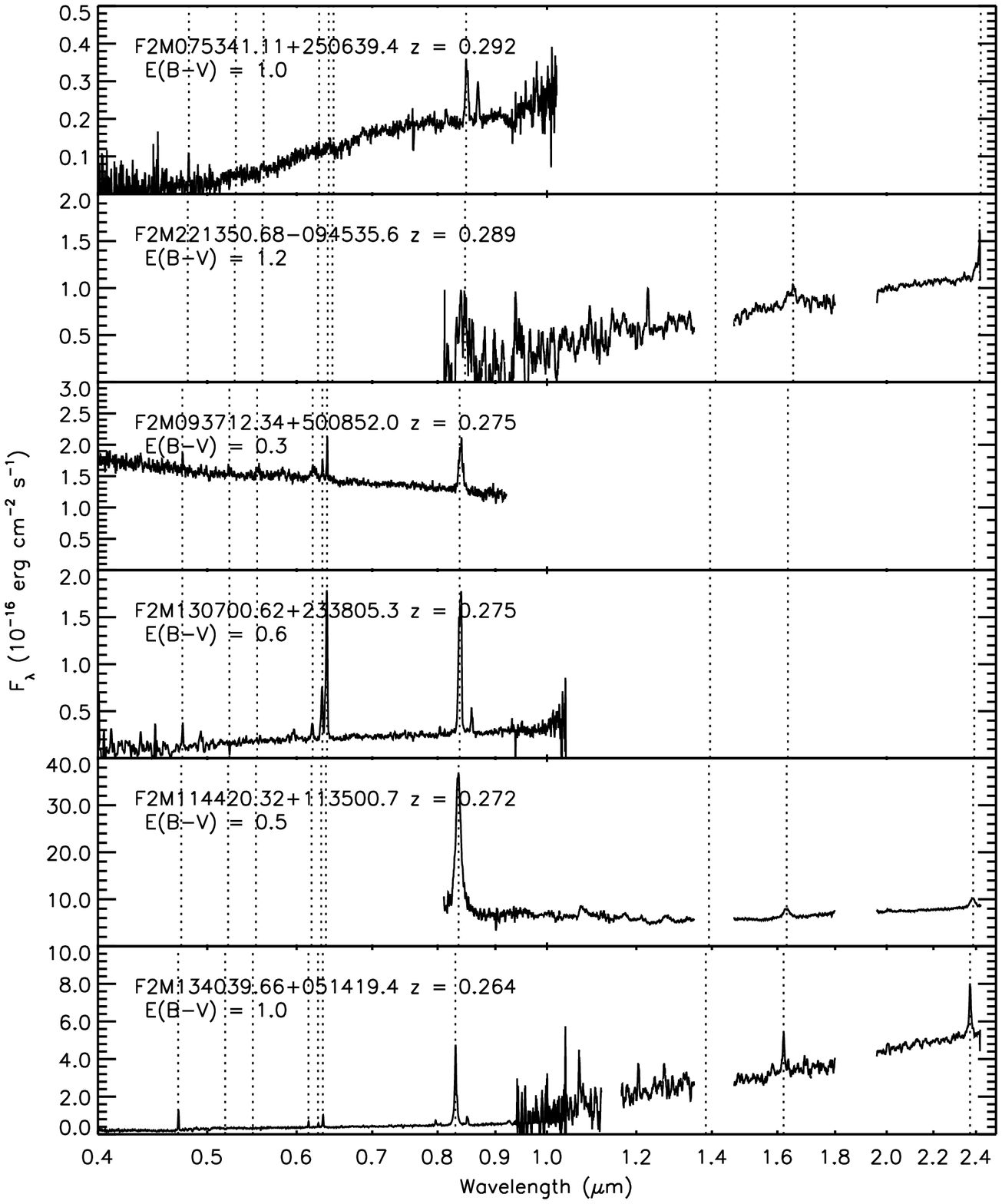}
\caption{{\it Continued.} Spectra of F2M quasars.}
\end{figure}

\begin{figure}
\figurenum{6.19}
\plotone{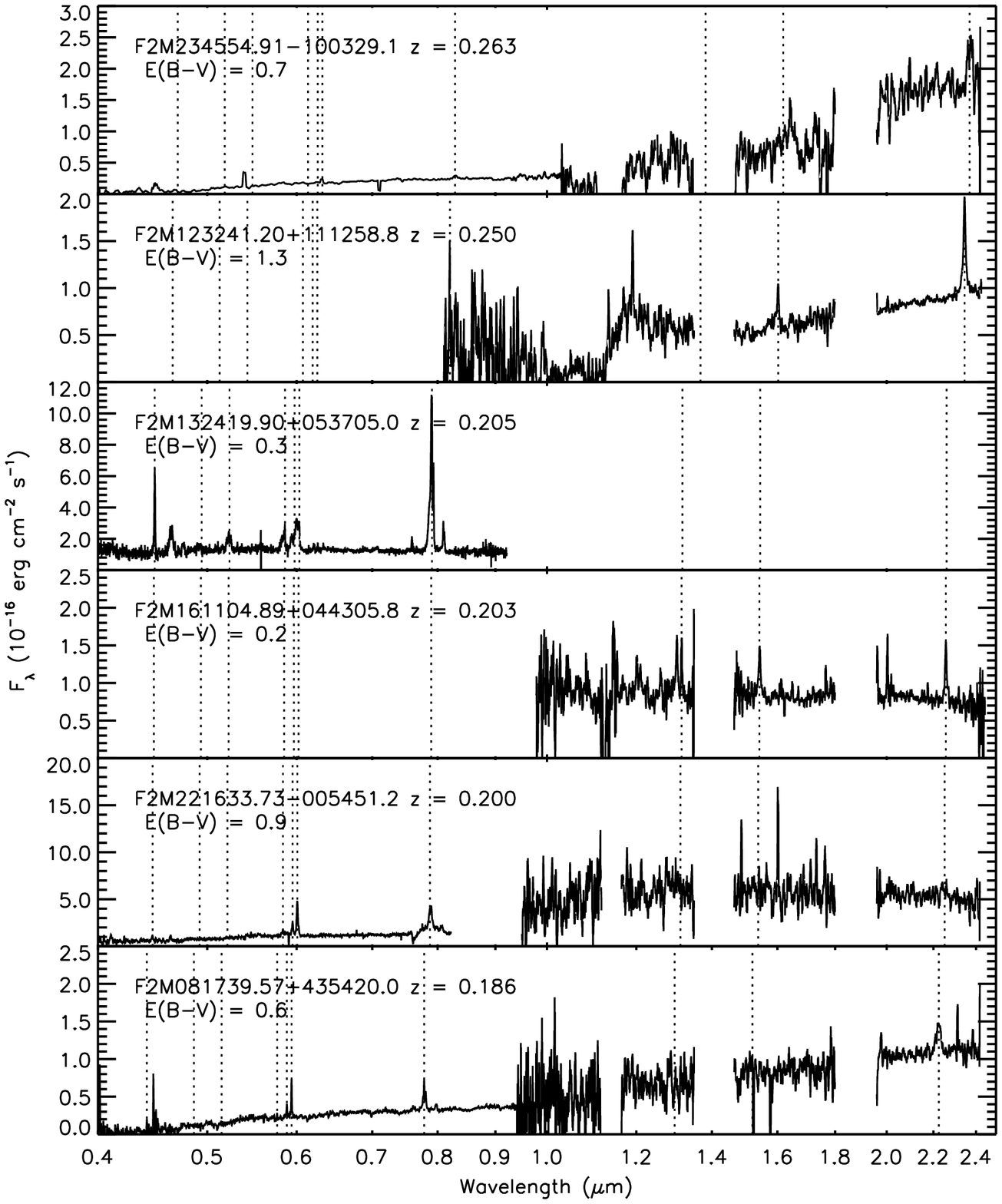}
\caption{{\it Continued.} Spectra of F2M quasars.}
\end{figure}

\begin{figure}
\figurenum{6.20}
\plotone{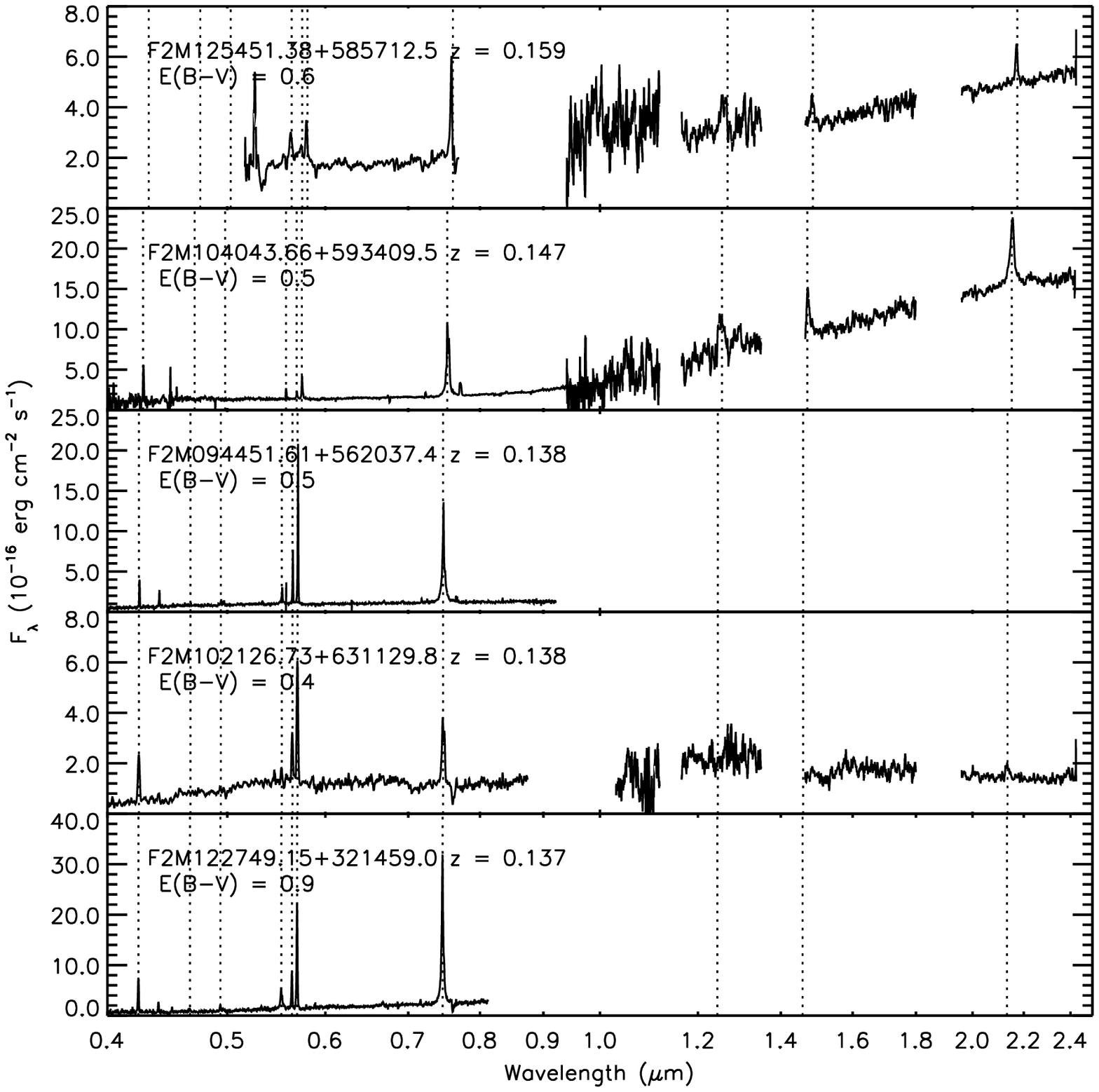}
\caption{{\it Continued.} Spectra of F2M quasars.}
\end{figure}

\end{document}

%% file: tab1.tex



\begin{deluxetable}{cccccccccccccccc}


\tabletypesize{\scriptsize}

\tablewidth{0pt}
\tablecaption{F2M Red Quasar Candidates \label{tab:candidates}}


\tablehead{\colhead{R.A.} & \colhead{Dec} & \colhead{$B$} & \colhead{$R$} & \colhead{$J$} & \colhead{$H$} & \colhead{$K_s$\tablenotemark{a} } & \colhead{$F_{pk}$\tablenotemark{a}} & \colhead{$F_{int}$} & \colhead{$J-K_s$} & \colhead{$R-K_s$} & \colhead{$z$} & \colhead{Class} & \colhead{Optical\tablenotemark{b}} & \colhead{Near$-$IR\tablenotemark{c}} & \colhead{Ref.} \\ 
\colhead{(J2000)} & \colhead{(J2000)} & \colhead{(mag)} & \colhead{(mag)} & \colhead{(mag)} & \colhead{(mag)} & \colhead{(mag)} & \colhead{(mJy)} & \colhead{(mJy)} & \colhead{(mag)} & \colhead{(mag)} & \colhead{} & \colhead{} & \colhead{} & \colhead{} & \colhead{} \\
\colhead{(1)} & \colhead{(2)} & \colhead{(3)} & \colhead{(4)} & \colhead{(5)} & \colhead{(6)} & \colhead{(7)} & \colhead{(8)} & \colhead{(9)} & \colhead{(10)} & \colhead{(11)} & \colhead{(12)} & \colhead{(13)} & \colhead{(14)} & \colhead{(15)} & \colhead{(16)} 
} 

\startdata
00:23:47.96 & $-$01:45:42.7 & 21.44 & 19.71 & 17.25 & 16.05 & 15.07 & 1.09 & 0.79 & 2.18 & 4.64 & 0.342 & Galaxy & O10 & \ldots & . \\
00:25:34.91 & $-$00:52:51.6 & 20.90 & 19.68 & 17.12 & 16.56 & 15.31 & 3.11 & 3.47 & 1.82 & 4.37 & \ldots & ? & O22 & IR4 & . \\
00:30:09.10 & $-$00:27:44.4 & 19.68 & 18.36 & 16.50 & 15.34 & 14.20 & 2.95 & 2.52 & 2.31 & 4.16 & 0.242 & Starb?ULIRG & \ldots & \ldots & C96 \\
00:31:32.62 & +01:30:01.1 & 21.84 & 19.79 & 17.52 & 16.79 & 15.75 & 1.03 & 0.55 & 1.77 & 4.04 & 0.38 & Galaxy & O22 & \ldots & . \\
00:36:59.85 & $-$01:13:32.3 & 22.50 & 19.53 & 16.57 & 15.10 & 13.63 & 1.92 & 0.82 & 2.94 & 5.90 & 0.294 & QSO & O5 & \ldots & . \\
00:41:12.35 & $-$00:04:37.0 & 22.50 & 19.89 & 17.47 & 16.45 & 15.62 & 1.11 & 1.55 & 1.85 & 4.27 & 0.579 & Galaxy & O22 & \ldots & . \\
00:44:02.81 & $-$10:54:18.9 & 22.34 & 19.10 & 17.68 & 16.07 & 15.07 & 38.60 & 42.57 & 2.61 & 4.03 & 0.431 & NLAGN & O7 & \ldots & . \\
\enddata

\end{deluxetable}

%% file: tab2.tex

\begin{deluxetable}{cccccccc}

\tabletypesize{\footnotesize}

\tablewidth{0pt}

\tablecaption{Line Diagnostics of Narrow-Line Spectra\label{tab:diagnostics}}


\tablehead{\colhead{Object} & \colhead{Log([\ion{O}{3}]/H$\beta$)} & \colhead{Log([\ion{N}{2}]/H$\alpha$)} & \colhead{Log([\ion{S}{2}]/H$\alpha$)} & \colhead{Log([\ion{O}{1}]/H$\alpha$)} & \colhead{Diagram 1} & \colhead{Diagram 2} & \colhead{Diagram 3} \\ 
\colhead{(1)} & \colhead{(2)} & \colhead{(3)} & \colhead{(4)} & \colhead{(5)} & \colhead{(6)} & \colhead{(7)} & \colhead{(8)} } 

\startdata
    F2M0044$-$1054 &   0.95 &\nodata &$-$0.31 &$-$1.34 & \nodata &     agn &     agn\\
{\em F2M0730+2538} &   0.73 &$-$0.68 &$-$1.02 &$-$1.91 &     agn &   starb &   starb\\ 
{\em F2M0738+2141} &   1.31 &$-$0.29 &$-$0.61 &$-$1.18 &     agn &     agn &     agn\\ 
{\em F2M0738+3156} &   1.04 &$-$0.41 &$-$0.86 &$-$1.53 &     agn &     agn &     agn\\ 
      F2M0809+6407 &   0.68 &$-$0.50 &$-$0.87 &$-$1.38 &     agn &   starb &     agn\\ 
{\em F2M0810+1603} &$-$0.02 &$-$0.22 &$-$0.56 &$-$1.31 &   starb &   starb &   starb\\ 
{\em F2M0817+1958} &$-$1.10 &$-$1.07 &$-$1.60 &$-$1.55 &   starb &   starb &   starb\\  
{\em F2M0818+1305} &   1.00 &$-$0.35 &$-$0.54 &$-$1.16 &     agn &     agn &     agn\\ 
     F2M0820+4728  &$-$0.29 &$-$0.13 &$-$0.45 &$-$1.12 &   starb &   starb &   starb\\
     F2M0828+6420  &   0.86 &   0.17 &$-$0.28 &$-$1.14 &     agn &     agn &     agn\\
F2M0908+2235\tablenotemark{b} &   0.77 & \nodata& \nodata& \nodata& \nodata & \nodata & \nodata\\
{\em F2M0915+1819} &$-$0.16 &$-$0.39 &$-$0.47 &$-$1.28 &   starb &   starb &   starb\\ 
{\em F2M0915+2850} &   1.42 &   0.23 &$-$0.08 &$-$0.83 &     agn &     agn &     agn\\ 
     F2M0917+2459  &   1.08 &$-$0.27 &$-$0.31 &$-$1.08 &     agn &     agn &     agn\\
{\em F2M0919+1531} &$-$0.09 &$-$0.33 &$-$0.44 &$-$1.32 &   starb &   starb &   starb\\ 
      F2M0929+6350 &   0.16 &$-$0.63 &$-$0.39 &$-$1.50 &   starb &   starb &   starb\\
{\em F2M0949+2325} &   1.76 &$-$0.24 &$-$0.31 &$-$1.33 &     agn &     agn &     agn\\ 
{\em F2M1004+2943} &   0.74 &$-$0.38 &$-$0.77 &$-$1.49 &     agn &     agn &     agn\\ 
{\em F2M1015+1207} &   0.98 &$-$0.46 &$-$0.28 &$-$0.88 &     agn &     agn &     agn\\ 
{\em F2M1018+2135} &   0.50 &$-$1.24 &$-$1.09 &$-$1.58 &   starb &   starb &   starb\\ 
{\em F2M1022+1628}\tablenotemark{a} &   0.98 & \nodata& \nodata& \nodata&agn&agn&agn\\
F2M1049+4010\tablenotemark{b}  &   0.75 & \nodata& \nodata& \nodata& \nodata & \nodata & \nodata\\
     F2M1052+0650  &   1.99 & \nodata&$-$1.17 &$-$1.05 &     agn &     agn &     agn\\
     F2M1125+0754  &   1.12 &$-$0.65 &$-$0.50 &$-$1.42 &     agn &     agn &     agn\\
     F2M1130+2548  &$-$0.06 &$-$1.01 &$-$1.13 &$-$1.82 &   starb &   starb &   starb\\
{\em F2M1139+2743} &   0.47 &$-$0.56 &$-$0.41 &$-$0.82 &   starb &     agn &     agn\\ 
     F2M1155+0829  &   2.17 & \nodata&   1.56 &   2.84 &     agn &     agn &     agn\\
F2M1201m0332\tablenotemark{b}  &   0.89 & \nodata& \nodata& \nodata& \nodata & \nodata & \nodata\\
{\em F2M1202+2615}\tablenotemark{a} &   1.10 & \nodata& \nodata& \nodata&agn&agn&agn\\
     F2M1206p2857  & \nodata&$-$1.36 &$-$1.64 &\nodata &   Starb &   Starb & \nodata\\
    F2M1331$-$0232 &   1.14 &$-$0.27 &$-$0.49 &$-$1.03 &     agn &     agn &     agn\\
     F2M1354+0011  &   0.55 & \nodata&$-$1.33 &$-$1.80 & \nodata &   starb &   starb\\
    F2M1446$-$0050 &$-$2.54 &$-$0.23 &   0.60 &$-$0.54 &   starb &   starb &   starb\\
      F2M1448+3056 &   0.58 &$-$0.03 &$-$0.26 &$-$0.71 &     agn &     agn &     agn\\
     F2M1518+2336  &$-$0.01 &$-$0.41 &$-$0.62 &$-$1.71 &   starb &   starb &   starb\\
     F2M1612+3850\tablenotemark{a}  &  1.31 & \nodata& \nodata& \nodata&agn&agn&agn\\
     F2M1638+1058  &   0.71 &$-$0.33 &$-$0.63 &$-$1.02 &     agn &     agn &     agn\\
{\em F2M2328$-$1107} &   0.63 &$-$0.11 &$-$0.45 &$-$0.99 &     agn &     agn &     agn\\ 
\enddata 
\tablenotetext{a}{These objects were at redshifts higher than 0.55 and H$\alpha$ was outside our spectroscopic range.  Nevertheless, we classify these sources as AGN since their [\ion{O}{3}]/H$\beta$ ratios were above the extreme starburst dividing line shown in Figure \ref{fig:bpt}}
\tablenotetext{b}{These objects are similar to those in $^{\rm a}$, except their [\ion{O}{3}]/H$\beta$ ratios were too ambiguous to determine a clear classification.}
\tablecomments{Diagram 1 is the classification determined from the [\ion{N}{2}]/H$\alpha$ versus [\ion{O}{3}]/H$\beta$ diagnostic.\\
Diagram 2 is the classification determined from the  [\ion{S}{2}] /H$\alpha$ versus [\ion{O}{3}]/H$\beta$ diagnostic.
Diagram 3 is the classification determined from the  [\ion{O}{1}] /H$\alpha$ versus [\ion{O}{3}]/H$\beta$ diagnostic.}


\end{deluxetable}

%% file: tab3.tex



\begin{deluxetable}{cccrrrc}


\tabletypesize{\footnotesize}

\tablewidth{0pt} 

\tablecaption{Extinction Parameters for F2M Quasars \label{tab:ebv}}

\tablenum{3}

\tablehead{\colhead{Name} & \colhead{$z$} & \colhead{$K_s$} & \colhead{$A_K$} & \colhead{$R-K_s$} & \colhead{$E(B-V)$} & \colhead{Spectrum } \\ 
\colhead{} & \colhead{} & \colhead{(mag)} & \colhead{(mag)} & \colhead{(mag)} & \colhead{(mag)} & \colhead{(Used)} \\
 \colhead{(1)} & \colhead{(2)} & \colhead{(3)} & \colhead{(4)} & \colhead{(5)} & \colhead{(6)} & \colhead{(7)} } 

\startdata
F2M003659.85$-$011332.3 &  0.294 &  13.63 &     0.40 &     5.90 &     0.79 &  Comb.\\ 
F2M013435.68$-$093103.0 &  2.220 &  13.58 &     1.30 &  $>$7.22 &     0.57 &  Comb.\\ 
F2M013613.46$-$005223.3 &  0.718 &  15.51 &     0.51 &     4.64 &     0.64 &  Comb.\\ 
F2M015647.61$-$005807.4 &  0.507 &  14.87 &     0.30 &     5.01 &     0.47 &  Opt.\\ 
F2M022950.66$-$084235.5 &  0.440 &  14.86 &     0.09 &     4.32 &     0.15 &  Opt.\\ 
F2M072910.35$+$333634.0 &  0.957 &  14.52 &     0.83 &     5.64 &     0.84 &  Opt.\\ 
\enddata




\end{deluxetable}

%% file: tab4.tex



\begin{deluxetable}{cccccc}


\tabletypesize{\footnotesize}

\tablewidth{0pt} 

\tablecaption{QSO Counts and Surface Densities in K Band  \label{tab:surfdens}}

\tablenum{3}

\tablehead{\colhead{$K$ Magnitude Range} & \colhead{FBQS II} & \colhead{FBQS III} & \colhead{SDSS\tablenotemark{a}} & \colhead{F2M} & \colhead{Absorption Corrected F2M } \\ 
\colhead{(mag)} & \colhead{(deg$^{-2}$ 0.5 mag$^{-1}$)} & \colhead{(deg$^{-2}$ 0.5 mag$^{-1}$)} & \colhead{(deg$^{-2}$ 0.5 mag$^{-1}$)} & \colhead{(deg$^{-2}$ 0.5 mag$^{-1}$)} & \colhead{(deg$^{-2}$ 0.5 mag$^{-1}$)} } 

\startdata
$11.0-11.5$ &  $0.0012\pm0.0007$ &  \ldots          &  $0.0003\pm0.0002$ &  \ldots                     &  \ldots\\ 
$11.5-12.0$ &  $0.0004\pm0.0004$ &  $0.002\pm0.002$ &  $0.0003\pm0.0002$ &  $0.0003\pm0.0002$          &  $0.0004\pm0.0002$\\ 
$12.0-12.5$ &  $0.007\pm0.001$   &  \ldots          &  $0.0031\pm0.0007$ &  $0.0001\pm0.0001$          &  $0.0001\pm0.0001$\\ 
$12.5-13.0$ &  $0.004\pm0.001$   &  \ldots          &  $0.0053\pm0.0009$ &  $0.0001\pm0.0001$          &  $0.0006\pm0.0003$\\ 
$13.0-13.5$ &  $0.00655\pm0.002$ &  \dots           &  $0.011\pm0.001$   &  $0.0006\pm0.0003$          &  $0.0023\pm0.0006$\\ 
$13.5-14.0$ &  $0.0152\pm0.002$  &  $0.008\pm0.004$ &  $0.018\pm0.002$   &  $0.0022^{+0.0007}_{-0.0006}$ &  $0.0033\pm0.0007$\\ 
$14.0-14.5$ &  $0.033\pm0.004$   &  $0.036\pm0.008$ &  $0.038\pm0.002$   &  $0.0028^{+0.0008}_{-0.0006}$ &  $0.0046\pm0.0008$\\ 
$14.5-15.0$ &  $0.049\pm0.004$   &  $0.05\pm0.01$   &  $0.068\pm0.003$   &  $0.0046^{+0.0046}_{-0.0008}$ &  $0.0045\pm0.0008$\\ 
$15.0-15.5$ &  $0.060\pm0.005$   &  $0.08\pm0.01$   &  $0.123\pm0.004$   &  $0.007^{+0.008}_{-0.001}$    &  $0.0013\pm0.0004$\\ 
$15.5-16.0$ &  $0.040\pm0.004$   &  $0.10\pm0.02$   &  $0.124\pm0.005$   &  $0.0003^{+0.0025}_{-0.0002}$ &  \ldots\\ 
\hline
\phm{0.00}$-14.5$ & $0.064\pm0.005$ & $0.046\pm0.009$ &  $0.076\pm0.003$ & $0.0061^{+0.0011}_{-0.0009}$ & $0.011\pm0.001$ \\
\hline
\phm{0.00}$-15.0$ & $0.113\pm0.007$ & $0.10\pm0.01$ &  $0.144\pm0.005$ & $0.011^{+0.005}_{-0.001}$ & $0.016\pm0.002$ \\
\enddata

\tablenotetext{a}{This sample is made up of FIRST-detected quasars from the SDSS.  See \S4 for how we derive the space densities for these quasars.}


\end{deluxetable}

%% file: tab5.tex



\begin{deluxetable}{ccccccccc}


\tabletypesize{\small}

\tablewidth{0pt} 

\tablecaption{Absorption Properties of F2M Quasars with $z > 0.9$ \label{tab:bals}}

\tablenum{4}

\tablehead{\colhead{} & \colhead{} & \multicolumn{2}{c}{Balnicity Index} & \multicolumn{2}{c}{Absorption Index} & \multicolumn{2}{c}{$\chi^2$} & \colhead{} \\
\colhead{Name} & \colhead{Redshift} & \colhead{\ion{C}{4}} & \colhead{\ion{Mg}{2}} & \colhead{\ion{C}{4}} & \colhead{\ion{Mg}{2}} & \colhead{\ion{C}{4}} & \colhead{\ion{Mg}{2}} & \colhead{Comment} \\
\colhead{(1)} & \colhead{(2)} & \colhead{(3)} & \colhead{(4)} & \colhead{(5)} & \colhead{(6)} & \colhead{(7)} & \colhead{(8)} & \colhead{(9)} } 
\startdata
F2M0134$-$0931 & 2.220 &     1119 &          0 &      6112 &          0 &   0.199 &          0 &   non-BAL\\ 
F2M0729+3336 &   0.957 &  \nodata &           0 & \nodata &      390 &\nodata&  4.419 &   LoBAL?    \\ 
F2M0738+2750 &   1.985 &             0 &           0 &        402 &      778 &   0.025 &  83.78 &   mini-BAL/LoBAL\\ 
F2M0832+0509 &   1.070 &  \nodata &           0 & \nodata &           0 &\nodata&           0 &   non-BAL\\ 
F2M0834+5112 &   2.391 &        798 & \nodata &     2728 &\nodata&   0.249 &\nodata&   FeLoBAL\tablenotemark{a}\\
F2M0854+3425 &   3.050 &        535 & \nodata &     2077 &\nodata&   11.65 &\nodata&   HiBAL \\
F2M0904$-$0145 & 1.005 & \nodata &           0 & \nodata &          0 &\nodata&           0 &   \nodata \\ 
F2M0916$-$0319 & 1.560 & \nodata &           0 & \nodata &          0 &\nodata&           0 &   non-BAL \\ 
F2M0916+3854 &   1.265 &  \nodata &            0 & \nodata &          0 &\nodata&           0 &   non-BAL \\ 
F2M0921+1918 &   1.791 &        958 &            0 &     5203 &      371 &   0.357 &   49.59 &   miniBAL/FeLoBAL \\ 
F2M0925+4217 &   1.879 &        232 &           0 &     2312 &    1033 &    0.977 &   31.36 &   miniBAL/FeLoBAL\\ 
F2M0927+3930 &   1.170 & \nodata &           0 & \nodata &    1320 &\nodata&    143.0 &   miniBAL/FeLoBAL\tablenotemark{b}\\ 
F2M0943+5417 &   2.232 &       593 &    2539 &     4699 &     8467 &     1.90 &  22310 &   FeLoBAL\\ 
F2M1004+1229 &   2.658 &     6121 &  11540 &     8711 &  17374 &     4.49 &    3062 &   FeLoBAL\\ 
F2M1005+6357 &   1.280 & \nodata &    6351 & \nodata &    7309 &\nodata&   4.512 &  \nodata \\ 
F2M1012+2825 &   0.937 & \nodata &      428 & \nodata &       452 &\nodata&   0.611 &   non-BAL\\ 
F2M1025$-$0739 & 2.340 &          0 &           0 &       775 &            0 &   1.369 &          0 &   non-BAL\tablenotemark{c}\\ 
F2M1033+6051 &   1.401 & \nodata &           0 & \nodata &           0 &\nodata&          0 &   non-BAL \\ 
F2M1036+2828 &   1.762 &       541 &           0 &     2976 &       363 &   3.921 &   1241 &   mini-BAL/FeLoBAL \\ 
F2M1222+4223 &   2.108 &     1409 &           0 &     2652 &     1427 &  33.93 &  13.48 &   mini-BAL/FeLoBAL \\ 
F2M1310$-$0030 & 2.650 &   8434 & \nodata &   10752 &\nodata&  19.31 &\nodata&   FeLoBAL\\ 
F2M1311+5513 &   0.926 & \nodata &           0 & \nodata &           0 &\nodata&          0 &   non-BAL \\ 
F2M1341+3301 &   1.715 &            0 &     1725 &       979 &    6368 &  0.729 &   1031 &   FeLoBAL\\ 
F2M1353+3657 &   1.311 & \nodata &     4503 & \nodata &   3111 &\nodata&  1.166 &  \nodata\\ 
F2M1427+3723 &   2.168 &       117 &             0 &     1507 &     402 &   33.47 &   1519 &   FeLoBAL\tablenotemark{d} \\ 
F2M1456+0114 &   2.363 &     3991 & \nodata &     6422 &\nodata&     5.84 &\nodata&   FeLoBAL\tablenotemark{a}  \\ 
F2M1500+0231 &   1.516 & \nodata &       355 & \nodata &   4232 &\nodata&   10.61 &   FeLoBAL\\ 
F2M1531+2423 &   2.287 &          46 &            0 &     1383 &     351 &   13.25 &   24.41 &   mini-BAL/LoBAL\\ 
F2M1548+0913 &   1.509 & \nodata &            0 & \nodata &         0 &\nodata&           0 &   non-BAL\\ 
F2M1549+1245 &   2.373 &     1604 &            0 &     4763 &     778 &   2098  &   176.2 &   FeLoBAL\tablenotemark{d}\\ 
F2M1638+2707 &   1.689 &            0 &            0 &         49 &      579 &   0.011 &   217.1 &   mini-BAL/FeLoBAL\\ 
F2M1650+3242 &   1.059 & \nodata &            0 & \nodata &         0 &\nodata&            0 &   non-BAL\\ 
F2M2222$-$0202 & 2.252 &          0 &            0 &     1054 &   1184 &   69.36 &   144.7 &   mini-BAL/(Lo)BAL \\ 
\enddata

\tablenotetext{a}{Classification is based on \ion{Fe}{2} absorption.}
\tablenotetext{b}{Absorption occurs at the quasar redshift, with no velocity shift.}
\tablenotetext{c}{Three narrow FeLoBAL lines.}
\tablenotetext{d}{\ion{Mg}{2} mini-BAL; clear \ion{C}{4} BAL,  \ion{Fe}{2} absorption.}
\tablenotetext{key}{text}



\end{deluxetable}

%% file: ems_arXiv.bbl
\begin{thebibliography}{139}
\expandafter\ifx\csname natexlab\endcsname\relax\def\natexlab#1{#1}\fi

\bibitem[{{Adelman-McCarthy} {et~al.}(2007){Adelman-McCarthy}, {Ag{\"u}eros},
  {Allam}, {Anderson}, {Anderson}, {Annis}, {Bahcall}, {Bailer-Jones},
  {Baldry}, {Barentine}, {Beers}, {Belokurov}, {Berlind}, {Bernardi},
  {Blanton}, {Bochanski}, {Boroski}, {Bramich}, {Brewington}, {Brinchmann},
  {Brinkmann}, {Brunner}, {Budav{\'a}ri}, {Carey}, {Carliles}, {Carr},
  {Castander}, {Connolly}, {Cool}, {Cunha}, {Csabai}, {Dalcanton}, {Doi},
  {Eisenstein}, {Evans}, {Evans}, {Fan}, {Finkbeiner}, {Friedman}, {Frieman},
  {Fukugita}, {Gillespie}, {Gilmore}, {Glazebrook}, {Gray}, {Grebel}, {Gunn},
  {de Haas}, {Hall}, {Harvanek}, {Hawley}, {Hayes}, {Heckman}, {Hendry},
  {Hennessy}, {Hindsley}, {Hirata}, {Hogan}, {Hogg}, {Holtzman}, {Ichikawa},
  {Ichikawa}, {Ivezi{\'c}}, {Jester}, {Johnston}, {Jorgensen}, {Juri{\'c}},
  {Kauffmann}, {Kent}, {Kleinman}, {Knapp}, {Kniazev}, {Kron}, {Krzesinski},
  {Kuropatkin}, {Lamb}, {Lampeitl}, {Lee}, {Leger}, {Lima}, {Lin}, {Long},
  {Loveday}, {Lupton}, {Mandelbaum}, {Margon}, {Mart{\'{\i}}nez-Delgado},
  {Matsubara}, {McGehee}, {McKay}, {Meiksin}, {Munn}, {Nakajima}, {Nash},
  {Neilsen}, {Newberg}, {Nichol}, {Nieto-Santisteban}, {Nitta}, {Oyaizu},
  {Okamura}, {Ostriker}, {Padmanabhan}, {Park}, {Peoples}, {Pier}, {Pope},
  {Pourbaix}, {Quinn}, {Raddick}, {Re Fiorentin}, {Richards}, {Richmond},
  {Rix}, {Rockosi}, {Schlegel}, {Schneider}, {Scranton}, {Seljak}, {Sheldon},
  {Shimasaku}, {Silvestri}, {Smith}, {Smol{\v c}i{\'c}}, {Snedden}, {Stebbins},
  {Stoughton}, {Strauss}, {SubbaRao}, {Suto}, {Szalay}, {Szapudi}, {Szkody},
  {Tegmark}, {Thakar}, {Tremonti}, {Tucker}, {Uomoto}, {Vanden Berk},
  {Vandenberg}, {Vidrih}, {Vogeley}, {Voges}, {Vogt}, {Weinberg}, {West},
  {White}, {Wilhite}, {Yanny}, {Yocum}, {York}, {Zehavi}, {Zibetti}, \&
  {Zucker}}]{Adelman-McCarthy07}
{Adelman-McCarthy}, J.~K. {et~al.} 2007, \apjs, 172, 634

\bibitem[{{Aihara} {et~al.}(2011){Aihara}, {Allende Prieto}, {An}, {Anderson},
  {Aubourg}, {Balbinot}, {Beers}, {Berlind}, {Bickerton}, {Bizyaev}, {Blanton},
  {Bochanski}, {Bolton}, {Bovy}, {Brandt}, {Brinkmann}, {Brown}, {Brownstein},
  {Busca}, {Campbell}, {Carr}, {Chen}, {Chiappini}, {Comparat}, {Connolly},
  {Cortes}, {Croft}, {Cuesta}, {da Costa}, {Davenport}, {Dawson}, {Dhital},
  {Ealet}, {Ebelke}, {Edmondson}, {Eisenstein}, {Escoffier}, {Esposito},
  {Evans}, {Fan}, {Femen{\'{\i}}a Castell{\'a}}, {Font-Ribera}, {Frinchaboy},
  {Ge}, {Gillespie}, {Gilmore}, {Gonz{\'a}lez Hern{\'a}ndez}, {Gott}, {Gould},
  {Grebel}, {Gunn}, {Hamilton}, {Harding}, {Harris}, {Hawley}, {Hearty}, {Ho},
  {Hogg}, {Holtzman}, {Honscheid}, {Inada}, {Ivans}, {Jiang}, {Johnson},
  {Jordan}, {Jordan}, {Kazin}, {Kirkby}, {Klaene}, {Knapp}, {Kneib},
  {Kochanek}, {Koesterke}, {Kollmeier}, {Kron}, {Lampeitl}, {Lang}, {Le Goff},
  {Lee}, {Lin}, {Long}, {Loomis}, {Lucatello}, {Lundgren}, {Lupton}, {Ma},
  {MacDonald}, {Mahadevan}, {Maia}, {Makler}, {Malanushenko}, {Malanushenko},
  {Mandelbaum}, {Maraston}, {Margala}, {Masters}, {McBride}, {McGehee},
  {McGreer}, {M{\'e}nard}, {Miralda-Escud{\'e}}, {Morrison}, {Mullally},
  {Muna}, {Munn}, {Murayama}, {Myers}, {Naugle}, {Fausti Neto}, {Cuong Nguyen},
  {Nichol}, {O'Connell}, {Ogando}, {Olmstead}, {Oravetz}, {Padmanabhan},
  {Palanque-Delabrouille}, {Pan}, {Pandey}, {P{\^a}ris}, {Percival},
  {Petitjean}, {Pfaffenberger}, {Pforr}, {Phleps}, {Pichon}, {Pieri}, {Prada},
  {Price-Whelan}, {Raddick}, {Ramos}, {Reyl{\'e}}, {Rich}, {Richards}, {Rix},
  {Robin}, {Rocha-Pinto}, {Rockosi}, {Roe}, {Rollinde}, {Ross}, {Ross},
  {Rossetto}, {S{\'a}nchez}, {Sayres}, {Schlegel}, {Schlesinger}, {Schmidt},
  {Schneider}, {Sheldon}, {Shu}, {Simmerer}, {Simmons}, {Sivarani}, {Snedden},
  {Sobeck}, {Steinmetz}, {Strauss}, {Szalay}, {Tanaka}, {Thakar}, {Thomas},
  {Tinker}, {Tofflemire}, {Tojeiro}, {Tremonti}, {Vandenberg}, {Vargas
  Maga{\~n}a}, {Verde}, {Vogt}, {Wake}, {Wang}, {Weaver}, {Weinberg}, {White},
  {White}, {Yanny}, {Yasuda}, {Yeche}, \& {Zehavi}}]{Aihara11}
{Aihara}, H. {et~al.} 2011, \apjs, 193, 29

\bibitem[{{Antonucci}(1993)}]{Antonucci93}
{Antonucci}, R. 1993, \araa, 31, 473

\bibitem[{{Bajtlik} {et~al.}(1988){Bajtlik}, {Duncan}, \&
  {Ostriker}}]{Bajtlik88}
{Bajtlik}, S., {Duncan}, R.~C., \& {Ostriker}, J.~P. 1988, \apj, 327, 570

\bibitem[{{Baldwin} {et~al.}(1981){Baldwin}, {Phillips}, \&
  {Terlevich}}]{Baldwin81}
{Baldwin}, J.~A., {Phillips}, M.~M., \& {Terlevich}, R. 1981, \pasp, 93, 5

\bibitem[{{Becker} {et~al.}(2000){Becker}, {White}, {Gregg}, {Brotherton},
  {Laurent-Muehleisen}, \& {Arav}}]{Becker00}
{Becker}, R.~H., {White}, R.~L., {Gregg}, M.~D., {Brotherton}, M.~S.,
  {Laurent-Muehleisen}, S.~A., \& {Arav}, N. 2000, \apj, 538, 72

\bibitem[{{Becker} {et~al.}(2001){Becker}, {White}, {Gregg},
  {Laurent-Muehleisen}, {Brotherton}, {Impey}, {Chaffee}, {Richards},
  {Helfand}, {Lacy}, {Courbin}, \& {Proctor}}]{Becker01}
{Becker}, R.~H. {et~al.} 2001, \apjs, 135, 227

\bibitem[{{Becker} {et~al.}(1995){Becker}, {White}, \& {Helfand}}]{Becker95}
{Becker}, R.~H., {White}, R.~L., \& {Helfand}, D.~J. 1995, \apj, 450, 559

\bibitem[{{Bennert} {et~al.}(2008){Bennert}, {Canalizo}, {Jungwiert},
  {Stockton}, {Schweizer}, {Peng}, \& {Lacy}}]{Bennert08}
{Bennert}, N., {Canalizo}, G., {Jungwiert}, B., {Stockton}, A., {Schweizer},
  F., {Peng}, C.~Y., \& {Lacy}, M. 2008, \apj, 677, 846

\bibitem[{{Brotherton} {et~al.}(2001){Brotherton}, {Tran}, {Becker}, {Gregg},
  {Laurent-Muehleisen}, \& {White}}]{Brotherton01}
{Brotherton}, M.~S., {Tran}, H.~D., {Becker}, R.~H., {Gregg}, M.~D.,
  {Laurent-Muehleisen}, S.~A., \& {White}, R.~L. 2001, \apj, 546, 775

\bibitem[{{Calzetti} {et~al.}(1994){Calzetti}, {Kinney}, \&
  {Storchi-Bergmann}}]{Calzetti94}
{Calzetti}, D., {Kinney}, A.~L., \& {Storchi-Bergmann}, T. 1994, \apj, 429, 582

\bibitem[{{Cardelli} {et~al.}(1989){Cardelli}, {Clayton}, \&
  {Mathis}}]{Cardelli89}
{Cardelli}, J.~A., {Clayton}, G.~C., \& {Mathis}, J.~S. 1989, \apj, 345, 245

\bibitem[{{Croom} {et~al.}(2009){Croom}, {Richards}, {Shanks}, {Boyle},
  {Strauss}, {Myers}, {Nichol}, {Pimbblet}, {Ross}, {Schneider}, {Sharp}, \&
  {Wake}}]{Croom09}
{Croom}, S.~M. {et~al.} 2009, \mnras, 399, 1755

\bibitem[{{Croom} {et~al.}(2001){Croom}, {Smith}, {Boyle}, {Shanks}, {Loaring},
  {Miller}, \& {Lewis}}]{Croom01}
{Croom}, S.~M., {Smith}, R.~J., {Boyle}, B.~J., {Shanks}, T., {Loaring}, N.~S.,
  {Miller}, L., \& {Lewis}, I.~J. 2001, \mnras, 322, L29

\bibitem[{{Croton} {et~al.}(2006){Croton}, {Springel}, {White}, {De Lucia},
  {Frenk}, {Gao}, {Jenkins}, {Kauffmann}, {Navarro}, \& {Yoshida}}]{Croton06}
{Croton}, D.~J. {et~al.} 2006, \mnras, 365, 11

\bibitem[{{Cushing} {et~al.}(2004){Cushing}, {Vacca}, \& {Rayner}}]{Cushing04}
{Cushing}, M.~C., {Vacca}, W.~D., \& {Rayner}, J.~T. 2004, \pasp, 116, 362

\bibitem[{{Cutri} {et~al.}(2001)}]{Cutri01}
{Cutri}, R.~M., {et~al.} 2001, in ASP Conf. Ser. 232: The New Era of Wide Field
  Astronomy, 78--+

\bibitem[{{Di Matteo} {et~al.}(2005){Di Matteo}, {Springel}, \&
  {Hernquist}}]{DiMatteo05}
{Di Matteo}, T., {Springel}, V., \& {Hernquist}, L. 2005, \nat, 433, 604

\bibitem[{{Drinkwater} {et~al.}(1997){Drinkwater}, {Webster}, {Francis},
  {Condon}, {Ellison}, {Jauncey}, {Lovell}, {Peterson}, \&
  {Savage}}]{Drinkwater97}
{Drinkwater}, M.~J. {et~al.} 1997, \mnras, 284, 85

\bibitem[{{Dunlop} {et~al.}(2003){Dunlop}, {McLure}, {Kukula}, {Baum}, {O'Dea},
  \& {Hughes}}]{Dunlop03}
{Dunlop}, J.~S., {McLure}, R.~J., {Kukula}, M.~J., {Baum}, S.~A., {O'Dea},
  C.~P., \& {Hughes}, D.~H. 2003, \mnras, 340, 1095

\bibitem[{{Eisenstein} {et~al.}(2011){Eisenstein}, {Weinberg}, {Agol},
  {Aihara}, {Allende Prieto}, {Anderson}, {Arns}, {Aubourg}, {Bailey},
  {Balbinot}, \& et~al.}]{Eisenstein11}
{Eisenstein}, D.~J. {et~al.} 2011, \aj, 142, 72

\bibitem[{{Elitzur}(2008)}]{Elitzur08}
{Elitzur}, M. 2008, \nat, 52, 274

\bibitem[{{Farrah} {et~al.}(2010){Farrah}, {Urrutia}, {Lacy}, {Lebouteiller},
  {Spoon}, {Bernard-Salas}, {Connolly}, {Afonso}, {Connolly}, \&
  {Houck}}]{Farrah10}
{Farrah}, D. {et~al.} 2010, \apj, 717, 868

\bibitem[{{Ferrarese} \& {Merritt}(2000)}]{Ferrarese00}
{Ferrarese}, L., \& {Merritt}, D. 2000, \apjl, 539, L9

\bibitem[{{Fitzpatrick}(1999)}]{Fitzpatrick99}
{Fitzpatrick}, E.~L. 1999, \pasp, 111, 63

\bibitem[{{Floyd} {et~al.}(2004){Floyd}, {Kukula}, {Dunlop}, {McLure},
  {Miller}, {Percival}, {Baum}, \& {O'Dea}}]{Floyd04}
{Floyd}, D.~J.~E., {Kukula}, M.~J., {Dunlop}, J.~S., {McLure}, R.~J., {Miller},
  L., {Percival}, W.~J., {Baum}, S.~A., \& {O'Dea}, C.~P. 2004, \mnras, 355,
  196

\bibitem[{{Foltz} {et~al.}(1986){Foltz}, {Weymann}, {Peterson}, {Sun},
  {Malkan}, \& {Chaffee}}]{Foltz86}
{Foltz}, C.~B., {Weymann}, R.~J., {Peterson}, B.~M., {Sun}, L., {Malkan},
  M.~A., \& {Chaffee}, F.~H. 1986, \apj, 307, 504

\bibitem[{{Francis} {et~al.}(1991){Francis}, {Hewett}, {Foltz}, {Chaffee},
  {Weymann}, \& {Morris}}]{Francis91}
{Francis}, P.~J., {Hewett}, P.~C., {Foltz}, C.~B., {Chaffee}, F.~H., {Weymann},
  R.~J., \& {Morris}, S.~L. 1991, \apj, 373, 465

\bibitem[{{Gallagher} {et~al.}(2007){Gallagher}, {Hines}, {Blaylock},
  {Priddey}, {Brandt}, \& {Egami}}]{Gallagher07}
{Gallagher}, S.~C., {Hines}, D.~C., {Blaylock}, M., {Priddey}, R.~S., {Brandt},
  W.~N., \& {Egami}, E.~E. 2007, \apj, 665, 157

\bibitem[{{Gebhardt} {et~al.}(2000){Gebhardt}, {Bender}, {Bower}, {Dressler},
  {Faber}, {Filippenko}, {Green}, {Grillmair}, {Ho}, {Kormendy}, {Lauer},
  {Magorrian}, {Pinkney}, {Richstone}, \& {Tremaine}}]{Gebhardt00}
{Gebhardt}, K. {et~al.} 2000, \apjl, 539, L13

\bibitem[{{Georgakakis} {et~al.}(2009){Georgakakis}, {Clements}, {Bendo},
  {Rowan-Robinson}, {Nandra}, \& {Brotherton}}]{Georgakakis09}
{Georgakakis}, A., {Clements}, D.~L., {Bendo}, G., {Rowan-Robinson}, M.,
  {Nandra}, K., \& {Brotherton}, M.~S. 2009, \mnras, 394, 533

\bibitem[{{Georgakakis} {et~al.}(2012){Georgakakis}, {Grossi}, {Afonso}, \&
  {Hopkins}}]{Georgakakis12}
{Georgakakis}, A., {Grossi}, M., {Afonso}, J., \& {Hopkins}, A.~M. 2012,
  \mnras, 421, 2223

\bibitem[{{Glikman} {et~al.}(2010){Glikman}, {Bogosavljevi{\'c}}, {Djorgovski},
  {Stern}, {Dey}, {Jannuzi}, \& {Mahabal}}]{Glikman10}
{Glikman}, E., {Bogosavljevi{\'c}}, M., {Djorgovski}, S.~G., {Stern}, D.,
  {Dey}, A., {Jannuzi}, B.~T., \& {Mahabal}, A. 2010, \apj, 710, 1498

\bibitem[{{Glikman} {et~al.}(2011){Glikman}, {Djorgovski}, {Stern}, {Dey},
  {Jannuzi}, \& {Lee}}]{Glikman11}
{Glikman}, E., {Djorgovski}, S.~G., {Stern}, D., {Dey}, A., {Jannuzi}, B.~T.,
  \& {Lee}, K.-S. 2011, \apjl, 728, L26

\bibitem[{{Glikman} {et~al.}(2004){Glikman}, {Gregg}, {Lacy}, {Helfand},
  {Becker}, \& {White}}]{Glikman04}
{Glikman}, E., {Gregg}, M.~D., {Lacy}, M., {Helfand}, D.~J., {Becker}, R.~H.,
  \& {White}, R.~L. 2004, \apj, 607, 60

\bibitem[{{Glikman} {et~al.}(2006){Glikman}, {Helfand}, \& {White}}]{Glikman06}
{Glikman}, E., {Helfand}, D.~J., \& {White}, R.~L. 2006, \apj, 640, 579

\bibitem[{{Glikman} {et~al.}(2007){Glikman}, {Helfand}, {White}, {Becker},
  {Gregg}, \& {Lacy}}]{Glikman07}
{Glikman}, E., {Helfand}, D.~J., {White}, R.~L., {Becker}, R.~H., {Gregg},
  M.~D., \& {Lacy}, M. 2007, \apj, 667, 673

\bibitem[{{Gordon} \& {Clayton}(1998)}]{Gordon98}
{Gordon}, K.~D., \& {Clayton}, G.~C. 1998, \apj, 500, 816

\bibitem[{{Gordon} {et~al.}(2003){Gordon}, {Clayton}, {Misselt}, {Landolt}, \&
  {Wolff}}]{Gordon03}
{Gordon}, K.~D., {Clayton}, G.~C., {Misselt}, K.~A., {Landolt}, A.~U., \&
  {Wolff}, M.~J. 2003, \apj, 594, 279

\bibitem[{{Grazian} {et~al.}(2004){Grazian}, {Negrello}, {Moscardini},
  {Cristiani}, {Haehnelt}, {Matarrese}, {Omizzolo}, \& {Vanzella}}]{Grazian04}
{Grazian}, A., {Negrello}, M., {Moscardini}, L., {Cristiani}, S., {Haehnelt},
  M.~G., {Matarrese}, S., {Omizzolo}, A., \& {Vanzella}, E. 2004, \aj, 127, 592

\bibitem[{{Gregg} {et~al.}(1996){Gregg}, {Becker}, {White}, {Helfand},
  {McMahon}, \& {Hook}}]{Gregg96}
{Gregg}, M.~D., {Becker}, R.~H., {White}, R.~L., {Helfand}, D.~J., {McMahon},
  R.~G., \& {Hook}, I.~M. 1996, \aj, 112, 407

\bibitem[{{Gregg} {et~al.}(2002){Gregg}, {Lacy}, {White}, {Glikman}, {Helfand},
  {Becker}, \& {Brotherton}}]{Gregg02}
{Gregg}, M.~D., {Lacy}, M., {White}, R.~L., {Glikman}, E., {Helfand}, D.,
  {Becker}, R.~H., \& {Brotherton}, M.~S. 2002, \apj, 564, 133

\bibitem[{{Gunn} {et~al.}(1998){Gunn}, {Carr}, {Rockosi}, {Sekiguchi}, {Berry},
  {Elms}, {de Haas}, {Ivezi{\'c}}, {Knapp}, {Lupton}, {Pauls}, {Simcoe},
  {Hirsch}, {Sanford}, {Wang}, {York}, {Harris}, {Annis}, {Bartozek},
  {Boroski}, {Bakken}, {Haldeman}, {Kent}, {Holm}, {Holmgren}, {Petravick},
  {Prosapio}, {Rechenmacher}, {Doi}, {Fukugita}, {Shimasaku}, {Okada}, {Hull},
  {Siegmund}, {Mannery}, {Blouke}, {Heidtman}, {Schneider}, {Lucinio}, \&
  {Brinkman}}]{Gunn98}
{Gunn}, J.~E. {et~al.} 1998, \aj, 116, 3040

\bibitem[{{Gunn} {et~al.}(2006){Gunn}, {Siegmund}, {Mannery}, {Owen}, {Hull},
  {Leger}, {Carey}, {Knapp}, {York}, {Boroski}, {Kent}, {Lupton}, {Rockosi},
  {Evans}, {Waddell}, {Anderson}, {Annis}, {Barentine}, {Bartoszek}, {Bastian},
  {Bracker}, {Brewington}, {Briegel}, {Brinkmann}, {Brown}, {Carr},
  {Czarapata}, {Drennan}, {Dombeck}, {Federwitz}, {Gillespie}, {Gonzales},
  {Hansen}, {Harvanek}, {Hayes}, {Jordan}, {Kinney}, {Klaene}, {Kleinman},
  {Kron}, {Kresinski}, {Lee}, {Limmongkol}, {Lindenmeyer}, {Long}, {Loomis},
  {McGehee}, {Mantsch}, {Neilsen}, {Neswold}, {Newman}, {Nitta}, {Peoples},
  {Pier}, {Prieto}, {Prosapio}, {Rivetta}, {Schneider}, {Snedden}, \&
  {Wang}}]{Gunn06}
---. 2006, \aj, 131, 2332

\bibitem[{{Haiman} \& {Cen}(2002)}]{Haiman02}
{Haiman}, Z., \& {Cen}, R. 2002, \apj, 578, 702

\bibitem[{{Hall} {et~al.}(2002){Hall}, {Anderson}, {Strauss}, {York},
  {Richards}, {Fan}, {Knapp}, {Schneider}, {Vanden Berk}, {Geballe}, {Bauer},
  {Becker}, {Davis}, {Rix}, {Nichol}, {Bahcall}, {Brinkmann}, {Brunner},
  {Connolly}, {Csabai}, {Doi}, {Fukugita}, {Gunn}, {Haiman}, {Harvanek},
  {Heckman}, {Hennessy}, {Inada}, {Ivezi{\' c}}, {Johnston}, {Kleinman},
  {Krolik}, {Krzesinski}, {Kunszt}, {Lamb}, {Long}, {Lupton}, {Miknaitis},
  {Munn}, {Narayanan}, {Neilsen}, {Newman}, {Nitta}, {Okamura}, {Pentericci},
  {Pier}, {Schlegel}, {Snedden}, {Szalay}, {Thakar}, {Tsvetanov}, {White}, \&
  {Zheng}}]{Hall02b}
{Hall}, P.~B. {et~al.} 2002, \apjs, 141, 267

\bibitem[{{Hazard} {et~al.}(1987){Hazard}, {McMahon}, {Webb}, \&
  {Morton}}]{Hazard87}
{Hazard}, C., {McMahon}, R.~G., {Webb}, J.~K., \& {Morton}, D.~C. 1987, \apj,
  323, 263

\bibitem[{{Herter} {et~al.}(2008){Herter}, {Henderson}, {Wilson}, {Matthews},
  {Rahmer}, {Bonati}, {Muirhead}, {Adams}, {Lloyd}, {Skrutskie}, {Moon},
  {Parshley}, {Nelson}, {Martinache}, \& {Gull}}]{Herter08}
{Herter}, T.~L. {et~al.} 2008, in Society of Photo-Optical Instrumentation
  Engineers (SPIE) Conference Series, Vol. 7014, Society of Photo-Optical
  Instrumentation Engineers (SPIE) Conference Series

\bibitem[{{Hewett} {et~al.}(2006){Hewett}, {Warren}, {Leggett}, \&
  {Hodgkin}}]{Hewett06}
{Hewett}, P.~C., {Warren}, S.~J., {Leggett}, S.~K., \& {Hodgkin}, S.~T. 2006,
  \mnras, 367, 454

\bibitem[{{Hopkins} {et~al.}(2005{\natexlab{a}}){Hopkins}, {Hernquist}, {Cox},
  {Di Matteo}, {Martini}, {Robertson}, \& {Springel}}]{Hopkins05c}
{Hopkins}, P.~F., {Hernquist}, L., {Cox}, T.~J., {Di Matteo}, T., {Martini},
  P., {Robertson}, B., \& {Springel}, V. 2005{\natexlab{a}}, \apj, 630, 705

\bibitem[{{Hopkins} {et~al.}(2006){Hopkins}, {Hernquist}, {Cox}, {Di Matteo},
  {Robertson}, \& {Springel}}]{Hopkins06a}
{Hopkins}, P.~F., {Hernquist}, L., {Cox}, T.~J., {Di Matteo}, T., {Robertson},
  B., \& {Springel}, V. 2006, \apjs, 163, 1

\bibitem[{{Hopkins} {et~al.}(2008){Hopkins}, {Hernquist}, {Cox}, \& {Kere{\v
  s}}}]{Hopkins08}
{Hopkins}, P.~F., {Hernquist}, L., {Cox}, T.~J., \& {Kere{\v s}}, D. 2008,
  \apjs, 175, 356

\bibitem[{{Hopkins} {et~al.}(2005{\natexlab{b}}){Hopkins}, {Hernquist},
  {Martini}, {Cox}, {Robertson}, {Di Matteo}, \& {Springel}}]{Hopkins05}
{Hopkins}, P.~F., {Hernquist}, L., {Martini}, P., {Cox}, T.~J., {Robertson},
  B., {Di Matteo}, T., \& {Springel}, V. 2005{\natexlab{b}}, \apjl, 625, L71

\bibitem[{{Hopkins} {et~al.}(2004){Hopkins}, {Strauss}, {Hall}, {Richards},
  {Cooper}, {Schneider}, {Vanden Berk}, {Jester}, {Brinkmann}, \&
  {Szokoly}}]{Hopkins04}
{Hopkins}, P.~F. {et~al.} 2004, \aj, 128, 1112

\bibitem[{{Ivezi{\' c}} {et~al.}(2002){Ivezi{\' c}}, {Menou}, {Knapp},
  {Strauss}, {Lupton}, {Vanden Berk}, {Richards}, {Tremonti}, {Weinstein},
  {Anderson}, {Bahcall}, {Becker}, {Bernardi}, {Blanton}, {Eisenstein}, {Fan},
  {Finkbeiner}, {Finlator}, {Frieman}, {Gunn}, {Hall}, {Kim}, {Kinkhabwala},
  {Narayanan}, {Rockosi}, {Schlegel}, {Schneider}, {Strateva}, {SubbaRao},
  {Thakar}, {Voges}, {White}, {Yanny}, {Brinkmann}, {Doi}, {Fukugita},
  {Hennessy}, {Munn}, {Nichol}, \& {York}}]{Ivezic02}
{Ivezi{\' c}}, {\v Z}. {et~al.} 2002, \aj, 124, 2364

\bibitem[{{Jakobsen} {et~al.}(2003){Jakobsen}, {Jansen}, {Wagner}, \&
  {Reimers}}]{Jakobsen03}
{Jakobsen}, P., {Jansen}, R.~A., {Wagner}, S., \& {Reimers}, D. 2003, \aap,
  397, 891

\bibitem[{{Kauffmann} \& {Haehnelt}(2000)}]{Kauffmann00}
{Kauffmann}, G., \& {Haehnelt}, M. 2000, \mnras, 311, 576

\bibitem[{{Kewley} {et~al.}(2006){Kewley}, {Groves}, {Kauffmann}, \&
  {Heckman}}]{Kewley06}
{Kewley}, L.~J., {Groves}, B., {Kauffmann}, G., \& {Heckman}, T. 2006, \mnras,
  372, 961

\bibitem[{{Kim} \& {Elvis}(1999)}]{Kim99}
{Kim}, D., \& {Elvis}, M. 1999, \apj, 516, 9

\bibitem[{{Kinney} {et~al.}(1996){Kinney}, {Calzetti}, {Bohlin}, {McQuade},
  {Storchi-Bergmann}, \& {Schmitt}}]{Kinney96}
{Kinney}, A.~L., {Calzetti}, D., {Bohlin}, R.~C., {McQuade}, K.,
  {Storchi-Bergmann}, T., \& {Schmitt}, H.~R. 1996, \apj, 467, 38

\bibitem[{{Kocevski} {et~al.}(2011){Kocevski}, {Faber}, {Mozena}, {Koekemoer},
  {Nandra}, {Rangel}, {Laird}, {Brusa}, {Wuyts}, {Trump}, {Koo}, {Somerville},
  {Bell}, {Lotz}, {Alexander}, {Bournaud}, {Conselice}, {Dahlen}, {Dekel},
  {Donley}, {Dunlop}, {Finoguenov}, {Georgakakis}, {Giavalisco}, {Guo},
  {Grogin}, {Hathi}, {Juneau}, {Kartaltepe}, {Lucas}, {McGrath}, {McIntosh},
  {Mobasher}, {Robaina}, {Rosario}, {Straughn}, {van der Wel}, \&
  {Villforth}}]{Kocevski11}
{Kocevski}, D.~D. {et~al.} 2011, ArXiv e-prints

\bibitem[{{Lacy} {et~al.}(2007){Lacy}, {Petric}, {Sajina}, {Canalizo},
  {Storrie-Lombardi}, {Armus}, {Fadda}, \& {Marleau}}]{Lacy07}
{Lacy}, M., {Petric}, A.~O., {Sajina}, A., {Canalizo}, G., {Storrie-Lombardi},
  L.~J., {Armus}, L., {Fadda}, D., \& {Marleau}, F.~R. 2007, \aj, 133, 186

\bibitem[{{Lacy} {et~al.}(2004){Lacy}, {Storrie-Lombardi}, {Sajina},
  {Appleton}, {Armus}, {Chapman}, {Choi}, {Fadda}, {Fang}, {Frayer},
  {Heinrichsen}, {Helou}, {Im}, {Marleau}, {Masci}, {Shupe}, {Soifer},
  {Surace}, {Teplitz}, {Wilson}, \& {Yan}}]{Lacy04}
{Lacy}, M. {et~al.} 2004, \apjs, 154, 166

\bibitem[{{Lasker} {et~al.}(2008){Lasker}, {Lattanzi}, {McLean}, {Bucciarelli},
  {Drimmel}, {Garcia}, {Greene}, {Guglielmetti}, {Hanley}, {Hawkins},
  {Laidler}, {Loomis}, {Meakes}, {Mignani}, {Morbidelli}, {Morrison},
  {Pannunzio}, {Rosenberg}, {Sarasso}, {Smart}, {Spagna}, {Sturch},
  {Volpicelli}, {White}, {Wolfe}, \& {Zacchei}}]{Lasker08}
{Lasker}, B.~M. {et~al.} 2008, \aj, 136, 735

\bibitem[{{Lawrence}(1991)}]{Lawrence91}
{Lawrence}, A. 1991, \mnras, 252, 586

\bibitem[{{Lawrence} \& {Elvis}(2010)}]{Lawrence10}
{Lawrence}, A., \& {Elvis}, M. 2010, \apj, 714, 561

\bibitem[{{Lawrence} {et~al.}(2007){Lawrence}, {Warren}, {Almaini}, {Edge},
  {Hambly}, {Jameson}, {Lucas}, {Casali}, {Adamson}, {Dye}, {Emerson},
  {Foucaud}, {Hewett}, {Hirst}, {Hodgkin}, {Irwin}, {Lodieu}, {McMahon},
  {Simpson}, {Smail}, {Mortlock}, \& {Folger}}]{Lawrence07}
{Lawrence}, A. {et~al.} 2007, \mnras, 379, 1599

\bibitem[{{Maddox} {et~al.}(2008){Maddox}, {Hewett}, {Warren}, \&
  {Croom}}]{Maddox08}
{Maddox}, N., {Hewett}, P.~C., {Warren}, S.~J., \& {Croom}, S.~M. 2008, \mnras,
  386, 1605

\bibitem[{{Magorrian} {et~al.}(1998){Magorrian}, {Tremaine}, {Richstone},
  {Bender}, {Bower}, {Dressler}, {Faber}, {Gebhardt}, {Green}, {Grillmair},
  {Kormendy}, \& {Lauer}}]{Magorrian98}
{Magorrian}, J. {et~al.} 1998, \aj, 115, 2285

\bibitem[{{Maiolino} {et~al.}(2001){Maiolino}, {Marconi}, {Salvati},
  {Risaliti}, {Severgnini}, {Oliva}, {La Franca}, \& {Vanzi}}]{Maiolino01}
{Maiolino}, R., {Marconi}, A., {Salvati}, M., {Risaliti}, G., {Severgnini}, P.,
  {Oliva}, E., {La Franca}, F., \& {Vanzi}, L. 2001, \aap, 365, 28

\bibitem[{{Mannucci} {et~al.}(2001){Mannucci}, {Basile}, {Poggianti},
  {Cimatti}, {Daddi}, {Pozzetti}, \& {Vanzi}}]{Mannucci01}
{Mannucci}, F., {Basile}, F., {Poggianti}, B.~M., {Cimatti}, A., {Daddi}, E.,
  {Pozzetti}, L., \& {Vanzi}, L. 2001, \mnras, 326, 745

\bibitem[{{Marconi} \& {Hunt}(2003)}]{Marconi03}
{Marconi}, A., \& {Hunt}, L.~K. 2003, \apjl, 589, L21

\bibitem[{{Martini}(2004)}]{Martini04}
{Martini}, P. 2004, Coevolution of Black Holes and Galaxies, 169

\bibitem[{{McMahon} {et~al.}(2002){McMahon}, {White}, {Helfand}, \&
  {Becker}}]{McMahon02}
{McMahon}, R.~G., {White}, R.~L., {Helfand}, D.~J., \& {Becker}, R.~H. 2002,
  \apjs, 143, 1

\bibitem[{{Menou} {et~al.}(2001){Menou}, {Vanden Berk}, {Ivezi{\'c}}, {Kim},
  {Knapp}, {Richards}, {Strateva}, {Fan}, {Gunn}, {Hall}, {Heckman}, {Krolik},
  {Lupton}, {Schneider}, {York}, {Anderson}, {Bahcall}, {Brinkmann}, {Brunner},
  {Csabai}, {Fukugita}, {Hennessy}, {Kunszt}, {Lamb}, {Munn}, {Nichol}, \&
  {Szokoly}}]{Menou01}
{Menou}, K. {et~al.} 2001, \apj, 561, 645

\bibitem[{{Misselt} {et~al.}(1999){Misselt}, {Clayton}, \&
  {Gordon}}]{Misselt99}
{Misselt}, K.~A., {Clayton}, G.~C., \& {Gordon}, K.~D. 1999, \apj, 515, 128

\bibitem[{{Netzer}(1975)}]{Netzer75}
{Netzer}, H. 1975, \mnras, 171, 395

\bibitem[{{Noterdaeme} {et~al.}(2010){Noterdaeme}, {Petitjean}, {Ledoux},
  {L{\'o}pez}, {Srianand}, \& {Vergani}}]{Noterdaeme10}
{Noterdaeme}, P., {Petitjean}, P., {Ledoux}, C., {L{\'o}pez}, S., {Srianand},
  R., \& {Vergani}, S.~D. 2010, \aap, 523, A80

\bibitem[{{O'Donnell}(1994)}]{Odonnell94}
{O'Donnell}, J.~E. 1994, \apj, 422, 158

\bibitem[{{Oke} {et~al.}(1995){Oke}, {Cohen}, {Carr}, {Cromer}, {Dingizian},
  {Harris}, {Labrecque}, {Lucinio}, {Schaal}, {Epps}, \& {Miller}}]{Oke95}
{Oke}, J.~B. {et~al.} 1995, \pasp, 107, 375

\bibitem[{{Osterbrock}(1989)}]{Osterbrock89}
{Osterbrock}, D.~E. 1989, {Astrophysics of gaseous nebulae and active galactic
  nuclei} (Research supported by the University of California, John Simon
  Guggenheim Memorial Foundation, University of Minnesota, et al.~Mill Valley,
  CA, University Science Books, 1989, 422 p.)

\bibitem[{{Oyabu} {et~al.}(2009){Oyabu}, {Kawara}, {Tsuzuki}, {Matsuoka},
  {Sameshima}, {Asami}, \& {Ohyama}}]{Oyabu09}
{Oyabu}, S., {Kawara}, K., {Tsuzuki}, Y., {Matsuoka}, Y., {Sameshima}, H.,
  {Asami}, N., \& {Ohyama}, Y. 2009, \apj, 697, 452

\bibitem[{{Polletta} {et~al.}(2008){Polletta}, {Weedman}, {H{\"o}nig},
  {Lonsdale}, {Smith}, \& {Houck}}]{Polletta08}
{Polletta}, M., {Weedman}, D., {H{\"o}nig}, S., {Lonsdale}, C.~J., {Smith},
  H.~E., \& {Houck}, J. 2008, \apj, 675, 960

\bibitem[{{Polletta} {et~al.}(2006){Polletta}, {Wilkes}, {Siana}, {Lonsdale},
  {Kilgard}, {Smith}, {Kim}, {Owen}, {Efstathiou}, {Jarrett}, {Stacey},
  {Franceschini}, {Rowan-Robinson}, {Babbedge}, {Berta}, {Fang}, {Farrah},
  {Gonz{\'a}lez-Solares}, {Morrison}, {Surace}, \& {Shupe}}]{Polletta06}
{Polletta}, M.~d.~C. {et~al.} 2006, \apj, 642, 673

\bibitem[{{Porciani} {et~al.}(2004){Porciani}, {Magliocchetti}, \&
  {Norberg}}]{Porciani04}
{Porciani}, C., {Magliocchetti}, M., \& {Norberg}, P. 2004, \mnras, 355, 1010

\bibitem[{{Rawlings} {et~al.}(1995){Rawlings}, {Lacy}, {Sivia}, \&
  {Eales}}]{Rawlings95}
{Rawlings}, S., {Lacy}, M., {Sivia}, D.~S., \& {Eales}, S.~A. 1995, \mnras,
  274, 428

\bibitem[{{Rayner} {et~al.}(2003){Rayner}, {Toomey}, {Onaka}, {Denault},
  {Stahlberger}, {Vacca}, {Cushing}, \& {Wang}}]{Rayner03}
{Rayner}, J.~T., {Toomey}, D.~W., {Onaka}, P.~M., {Denault}, A.~J.,
  {Stahlberger}, W.~E., {Vacca}, W.~D., {Cushing}, M.~C., \& {Wang}, S. 2003,
  \pasp, 115, 362

\bibitem[{{Rees} {et~al.}(1989){Rees}, {Netzer}, \& {Ferland}}]{Rees89}
{Rees}, M.~J., {Netzer}, H., \& {Ferland}, G.~J. 1989, \apj, 347, 640

\bibitem[{{Reid} {et~al.}(1991){Reid}, {Brewer}, {Brucato}, {McKinley},
  {Maury}, {Mendenhall}, {Mould}, {Mueller}, {Neugebauer}, {Phinney},
  {Sargent}, {Schombert}, \& {Thicksten}}]{Reid91}
{Reid}, I.~N. {et~al.} 1991, \pasp, 103, 661

\bibitem[{{Richards}(2001)}]{Richards01}
{Richards}, G.~T. 2001, \apjs, 133, 53

\bibitem[{{Richards} {et~al.}(2002){Richards}, {Fan}, {Newberg}, {Strauss},
  {Vanden Berk}, {Schneider}, {Yanny}, {Boucher}, {Burles}, {Frieman}, {Gunn},
  {Hall}, {Ivezi{\'c}}, {Kent}, {Loveday}, {Lupton}, {Rockosi}, {Schlegel},
  {Stoughton}, {SubbaRao}, \& {York}}]{Richards02}
{Richards}, G.~T. {et~al.} 2002, \aj, 123, 2945

\bibitem[{{Richards} {et~al.}(2003){Richards}, {Hall}, {Vanden Berk},
  {Strauss}, {Schneider}, {Weinstein}, {Reichard}, {York}, {Knapp}, {Fan},
  {Ivezi{\'c}}, {Brinkmann}, {Budav{\'a}ri}, {Csabai}, \&
  {Nichol}}]{Richards03}
---. 2003, \aj, 126, 1131

\bibitem[{{Richards} {et~al.}(2004){Richards}, {Nichol}, {Gray}, {Brunner},
  {Lupton}, {Vanden Berk}, {Chong}, {Weinstein}, {Schneider}, {Anderson},
  {Munn}, {Harris}, {Strauss}, {Fan}, {Gunn}, {Ivezi{\'c}}, {York},
  {Brinkmann}, \& {Moore}}]{Richards04}
---. 2004, \apjs, 155, 257

\bibitem[{{Richards} {et~al.}(2006){Richards}, {Strauss}, {Fan}, {Hall},
  {Jester}, {Schneider}, {Vanden Berk}, {Stoughton}, {Anderson}, {Brunner},
  {Gray}, {Gunn}, {Ivezi{\'c}}, {Kirkland}, {Knapp}, {Loveday}, {Meiksin},
  {Pope}, {Szalay}, {Thakar}, {Yanny}, {York}, {Barentine}, {Brewington},
  {Brinkmann}, {Fukugita}, {Harvanek}, {Kent}, {Kleinman}, {Krzesi{\'n}ski},
  {Long}, {Lupton}, {Nash}, {Neilsen}, {Nitta}, {Schlegel}, \&
  {Snedden}}]{Richards06}
---. 2006, \aj, 131, 2766

\bibitem[{{Ross} {et~al.}(2011){Ross}, {Myers}, {Sheldon}, {Y{\`e}che},
  {Strauss}, {Bovy}, {Kirkpatrick}, {Richards}, {Aubourg}, {Blanton}, {Brandt},
  {Carithers}, {Croft}, {da Silva}, {Dawson}, {Eisenstein}, {Hennawi}, {Ho},
  {Hogg}, {Lee}, {Lundgren}, {McMahon}, {Miralda-Escude},
  {Palanque-Delabrouille}, {Paris}, {Petitjean}, {Pieri}, {Rich}, {Roe},
  {Schiminovich}, {Schlegel}, {Schneider}, {Slosar}, {Suzuki}, {Tinker},
  {Weinberg}, {Weyant}, {White}, \& {Wood-Vasey}}]{Ross11}
{Ross}, N.~P. {et~al.} 2011, ArXiv e-prints

\bibitem[{{Sandage}(1965)}]{Sandage65}
{Sandage}, A. 1965, \apj, 141, 1560

\bibitem[{{Sanders} \& {Mirabel}(1996)}]{Sanders96}
{Sanders}, D.~B., \& {Mirabel}, I.~F. 1996, \araa, 34, 749

\bibitem[{{Sanders} {et~al.}(1988){Sanders}, {Soifer}, {Elias}, {Madore},
  {Matthews}, {Neugebauer}, \& {Scoville}}]{Sanders88a}
{Sanders}, D.~B., {Soifer}, B.~T., {Elias}, J.~H., {Madore}, B.~F., {Matthews},
  K., {Neugebauer}, G., \& {Scoville}, N.~Z. 1988, \apj, 325, 74

\bibitem[{{Schawinski} {et~al.}(2011){Schawinski}, {Treister}, {Urry},
  {Cardamone}, {Simmons}, \& {Yi}}]{Schawinski11}
{Schawinski}, K., {Treister}, E., {Urry}, C.~M., {Cardamone}, C.~N., {Simmons},
  B., \& {Yi}, S.~K. 2011, \apjl, 727, L31

\bibitem[{{Schmidt}(1963)}]{Schmidt63}
{Schmidt}, M. 1963, \nat, 197, 1040

\bibitem[{{Schneider} {et~al.}(2010){Schneider}, {Richards}, {Hall}, {Strauss},
  {Anderson}, {Boroson}, {Ross}, {Shen}, {Brandt}, {Fan}, {Inada}, {Jester},
  {Knapp}, {Krawczyk}, {Thakar}, {Vanden Berk}, {Voges}, {Yanny}, {York},
  {Bahcall}, {Bizyaev}, {Blanton}, {Brewington}, {Brinkmann}, {Eisenstein},
  {Frieman}, {Fukugita}, {Gray}, {Gunn}, {Hibon}, {Ivezi{\'c}}, {Kent}, {Kron},
  {Lee}, {Lupton}, {Malanushenko}, {Malanushenko}, {Oravetz}, {Pan}, {Pier},
  {Price}, {Saxe}, {Schlegel}, {Simmons}, {Snedden}, {SubbaRao}, {Szalay}, \&
  {Weinberg}}]{Schneider10}
{Schneider}, D.~P. {et~al.} 2010, \aj, 139, 2360

\bibitem[{{Shaver} {et~al.}(1996){Shaver}, {Wall}, {Kellermann}, {Jackson}, \&
  {Hawkins}}]{Shaver96}
{Shaver}, P.~A., {Wall}, J.~V., {Kellermann}, K.~I., {Jackson}, C.~A., \&
  {Hawkins}, M.~R.~S. 1996, \nat, 384, 439

\bibitem[{{Sheinis} {et~al.}(2002){Sheinis}, {Bolte}, {Epps}, {Kibrick},
  {Miller}, {Radovan}, {Bigelow}, \& {Sutin}}]{Sheinis02}
{Sheinis}, A.~I., {Bolte}, M., {Epps}, H.~W., {Kibrick}, R.~I., {Miller},
  J.~S., {Radovan}, M.~V., {Bigelow}, B.~C., \& {Sutin}, B.~M. 2002, \pasp,
  114, 851

\bibitem[{{Shen} \& {M{\'e}nard}(2011)}]{Shen11}
{Shen}, Y., \& {M{\'e}nard}, B. 2011, ArXiv e-prints

\bibitem[{{Simmons} {et~al.}(2011){Simmons}, {Van Duyne}, {Urry}, {Treister},
  {Koekemoer}, {Grogin}, \& {The GOODS Team}}]{Simmons11}
{Simmons}, B.~D., {Van Duyne}, J., {Urry}, C.~M., {Treister}, E., {Koekemoer},
  A.~M., {Grogin}, N.~A., \& {The GOODS Team}. 2011, \apj, 734, 121

\bibitem[{{Skrutskie} {et~al.}(2006){Skrutskie}, {Cutri}, {Stiening},
  {Weinberg}, {Schneider}, {Carpenter}, {Beichman}, {Capps}, {Chester},
  {Elias}, {Huchra}, {Liebert}, {Lonsdale}, {Monet}, {Price}, {Seitzer},
  {Jarrett}, {Kirkpatrick}, {Gizis}, {Howard}, {Evans}, {Fowler}, {Fullmer},
  {Hurt}, {Light}, {Kopan}, {Marsh}, {McCallon}, {Tam}, {Van Dyk}, \&
  {Wheelock}}]{Skrutskie06}
{Skrutskie}, M.~F. {et~al.} 2006, \aj, 131, 1163

\bibitem[{{Smith} {et~al.}(2002){Smith}, {Schmidt}, {Hines}, {Cutri}, \&
  {Nelson}}]{Smith02}
{Smith}, P.~S., {Schmidt}, G.~D., {Hines}, D.~C., {Cutri}, R.~M., \& {Nelson},
  B.~O. 2002, \apj, 569, 23

\bibitem[{{Sprayberry} \& {Foltz}(1992)}]{Sprayberry92}
{Sprayberry}, D., \& {Foltz}, C.~B. 1992, \apj, 390, 39

\bibitem[{{Srianand} {et~al.}(2008){Srianand}, {Gupta}, {Petitjean},
  {Noterdaeme}, \& {Saikia}}]{Srianand08}
{Srianand}, R., {Gupta}, N., {Petitjean}, P., {Noterdaeme}, P., \& {Saikia},
  D.~J. 2008, \mnras, 391, L69

\bibitem[{{Stocke} {et~al.}(1992){Stocke}, {Morris}, {Weymann}, \&
  {Foltz}}]{Stocke92}
{Stocke}, J.~T., {Morris}, S.~L., {Weymann}, R.~J., \& {Foltz}, C.~B. 1992,
  \apj, 396, 487

\bibitem[{{Stoughton} {et~al.}(2002){Stoughton}, {Lupton}, {Bernardi},
  {Blanton}, {Burles}, {Castander}, {Connolly}, {Eisenstein}, {Frieman},
  {Hennessy}, {Hindsley}, {Ivezi{\'c}}, {Kent}, {Kunszt}, {Lee}, {Meiksin},
  {Munn}, {Newberg}, {Nichol}, {Nicinski}, {Pier}, {Richards}, {Richmond},
  {Schlegel}, {Smith}, {Strauss}, {SubbaRao}, {Szalay}, {Thakar}, {Tucker},
  {Vanden Berk}, {Yanny}, {Adelman}, {Anderson}, {Anderson}, {Annis},
  {Bahcall}, {Bakken}, {Bartelmann}, {Bastian}, {Bauer}, {Berman},
  {B{\"o}hringer}, {Boroski}, {Bracker}, {Briegel}, {Briggs}, {Brinkmann},
  {Brunner}, {Carey}, {Carr}, {Chen}, {Christian}, {Colestock}, {Crocker},
  {Csabai}, {Czarapata}, {Dalcanton}, {Davidsen}, {Davis}, {Dehnen},
  {Dodelson}, {Doi}, {Dombeck}, {Donahue}, {Ellman}, {Elms}, {Evans}, {Eyer},
  {Fan}, {Federwitz}, {Friedman}, {Fukugita}, {Gal}, {Gillespie}, {Glazebrook},
  {Gray}, {Grebel}, {Greenawalt}, {Greene}, {Gunn}, {de Haas}, {Haiman},
  {Haldeman}, {Hall}, {Hamabe}, {Hansen}, {Harris}, {Harris}, {Harvanek},
  {Hawley}, {Hayes}, {Heckman}, {Helmi}, {Henden}, {Hogan}, {Hogg}, {Holmgren},
  {Holtzman}, {Huang}, {Hull}, {Ichikawa}, {Ichikawa}, {Johnston}, {Kauffmann},
  {Kim}, {Kimball}, {Kinney}, {Klaene}, {Kleinman}, {Klypin}, {Knapp},
  {Korienek}, {Krolik}, {Kron}, {Krzesi{\'n}ski}, {Lamb}, {Leger},
  {Limmongkol}, {Lindenmeyer}, {Long}, {Loomis}, {Loveday}, {MacKinnon},
  {Mannery}, {Mantsch}, {Margon}, {McGehee}, {McKay}, {McLean}, {Menou},
  {Merelli}, {Mo}, {Monet}, {Nakamura}, {Narayanan}, {Nash}, {Neilsen},
  {Newman}, {Nitta}, {Odenkirchen}, {Okada}, {Okamura}, {Ostriker}, {Owen},
  {Pauls}, {Peoples}, {Peterson}, {Petravick}, {Pope}, {Pordes}, {Postman},
  {Prosapio}, {Quinn}, {Rechenmacher}, {Rivetta}, {Rix}, {Rockosi}, {Rosner},
  {Ruthmansdorfer}, {Sandford}, {Schneider}, {Scranton}, {Sekiguchi}, {Sergey},
  {Sheth}, {Shimasaku}, {Smee}, {Snedden}, {Stebbins}, {Stubbs}, {Szapudi},
  {Szkody}, {Szokoly}, {Tabachnik}, {Tsvetanov}, {Uomoto}, {Vogeley}, {Voges},
  {Waddell}, {Walterbos}, {Wang}, {Watanabe}, {Weinberg}, {White}, {White},
  {Wilhite}, {Wolfe}, {Yasuda}, {York}, {Zehavi}, \& {Zheng}}]{Stoughton02}
{Stoughton}, C. {et~al.} 2002, \aj, 123, 485

\bibitem[{{Sturm} {et~al.}(2006){Sturm}, {Hasinger}, {Lehmann}, {Mainieri},
  {Genzel}, {Lehnert}, {Lutz}, \& {Tacconi}}]{Sturm06}
{Sturm}, E., {Hasinger}, G., {Lehmann}, I., {Mainieri}, V., {Genzel}, R.,
  {Lehnert}, M.~D., {Lutz}, D., \& {Tacconi}, L.~J. 2006, \apj, 642, 81

\bibitem[{{Telfer} {et~al.}(2002){Telfer}, {Zheng}, {Kriss}, \&
  {Davidsen}}]{Telfer02}
{Telfer}, R.~C., {Zheng}, W., {Kriss}, G.~A., \& {Davidsen}, A.~F. 2002, \apj,
  565, 773

\bibitem[{{Tonry} \& {Davis}(1979)}]{Tonry79}
{Tonry}, J., \& {Davis}, M. 1979, \aj, 84, 1511

\bibitem[{{Treister} {et~al.}(2008){Treister}, {Krolik}, \&
  {Dullemond}}]{Treister08}
{Treister}, E., {Krolik}, J.~H., \& {Dullemond}, C. 2008, \apj, 679, 140

\bibitem[{{Treister} {et~al.}(2004){Treister}, {Urry}, {Chatzichristou},
  {Bauer}, {Alexander}, {Koekemoer}, {Van Duyne}, {Brandt}, {Bergeron},
  {Stern}, {Moustakas}, {Chary}, {Conselice}, {Cristiani}, \&
  {Grogin}}]{Treister04}
{Treister}, E. {et~al.} 2004, \apj, 616, 123

\bibitem[{{Treister} {et~al.}(2009){Treister}, {Virani}, {Gawiser}, {Urry},
  {Lira}, {Francke}, {Blanc}, {Cardamone}, {Damen}, {Taylor}, \&
  {Schawinski}}]{Treister09}
---. 2009, \apj, 693, 1713

\bibitem[{{Trump} {et~al.}(2006){Trump}, {Hall}, {Reichard}, {Richards},
  {Schneider}, {Vanden Berk}, {Knapp}, {Anderson}, {Fan}, {Brinkman},
  {Kleinman}, \& {Nitta}}]{Trump06}
{Trump}, J.~R. {et~al.} 2006, \apjs, 165, 1

\bibitem[{{Urrutia} {et~al.}(2009){Urrutia}, {Becker}, {White}, {Glikman},
  {Lacy}, {Hodge}, \& {Gregg}}]{Urrutia09}
{Urrutia}, T., {Becker}, R.~H., {White}, R.~L., {Glikman}, E., {Lacy}, M.,
  {Hodge}, J., \& {Gregg}, M.~D. 2009, \apj, 698, 1095

\bibitem[{{Urrutia} {et~al.}(2008){Urrutia}, {Lacy}, \& {Becker}}]{Urrutia08}
{Urrutia}, T., {Lacy}, M., \& {Becker}, R.~H. 2008, \apj, 674, 80

\bibitem[{{Urry} \& {Padovani}(1995)}]{Urry95}
{Urry}, C.~M., \& {Padovani}, P. 1995, \pasp, 107, 803

\bibitem[{{Vacca} {et~al.}(2003){Vacca}, {Cushing}, \& {Rayner}}]{Vacca03}
{Vacca}, W.~D., {Cushing}, M.~C., \& {Rayner}, J.~T. 2003, \pasp, 115, 389

\bibitem[{{Vanden Berk} {et~al.}(2001){Vanden Berk}, {Richards}, {Bauer},
  {Strauss}, {Schneider}, {Heckman}, {York}, {Hall}, {Fan}, {Knapp},
  {Anderson}, {Annis}, {Bahcall}, {Bernardi}, {Briggs}, {Brinkmann}, {Brunner},
  {Burles}, {Carey}, {Castander}, {Connolly}, {Crocker}, {Csabai}, {Doi},
  {Finkbeiner}, {Friedman}, {Frieman}, {Fukugita}, {Gunn}, {Hennessy},
  {Ivezi{\' c}}, {Kent}, {Kunszt}, {Lamb}, {Leger}, {Long}, {Loveday},
  {Lupton}, {Meiksin}, {Merelli}, {Munn}, {Newberg}, {Newcomb}, {Nichol},
  {Owen}, {Pier}, {Pope}, {Rockosi}, {Schlegel}, {Siegmund}, {Smee}, {Snir},
  {Stoughton}, {Stubbs}, {SubbaRao}, {Szalay}, {Szokoly}, {Tremonti}, {Uomoto},
  {Waddell}, {Yanny}, \& {Zheng}}]{VandenBerk01}
{Vanden Berk}, D.~E. {et~al.} 2001, \aj, 122, 549

\bibitem[{{Wall} {et~al.}(2005){Wall}, {Jackson}, {Shaver}, {Hook}, \&
  {Kellermann}}]{Wall05}
{Wall}, J.~V., {Jackson}, C.~A., {Shaver}, P.~A., {Hook}, I.~M., \&
  {Kellermann}, K.~I. 2005, \aap, 434, 133

\bibitem[{{Webster} {et~al.}(1995){Webster}, {Francis}, {Peterson},
  {Drinkwater}, \& {Masci}}]{Webster95}
{Webster}, R.~L., {Francis}, P.~J., {Peterson}, B.~A., {Drinkwater}, M.~J., \&
  {Masci}, F.~J. 1995, \nat, 375, 469

\bibitem[{{Weymann} {et~al.}(1991){Weymann}, {Morris}, {Foltz}, \&
  {Hewett}}]{Weymann91}
{Weymann}, R.~J., {Morris}, S.~L., {Foltz}, C.~B., \& {Hewett}, P.~C. 1991,
  \apj, 373, 23

\bibitem[{{White} {et~al.}(2000){White}, {Becker}, {Gregg},
  {Laurent-Muehleisen}, {Brotherton}, {Impey}, {Petry}, {Foltz}, {Chaffee},
  {Richards}, {Oegerle}, {Helfand}, {McMahon}, \& {Cabanela}}]{White00}
{White}, R.~L. {et~al.} 2000, \apjs, 126, 133

\bibitem[{{White} {et~al.}(1997){White}, {Becker}, {Helfand}, \&
  {Gregg}}]{White97}
{White}, R.~L., {Becker}, R.~H., {Helfand}, D.~J., \& {Gregg}, M.~D. 1997,
  \apj, 475, 479

\bibitem[{{White} {et~al.}(2007){White}, {Helfand}, {Becker}, {Glikman}, \& {de
  Vries}}]{White07}
{White}, R.~L., {Helfand}, D.~J., {Becker}, R.~H., {Glikman}, E., \& {de
  Vries}, W. 2007, \apj, 654, 99

\bibitem[{{White} {et~al.}(2003){White}, {Helfand}, {Becker}, {Gregg},
  {Postman}, {Lauer}, \& {Oegerle}}]{White03b}
{White}, R.~L., {Helfand}, D.~J., {Becker}, R.~H., {Gregg}, M.~D., {Postman},
  M., {Lauer}, T.~R., \& {Oegerle}, W. 2003, \aj, 126, 706

\bibitem[{{Whiting} {et~al.}(2001){Whiting}, {Webster}, \&
  {Francis}}]{Whiting01}
{Whiting}, M.~T., {Webster}, R.~L., \& {Francis}, P.~J. 2001, \mnras, 323, 718

\bibitem[{{Wright} {et~al.}(2010){Wright}, {Eisenhardt}, {Mainzer}, {Ressler},
  {Cutri}, {Jarrett}, {Kirkpatrick}, {Padgett}, {McMillan}, {Skrutskie},
  {Stanford}, {Cohen}, {Walker}, {Mather}, {Leisawitz}, {Gautier}, {McLean},
  {Benford}, {Lonsdale}, {Blain}, {Mendez}, {Irace}, {Duval}, {Liu}, {Royer},
  {Heinrichsen}, {Howard}, {Shannon}, {Kendall}, {Walsh}, {Larsen}, {Cardon},
  {Schick}, {Schwalm}, {Abid}, {Fabinsky}, {Naes}, \& {Tsai}}]{Wright10}
{Wright}, E.~L. {et~al.} 2010, \aj, 140, 1868

\bibitem[{{York} {et~al.}(2000){York}, {Adelman}, {Anderson}, {Anderson},
  {Annis}, {Bahcall}, {Bakken}, {Barkhouser}, {Bastian}, {Berman}, {Boroski},
  {Bracker}, {Briegel}, {Briggs}, {Brinkmann}, {Brunner}, {Burles}, {Carey},
  {Carr}, {Castander}, {Chen}, {Colestock}, {Connolly}, {Crocker}, {Csabai},
  {Czarapata}, {Davis}, {Doi}, {Dombeck}, {Eisenstein}, {Ellman}, {Elms},
  {Evans}, {Fan}, {Federwitz}, {Fiscelli}, {Friedman}, {Frieman}, {Fukugita},
  {Gillespie}, {Gunn}, {Gurbani}, {de Haas}, {Haldeman}, {Harris}, {Hayes},
  {Heckman}, {Hennessy}, {Hindsley}, {Holm}, {Holmgren}, {Huang}, {Hull},
  {Husby}, {Ichikawa}, {Ichikawa}, {Ivezi{\'c}}, {Kent}, {Kim}, {Kinney},
  {Klaene}, {Kleinman}, {Kleinman}, {Knapp}, {Korienek}, {Kron}, {Kunszt},
  {Lamb}, {Lee}, {Leger}, {Limmongkol}, {Lindenmeyer}, {Long}, {Loomis},
  {Loveday}, {Lucinio}, {Lupton}, {MacKinnon}, {Mannery}, {Mantsch}, {Margon},
  {McGehee}, {McKay}, {Meiksin}, {Merelli}, {Monet}, {Munn}, {Narayanan},
  {Nash}, {Neilsen}, {Neswold}, {Newberg}, {Nichol}, {Nicinski}, {Nonino},
  {Okada}, {Okamura}, {Ostriker}, {Owen}, {Pauls}, {Peoples}, {Peterson},
  {Petravick}, {Pier}, {Pope}, {Pordes}, {Prosapio}, {Rechenmacher}, {Quinn},
  {Richards}, {Richmond}, {Rivetta}, {Rockosi}, {Ruthmansdorfer}, {Sandford},
  {Schlegel}, {Schneider}, {Sekiguchi}, {Sergey}, {Shimasaku}, {Siegmund},
  {Smee}, {Smith}, {Snedden}, {Stone}, {Stoughton}, {Strauss}, {Stubbs},
  {SubbaRao}, {Szalay}, {Szapudi}, {Szokoly}, {Thakar}, {Tremonti}, {Tucker},
  {Uomoto}, {Vanden Berk}, {Vogeley}, {Waddell}, {Wang}, {Watanabe},
  {Weinberg}, {Yanny}, \& {Yasuda}}]{York00}
{York}, D.~G. {et~al.} 2000, \aj, 120, 1579

\bibitem[{{York} {et~al.}(2006){York}, {Khare}, {Vanden Berk}, {Kulkarni},
  {Crotts}, {Lauroesch}, {Richards}, {Schneider}, {Welty}, {Alsayyad}, {Kumar},
  {Lundgren}, {Shanidze}, {Smith}, {Vanlandingham}, {Baugher}, {Hall},
  {Jenkins}, {Menard}, {Rao}, {Tumlinson}, {Turnshek}, {Yip}, \&
  {Brinkmann}}]{York06}
---. 2006, \mnras, 367, 945

\bibitem[{{Yu} \& {Lu}(2005)}]{Yu05}
{Yu}, Q., \& {Lu}, Y. 2005, \apj, 620, 31

\bibitem[{{Zakamska} {et~al.}(2005){Zakamska}, {Schmidt}, {Smith}, {Strauss},
  {Krolik}, {Hall}, {Richards}, {Schneider}, {Brinkmann}, \&
  {Szokoly}}]{Zakamska05}
{Zakamska}, N.~L. {et~al.} 2005, \aj, 129, 1212

\bibitem[{{Zakamska} {et~al.}(2004){Zakamska}, {Strauss}, {Heckman}, {Ivezi{\'
  c}}, \& {Krolik}}]{Zakamska04}
{Zakamska}, N.~L., {Strauss}, M.~A., {Heckman}, T.~M., {Ivezi{\' c}}, {\v Z}.,
  \& {Krolik}, J.~H. 2004, \aj, 128, 1002

\bibitem[{{Zakamska} {et~al.}(2003){Zakamska}, {Strauss}, {Krolik}, {Collinge},
  {Hall}, {Hao}, {Heckman}, {Ivezi{\' c}}, {Richards}, {Schlegel}, {Schneider},
  {Strateva}, {Vanden Berk}, {Anderson}, \& {Brinkmann}}]{Zakamska03}
{Zakamska}, N.~L. {et~al.} 2003, \aj, 126, 2125

\bibitem[{{Zakamska} {et~al.}(2006){Zakamska}, {Strauss}, {Krolik}, {Ridgway},
  {Schmidt}, {Smith}, {Hao}, {Heckman}, \& {Schneider}}]{Zakamska06}
---. 2006, \nat, 50, 833

\end{thebibliography}
